\documentclass[showpacs,preprint,superscriptaddress,amsmath,amssymb]{revtex4-1}
\usepackage{graphicx}% Include figure files
\usepackage{dcolumn}% Align table columns on decimal point
\usepackage{bm}% bold math
\usepackage{epsfig}
\usepackage{subfigure}
\usepackage{graphics}
\usepackage{amssymb}
\usepackage{amsthm}
\usepackage{amsmath}
\usepackage{array}
\usepackage{color}

\begin{document}

\preprint{APS/123-QED}

r\title{An efficient three-dimensional multiple-relaxation-time lattice Boltzmann model for multiphase flows}

\author{H. Liang}
 \affiliation{Department of Physics, Hangzhou Dianzi University - Hangzhou 310018, China}
\author{B. C. Shi}
\email[Email:~]{shibc@hust.edu.cn. This work has been reported in $23^{\mathrm{rd}}$ DSFD, July 28-August 1, 2014, Pairs.}
\affiliation{School of Mathematics and Statistics, Huazhong University of Science and Technology, Wuhan, 430074, China}%
\author{Z. H. Chai}
\affiliation{School of Mathematics and Statistics, Huazhong University of Science and Technology, Wuhan, 430074, China}%
\date{\today}% It is always \today, today,

%%%%% Begin Abstract %%%%%%%%%%%
\begin{abstract}
In this paper, an efficient three-dimensional lattice
Boltzmann (LB) model with multiple-relaxation-time (MRT) collision operator is
developed for the simulation of multiphase
flows. This model is an extension of our previous two-dimensional
model (H. Liang, B. C. Shi, Z. L. Guo, and Z. H. Chai, Phys. Rev. E.
89, 053320 (2014)) to the three dimensions using the D3Q7 (seven discrete velocities in
three dimensions) lattice for the Chan-Hilliard equation (CHE) and the D3Q15 lattice for the
Navier-Stokes equations (NSEs). Due to the smaller lattice-velocity numbers used,
the computional efficiency can be significantly improved in simulating real three-dimensional flows,
and simultaneously the present model can recover to the CHE and NSEs correctly through the chapman-Enskog procedure. We compare
the present MRT model with the single-relaxation-time model and the previous three-dimensional LB model using
two benchmark interface-tracking problems, and numerical results show that the present MRT model can achieve a
significant improvement in the accuracy and stability of the interface capturing. The developed model is also able
to deal with multiphase fluids with very low viscosities due to the using of the MRT collision model,
which is demonstrated by the simulation of the classical Rayleigh-Taylor instability at various Reynolds numbers.
The maximum Reynolds number considered in this work reaches up to $4000$, which is larger than those of almost previous simulations.
It is found that the instabilty induces a more complex structure of the interface at a high Reynolds number.

\end{abstract}
%%%%% end %%%%%%%%%%%
% PACS
\maketitle

\section{Introduction}
In the pase two decades, the lattice Boltzmann (LB) method has received
great success in modeling various fluid systems, and the capability in simulating multiphase
flows can be recognized as its unique advantage that distinguishes from the traditional
numerical methods~\cite{Guo1}. One of most fundamental problems in LB method for multiphase flows is how to
describe the interfacial dynamics, which is a natural consequence of intermolecular interactions
between different phases. Up to now, based on different physical pictures
of the interactions, several types of LB multiphase models has been established, which are
substantially divided into four categories: the colour model~\cite{Gunstensen}, the
pseudo-potential model~\cite{Shan1, Shan2}, the free-energy model~\cite{Swift1, Swift2} and the phase-field based model
~\cite{He1, Lee1, Zhengl, Zu, Liang1}. In most of these LB multiphase models~\cite{Gunstensen, Shan1, Shan2, Swift1, Swift2},
the interface is not tracked explicitly, and the region with non-zero density gradient
is identified as the interface. Therefore the physics of interface tracking equation
is unknown. Fortunately, the phase-field theory provides a firm foundation on the interface physics, in
which the interface is tracked by an order parameter that mimics the Cahn-Hilliard equation (CHE)~\cite{Jacqmin, Ding},
\begin{equation}
{{\partial \phi } \over {\partial t}} + \nabla  \cdot \phi {\bf{u}}
= \nabla  \cdot {M}(\nabla {\mu} ),
\end{equation}
where $\phi$ represents an order variable, $M$ is the mobility coefficient, $\mu$ is the chemical potential and a function of $\phi$,
\begin{equation}
\mu  = 4\beta \phi (\phi - 1)(\phi + 1)-k{\nabla ^2}\phi,
\end{equation}
where $\beta$ and $k$ are related to the interface thickness $D$ and surface tension $\sigma$ by the relationships
$k=3D\sigma/8$ and $\beta=3\sigma/4D$ .
 $\mathbf{u}$ in Eq.(1) is the fluid
velocity and governed by the incompressible Navier-Stokes equations~\cite{Jacqmin, Ding}
\begin{subequations}
\begin{equation}
\nabla  \cdot {\bf{u}} = 0,
\end{equation}
\begin{equation}
 \rho ({{\partial {\bf{u}}} \over
{\partial t}} + {\bf{u}} \cdot \nabla {\bf{u}}) =  - \nabla p +
\nabla  \cdot \left[ {\nu\rho(\nabla {\bf{u}} + \nabla
{{\bf{u}}^T})} \right] + {{\bf{F}}_s} + {\bf{G}},
\end{equation}
\end{subequations}
where $\rho$ is the fluid density, $p$ is the pressure, $\nu$ is the kinematic
viscosity, $\textbf{F}_s$ is the surface tension, and $\textbf{G}$ is the external force.

Some researchers have constructed some LB multiphase models based on the phase-field theory, where the interface is needed
to be tracked explicitly by an index or order distribution function~\cite{He1, Zheng1, Zu, Liang1}. He {\it et al.} \cite{He1} proposed a LB model for incompressible multiphase flows, in which they adopted an index distribution function to track the interface and a pressure distribution function for solving the flow field.
Based on this model, they successfully simulated the two-dimensional Rayleigh-Taylor instability, and later the three-dimensional case using
D3Q15 lattice structure in LB equations for both the interface capturing and flow field~\cite{He2}. Although this model is rather robust, it suffers
from some limitations, one of which is that the recovered interface equation is inconsistent with the CHE noticed by Zheng {\it et al.}~\cite{Zheng1, Zheng2}.
To this end, they developed a LB model for the CHE, in which a source term on a spatial difference of the distribution function is introduced~\cite{Zheng2}. The model is subsequently extended to the three-dimension using D3Q7 lattice model~\cite{Zheng3}. Recently, Zu {\it et al.}~\cite{Zu} introduced another similar LB model for the CHE where a spatial difference term on the equilibrium distribution function was included. They also modified the equilibrium distribution function in LB equation for flow field such that the continuity equation (3a) can be derived. However, the computation of the macroscopic pressure and velocity in their scheme is implicit. More recently, we proposed a novel LB model for two-dimensional multiphase flows~\cite{Liang1}. On one hand, A simpler time-dependent source term is incorporated in LB equation for the interface capturing. As a result, the two-dimensional CHE can be recovered correctly. On the other hand, a equilibrium distribution function is delicately designed for flow field to derive the correct continuity equation while the hydrodynamic properties can be computed explicitly. This improved model is also extended to study axisymmetric multiphase flows~\cite{Liang2}

As a continuing work, in this paper an efficient three-dimensional LB model for incompressible
multiphase flow systems is developed based on the multiple-relaxation-time (MRT) method.
This model has some distinct advantages. Firstly, the model for the CHE requires only seven discrete velocities in
three dimensions (D3Q7), therefore the expenditure in data storage and computational time is smaller than that of other models
using D3Q15 lattice~\cite{He2, Zu}. Secondly, the MRT collision model is adopted, which has a better accuracy and stability than the
single-relaxation-time (SRT) model used commonly in other LB multiphase models~\cite{He2, Lee1, Zhengl, Zheng1, Zheng1, Zheng3, Zu}.
Finally, the present model is able to deal with fluid flows at a large Peclet number or a high Reynolds number.
The rest of this paper is organized as follows. Sec.~\ref{sec:method} presents our three-dimensional MRT LB model for multiphase flows. The model then is verified by several classical numerical experiments in Sec.~\ref{sec:Results}. Finally, we made a brief summary in Sec.~\ref{sec: sum}.

%%%%%%%%%%%%%%%%%%%%%%%%%%%%%%%%%%%%%%%%%%%%%%%%%%%%%%%%%%%%%%%%%%%%%%%%%%%%%%%%%%%%%%%%
%%
%% Model
%%
%%%%%%%%%%%%%%%%%%%%%%%%%%%%%%%%%%%%%%%%%%%%%%%%%%%%%%%%%%%%%%%%%%%%%%%%%%%%%%%%%%%%%%%%
\section{Three-dimensional MRT LB MODEL FOR MULTIPHASE FLOWS}\label{sec:method}
\subsection{Three-dimensional MRT LB model for the Cahn-Hilliard equation}
The LB equation with a MRT collision model for the CHE can be written as,
%Eq. (9)
\begin{equation}
{f_i}({\bf{x}} + {{\bf{c}}_i}\delta t,t + \delta t) -
{f_i}({\bf{x}},t) =  - ({{\bf{M}}^{ - 1}}{{\bf{S}}^f}{\bf{M}})_{ij}\left[ {{f_j}({\bf{x}},t) -
f_j^{eq}({\bf{x}},t)} \right] + {\delta_t}{F_i}({\bf{x}},t),
\end{equation}
where ${f_i}({\bf{x}},t)$ is the order distribution function used to track the interface, $f_i^{eq}({\bf{x}},t)$ is the equilibrium distribution
function defined as~\cite{Liang1, Shi1}
\begin{equation}
f_i^{eq}({\bf{x}},t) =\left\{
\begin{array}{ll}
 \phi  + ({\omega_i} - 1)\eta \mu,                                       & \textrm{ $i=0$}   \\
 {\omega_i}\eta\mu  + {\omega_i}{{{\textbf{c}_i} \cdot \phi {\bf{u}}} \over {c_s^2}}, & \textrm{ $i\neq0$},
\end{array}
\right.
\end{equation}
where the discrete velocities $\textbf{c}_i$, the weighting coefficients $\omega_i$  and the sound speed $c_s$ depend on the choice of the discrete-velocity model, $\eta$~is a parameter related to the mobility. In this work, an efficient D3Q7 discrete-velocity model is adopted for the CHE, and $\textbf{c}_i$ then can be given by
\begin{equation}
\mathbf{c}_{i}= c\left\{
\begin{array}{ccccccccccccccc}
0 & 1 & -1 & 0 & 0 & 0 & 0  \\
0 & 0 & 0 & 1 & -1 & 0 & 0  \\
0 & 0 & 0 & 0 & 0 & 1 & -1
\end{array} \right\},
\end{equation}
and further according to the following equations ($\mathbf{I}$ is the unit matrix)
\begin{equation}
\sum\limits_{i}{\omega_i}=1,~~\sum\limits_{i}{\omega_i}\mathbf{c}_i=\mathbf{0},~~\sum\limits_{i}{\omega_i}\mathbf{c}_i \mathbf{c}_i=c_s^2\mathbf{I},
\end{equation}
$\omega_i$ can be obtained as
\begin{equation}
\omega_0=1-3A,~~\omega_{1-6}=\frac{A}{2},
\end{equation}
where $A$ is a free parameter, $c_s^2=A c^2$. To ensure the positive weighting coefficients, the parameter $A$ should satisfy $0<A<\frac{1}{3}$.
The transformation matrix $\mathbf{M}$ of the D3Q7 model is defined by~\cite{Yoshida}
\begin{equation}
\bf{M}= \left(
\begin{array}{rrrrrrrrr}
  1 & 1 & 1 & 1 & 1 & 1 & 1  \\
  0 & 1 & -1 & 0 & 0 & 0 & 0  \\
  0 & 0 &  0 & 1 & -1 & 0 & 0  \\
  0 & 0 &  0 & 0 & 0  & 1 & -1 \\
  6 & -1 & -1 & -1 & -1 & -1 & -1 \\
  0 & 2 & 2 & -1 & -1 & -1 & -1 \\
  0 & 0 & 0 & 1 & 1 & -1 & -1 \\
\end{array}\right),
\end{equation}
which is constructed based on the polynomial set of the discrete velocities. ${\bf{S}}^f$ in Eq. (1) is a diagonal relaxation matrix,
\begin{equation}
{{\bf{S}}^f} =diag
(s_0,s_1,...,s_6),
\end{equation}
where $0<s_i<2$, and if the parameters $s_i$ equal to each other, the MRT model can reduce to the SRT model. To recover the correct CHE, the source term $F_i$ in Eq. (9) should be defined as
\begin{equation}
F_i = [{{\bf{M}}^{-1}}({\bf{I}} - {{\bf{S}}^f\over 2}){\bf{M}}]_{ij}R_j,
\end{equation}
where ${{R}_i}$ is given by~\cite{Liang1}
\begin{equation}
{{R}_i} = {{{\omega_i}{{\bf{c}}_i} \cdot {\partial _t}\phi
{\bf{u}}} \over {c_s^2}}.
\end{equation}
In the present model, the order parameter is computed by
\begin{equation}
\phi  = \sum\limits_i {{f_i}},
\end{equation}
and the density is taken as a linear function of the order parameter,
\begin{equation}
\rho  = {{1+\phi} \over {2}}{\rho
_l}  + {{1-\phi} \over {2}} {\rho _g},
\end{equation}
where $\rho_l$ and $\rho_g$ represent the densities of liquid and gas phases, respectively.

The evolution of LB equation (4) can be commonly divided into two steps, i.e., the collision process,
\begin{equation}
f_i^{+}={f_i}({\bf{x}},t) - ({{\bf{M}}^{ - 1}}{{\bf{S}}^f}{\bf{M}})_{ij}\left[ {{f_j}({\bf{x}},t) -
f_j^{eq}({\bf{x}},t)} \right] + {\delta_t}{F_i}({\bf{x}},t),
\end{equation}
and the propagation process,
\begin{equation}
{f_i}({\bf{x}} + {{\bf{c}}_i}\delta t,t + \delta t)=f_i^{+}.
\end{equation}
To reduce the matrix operations, it is wise that the collision process of MRT model is implemented in the moment space. By premultiplying
the transformation matrix, we can easily derive the equilibrium distribution function in moment space,
\begin{equation}
{\bf{mf}}^{eq}=( \phi,\frac{\phi u_x}{c},\frac{\phi u_y}{c}, \frac{\phi u_z}{c}, 6\phi-21A\eta\mu, 0 , 0)^{\rm T},
\end{equation}
where $u_x$, $u_y$ and $u_z$ are the x-, y- and z- components of macroscopic velocity $\mathbf{u}$, respectively. Similarly, the source term
$R_i$ in the moment space can be presented as
\begin{equation}
{\bf{mR}}=( 0, \frac{{\partial _t}{\phi u_x}}{c}, \frac{{\partial _t}{\phi u_y}}{c}, \frac{{\partial _t}{\phi u_z}}{c}, 0, 0, 0)^{\rm T}
\end{equation}

The Chapman-Enskog analysis is carried out on the LB evolution equation (4), and the results demonstrate that the CHE can be
derived correctly from the present MRT model, and the relationship between the mobility $M$ and the relaxation parameter is also derived as
\begin{equation}
 {M} =\eta c_s^2(\tau_f-0.5)\delta t,
\end{equation}
where $\tau_f =1/{s_1}$, and $s_1={s_2}={s_3}$.

%%%%%%%%%%%%%%%%%%%%%%%%%%%%%%%%%%%%%%%%%%%%%%%%%%%%%%%%%%%%%%%%%%%%%%%%%%%%%%%%%%%%%%%%
%%%%%%%%%%%%%%%%%%%%%%%%%%%%%%%%%%%%%%%%%%%%%%%%%%%%%%%%%%%%%%%%%%%%%%%%%%%%%%%%%%%%%%%%

\subsection{Three-dimensional MRT LB model for the Navier-Stokes equations}

The MRT LB equation with a source term for the NSEs reads as~\cite{Guo2}
\begin{equation}
{g}_i({\bf{x}} +{{\bf{c}}_i}{\delta _t},t + {\delta _t}) -
{g}_i({\bf{x}},t) =  - (\mathbf{\Gamma}^{-
1}{\mathbf{S}^g}{\mathbf{\Gamma}})_{ij}[{{g}_j({\bf{x}},t) -
g_j^{eq}({\bf{x}},t)}]+ {\delta _t}G_i,
\end{equation}
where $g_i$ is the density distribution function, $g_i^{eq}$ is the equilibrium distribution function and is defined by
\begin{equation}
 {g_i}^{eq}=\left\{
\begin{array}{ll}
{p \over {c_s^2}}({\omega _i} - 1) + \rho{s_i}({\bf{u}})              & \textrm{ $i=0$}    \\
{p \over {c_s^2}}{\omega _i} + \rho{s_i}({\bf{u}})                    & \textrm{ $i\neq0$} \\
\end{array}
\right.
\end{equation}
with
\begin{equation}
{s_i}({\bf{u}}) = {\omega _i}\left[
{{{{{\bf{c}}_i} \cdot {\bf{u}}} \over {c_s^2}} + {{{{({{\bf{c}}_i}
\cdot {\bf{u}})}^2}} \over {2c_s^4}} - {{{\bf{u}} \cdot {\bf{u}}}
\over {2c_s^2}}} \right].
\end{equation}
To simulate the fluid flows in three dimension, we can choice several types of lattice velocity models, such as D3Q15 or D3Q19~\cite{Qian1}. The
D3Q15 lattice model is used in this work due to its smaller data storage and higher computational efficiency. Following the work of Qian $et~al.$~\cite{Qian1},
the discrete velocities $\mathbf{c}_{i}$ of the D3Q15 lattice model can be given by
\begin{equation}
\mathbf{c}_{i}= c\left\{
\begin{array}{ccccccccccccccc}
0 & 1 & -1 & 0 & 0 & 0 & 0 & 1 & -1 & 1 & -1 & 1 & -1 & 1 & -1 \\
0 & 0 & 0 & 1 & -1 & 0 & 0 & 1 & -1 & 1 & -1 & -1 & 1 & -1 & 1\\
0 & 0 & 0 & 0 & 0 & 1 & -1 & 1 & -1 & -1 & 1 & 1 & -1 & -1 & 1
\end{array} \right\},
\end{equation}
where $c=\sqrt{3}c_s$. According to the ordering of $\mathbf{c}_{i}$, the weight coefficients are presented as
\begin{equation}
\omega_{0}=\frac{2}{9},~~\omega_{1-6}=\frac{1}{9},~~\omega_{7-14}=\frac{1}{72}.
\end{equation}
The transformation matrix $\mathbf{\Gamma}$ in Eq. (20) can be given by~\cite{Humieres}
\begin{equation}
\mathbf{\Gamma}= \left(
\begin{array}{rrrrrrrrrrrrrrr}
  1 &  1 &  1 &  1 &  1 &  1 &  1 & 1 & 1 &  1 &  1 & 1 & 1 &  1 &  1 \\
 -2 & -1 & -1 & -1 & -1 & -1 & -1 & 1 & 1 &  1 &  1 & 1 & 1 &  1 &  1 \\
 16 & -4 & -4 & -4 & -4 & -4 & -4 & 1 & 1 &  1 &  1 & 1 & 1 &  1 &  1 \\
  0 &  1 & -1 &  0 &  0 &  0 &  0 & 1 &-1 &  1 & -1 & 1 &-1 &  1 & -1 \\
  0 & -4 &  4 &  0 &  0 &  0 &  0 & 1 &-1 &  1 & -1 & 1 &-1 &  1 & -1 \\
  0 &  0 &  0 &  1 & -1 &  0 &  0 & 1 & 1 & -1 & -1 & 1 & 1 & -1 & -1 \\
  0 &  0 &  0 & -4 &  4 &  0 &  0 & 1 & 1 & -1 & -1 & 1 & 1 & -1 & -1 \\
  0 &  0 &  0 &  0 &  0 &  1 & -1 & 1 & 1 &  1 &  1 &-1 &-1 & -1 & -1 \\
  0 &  0 &  0 &  0 &  0 & -4 &  4 & 1 & 1 &  1 &  1 &-1 &-1 & -1 & -1 \\
  0 &  2 &  2 & -1 & -1 & -1 & -1 & 0 & 0 &  0 &  0 & 0 & 0 &  0 &  0 \\
  0 &  0 &  0 &  1 &  1 & -1 & -1 & 0 & 0 &  0 &  0 & 0 & 0 &  0 &  0 \\
  0 &  0 &  0 &  0 &  0 &  0 &  0 & 1 &-1 & -1 &  1 & 1 &-1 & -1 &  1 \\
  0 &  0 &  0 &  0 &  0 &  0 &  0 & 1 & 1 & -1 & -1 &-1 &-1 &  1 &  1 \\
  0 &  0 &  0 &  0 &  0 &  0 &  0 & 1 &-1 &  1 & -1 &-1 & 1 & -1 &  1 \\
  0 &  0 &  0 &  0 &  0 &  0 &  0 & 1 &-1 & -1 &  1 &-1 & 1 &  1 & -1
\end{array}\right),
\end{equation}
and the corresponding diagonal relaxation matrix $\mathbf{S}^g$ is denoted by
\begin{equation}
\mathbf{S}^g =diag(\lambda_0,\lambda_1,...,\lambda_{14}),
\end{equation}
where~$0<\lambda_i<2$. The source term $G_i$ in Eq. is defined as~\cite{Liang1}
\begin{equation}
G_i={[{{\mathbf{\Gamma}}^{ -1}}({\rm
\mathbf{I}} - {\mathbf{S}^g \over
2})\mathbf{\Gamma}}]_{ij}{T_j}
\end{equation}
where
\begin{equation}
{{T}_i}={{({{\bf{c}}_i} - {\bf{u}})} \over {c_s^2}} \cdot \left[
{{s_i}({\bf{u}})\nabla (\rho c_s^2)  +
({{\bf{F}}_s} + {{\bf{F}}_a} + {\bf{G}})({s _i}({\bf{u}})+\omega_i)}
\right],
\end{equation}
in which ${{\bf{F}}_a} = {0.5({\rho _A} - {\rho _B})}{M}\nabla^2\mu {\bf{u}}$ is an
interfacial force, and $\mathbf{F}_s$ is the surface tension taken the potential form $\mathbf{F}_s=\mu\nabla\phi$.
In the present model, the macroscopic pressure and velocity can be obtained from
\begin{equation}
{\bf{u}} = {{\sum\limits_i {{{\bf{c}}_i}\bar{g_i}}  + 0.5{\delta
_t}({{\bf{F}}_s} + {\bf{G}})} \over {\rho  - {{0.25({\rho _A} - {\rho
_B}){M}\nabla^2\mu }}}},
\end{equation}
\begin{equation}
 p = {{c_s^2} \over {(1 - {\omega
_0})}}\left[ {\sum\limits_{i \ne 0} {\bar{g_i}}  + {{{\delta _t}}
\over 2}{\bf{u}} \cdot \nabla \rho + \rho {s_0}(\textbf{u})}
\right].
\end{equation}

The collision process of MRT LB equation for NSEs is also carried out in the moment space. After some algebraic manipulations, the equilibrium distribution function $g_i^{eq}$ and source term $T_i$ in
the moment space can be respectively derived as
\begin{eqnarray}
{\bf{mg}}^{eq}=( 0, \frac{3p+\rho\mathbf{u}^2}{c^2}, -\frac{45p+5\rho\mathbf{u}^2}{c},\frac{\rho{u_x}}{c},-\frac{7\rho{u_x}}{3c},\frac{\rho{u_y}}{c},-\frac{7\rho{u_y}}{3c},\frac{\rho{u_z}}{c},-\frac{7\rho{u_z}}{3c}, \nonumber\\ \rho\frac{2{u_x}^2-{u_y}^2-{u_z}^2}{c^2}, \rho\frac{{u_y}^2-{u_z}^2}{c^2},\frac{{\rho}{u_x}{u_y}}{c^2},\frac{{\rho}{u_y}{u_z}}{c^2},\frac{{\rho}{u_x}{u_z}}{c^2},0)^{\rm T},
\end{eqnarray}
and
\begin{eqnarray}
{\bf{mT}}=[\mathbf{u}\cdot\nabla{\rho},\frac{\mathbf{u}\cdot(-\nabla{\rho}c_s^2+2\mathbf{F})}{c^2},-\frac{\mathbf{u}\cdot(7\nabla{\rho{c_s^2}}+10\mathbf{F})}{c^2},\frac{F_x}{c}, -\frac{7F_x}{3c},\frac{F_y}{c}, -\frac{7F_y}{3c},\frac{F_z}{c}, -\frac{7F_z}{3c}, \nonumber\\\frac{4u_x({\partial _x}{\rho}c_s^2+F_x)-2u_y({\partial _y}\rho{c_s^2}+F_y)-2u_z({\partial _z}\rho{c_s^2}+F_z)}{c^2}, \frac{2{u_y}({\partial _y}\rho{c_s^2}+F_y)-2{u_z}({\partial _z}\rho{c_s^2}+F_z)}{c^2},\nonumber\\ \frac{{u_x}({\partial _y}{\rho{c_s^2}+F_y})+{u_y}({\partial _x}{\rho{c_s^2}+F_x})}{c^2},\frac{{u_y}({\partial _z}{\rho{c_s^2}+F_z})+{u_z}({\partial _y}{\rho{c_s^2}+F_y})}{c^2},\nonumber\\ \frac{{u_x}({\partial _z}{\rho{c_s^2}+F_z})+{u_z}({\partial _x}{\rho{c_s^2}+F_x})}{c^2},0]^{\rm T}\nonumber\\
\end{eqnarray}

We also conduct the Chapman-Enskog analysis on LB Equation (20), and the results show that the present model can correctly recover to the
NSEs with the kinematic viscosity determined by
\begin{equation}
\nu  =  c_s^2(\tau_g - 0.5){\delta _t},
\end{equation}
where $\tau_g=1/{\lambda_9}$, and $\lambda_9=\lambda_{10}=\lambda_{11}=\lambda_{12}=\lambda_{13}$.

In practical applications, the derivative terms in present model should be discretized by suitable difference schemes. As widely adopted in the
references~\cite{Shi1, Shi2}, an explicit difference scheme
\begin{equation}
\partial_t\chi({\bf{x}},t)=\frac{\chi({\bf{x}},t)-\chi({\bf{x}},t-\delta_t)}{\delta_t}
\end{equation}
is used for computing the time derivative in Eq. (12), and the isotropic central schemes~\cite{Lou}
\begin{equation}
 \nabla \chi({\bf{x}},t)=\sum\limits_{i \ne 0}
{\frac{\omega_i\textbf{c}_i\chi({\bf{x}} +{{\bf{c}}_i}{\delta
_t},t)}{c_s^2 \delta_t}}
\end{equation}
and
\begin{equation}
\nabla^2\chi({\bf{x}},t)=\sum\limits_{i \ne 0}
{\frac{2\omega_i\textbf{c}_i[\chi({\bf{x}} +{{\bf{c}}_i}{\delta
_t},t)-\chi({\bf{x}},t)]}{c_s^2 \delta_t^2}}
\end{equation}
are employed for calculating the gradient and the Laplacian operator, respectively. In above equations, $\chi$ represents an arbitrary function.
It should be noted that the schemes (35) and (36) not only can preserve a secondary-order accuracy in space, but also can ensure the global
mass conservation of a multiphase system~\cite{Lou}.

%%%%%%%%%%%%%%%%%%%%%%%%%%%%%%%%%%%%%%%%%%%%%%%%%%%%%%%%%%%%%%%%%%%%%%%%%%%%%%%%%%%%%%%%
%%
%% Section 3 Numerical tests for interface-capturing LB MODEL
%%
%%%%%%%%%%%%%%%%%%%%%%%%%%%%%%%%%%%%%%%%%%%%%%%%%%%%%%%%%%%%%%%%%%%%%%%%%%%%%%%%%%%%%%%%
\section{Numerical Results and discussions}\label{sec:Results}
In this section, we will validate the present three-dimensional MRT LB model with several numerical examples.
We first test the performance of the three-dimensional LB model in interface capturing by simulating two classical benchmark problems:
rotation of the Zalesak's sphere~\cite{Enright} and deformation field flow~\cite{Leveque}. In these simulations, the evolution equation (4) is only adopted
in that the velocity distribution has been specified in advance. Next, we simulated the three-dimensional Rayleigh-Taylor instability
to show the ability of the present model for multiphase flows, where a comparison between the numerical results and
some available results is also conducted.

\begin{figure}
\centering
\includegraphics[width=1.2in,height=1.212in]{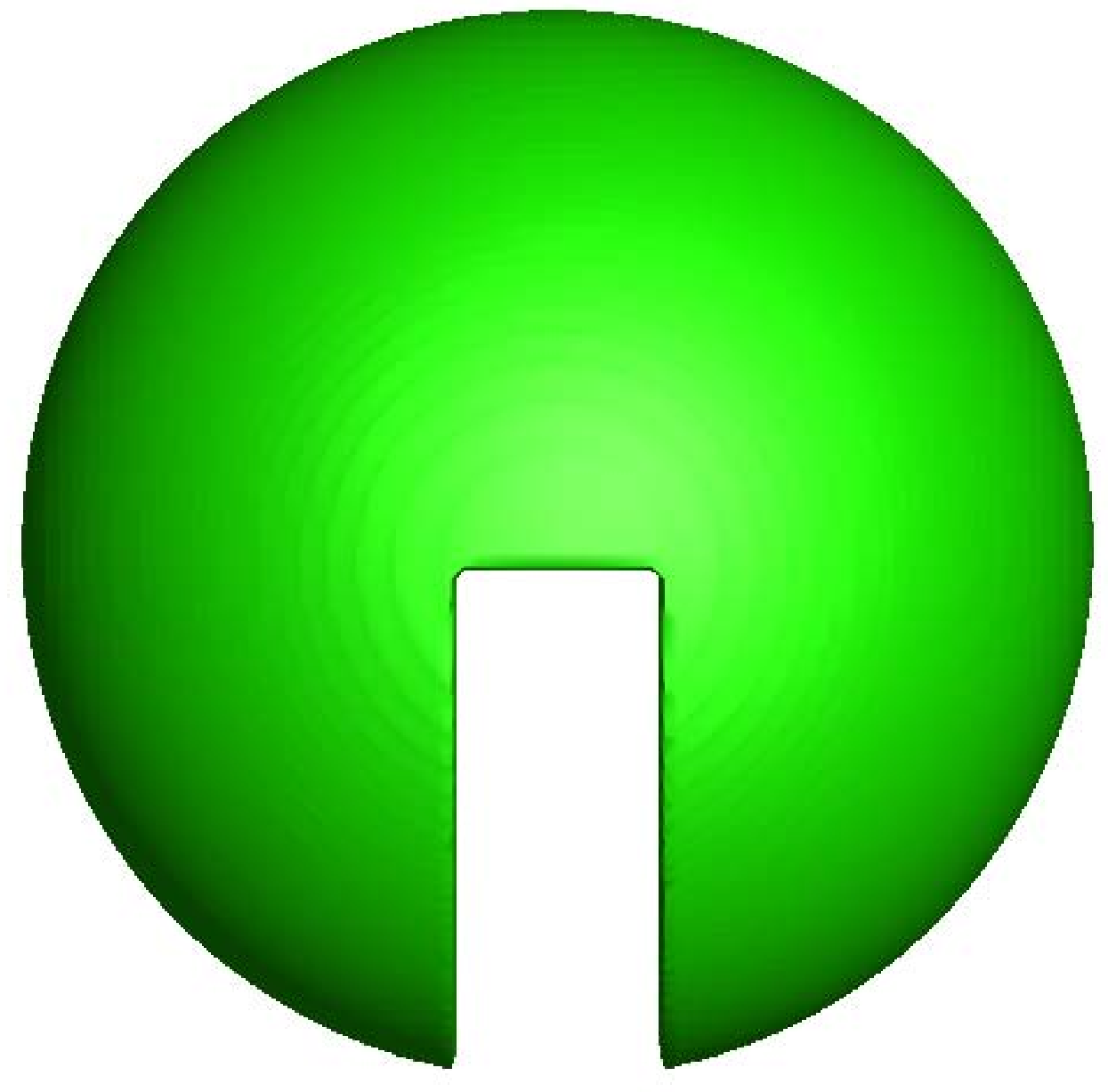}
\includegraphics[width=1.2in,height=1.212in]{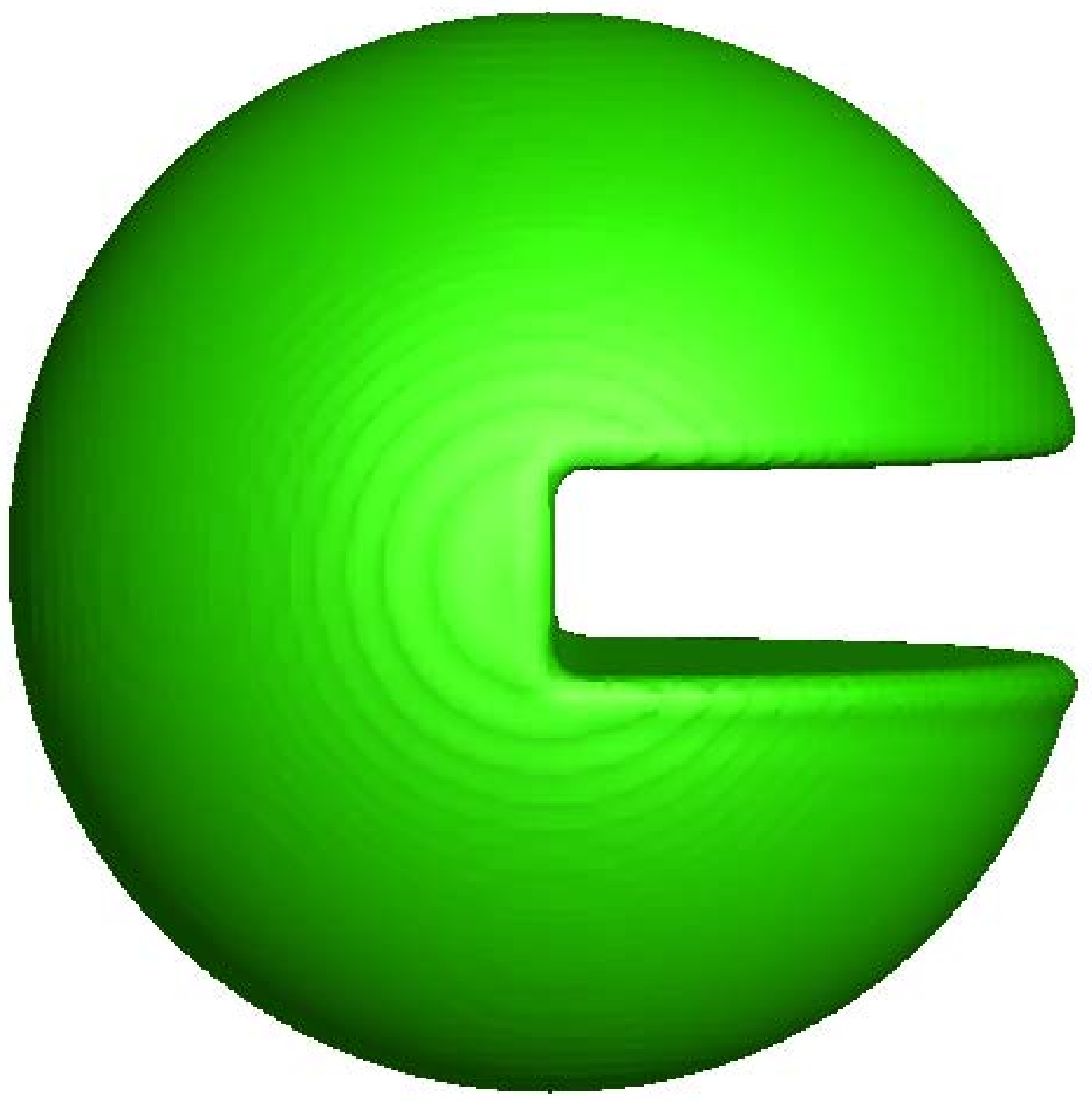}
\includegraphics[width=1.2in,height=1.212in]{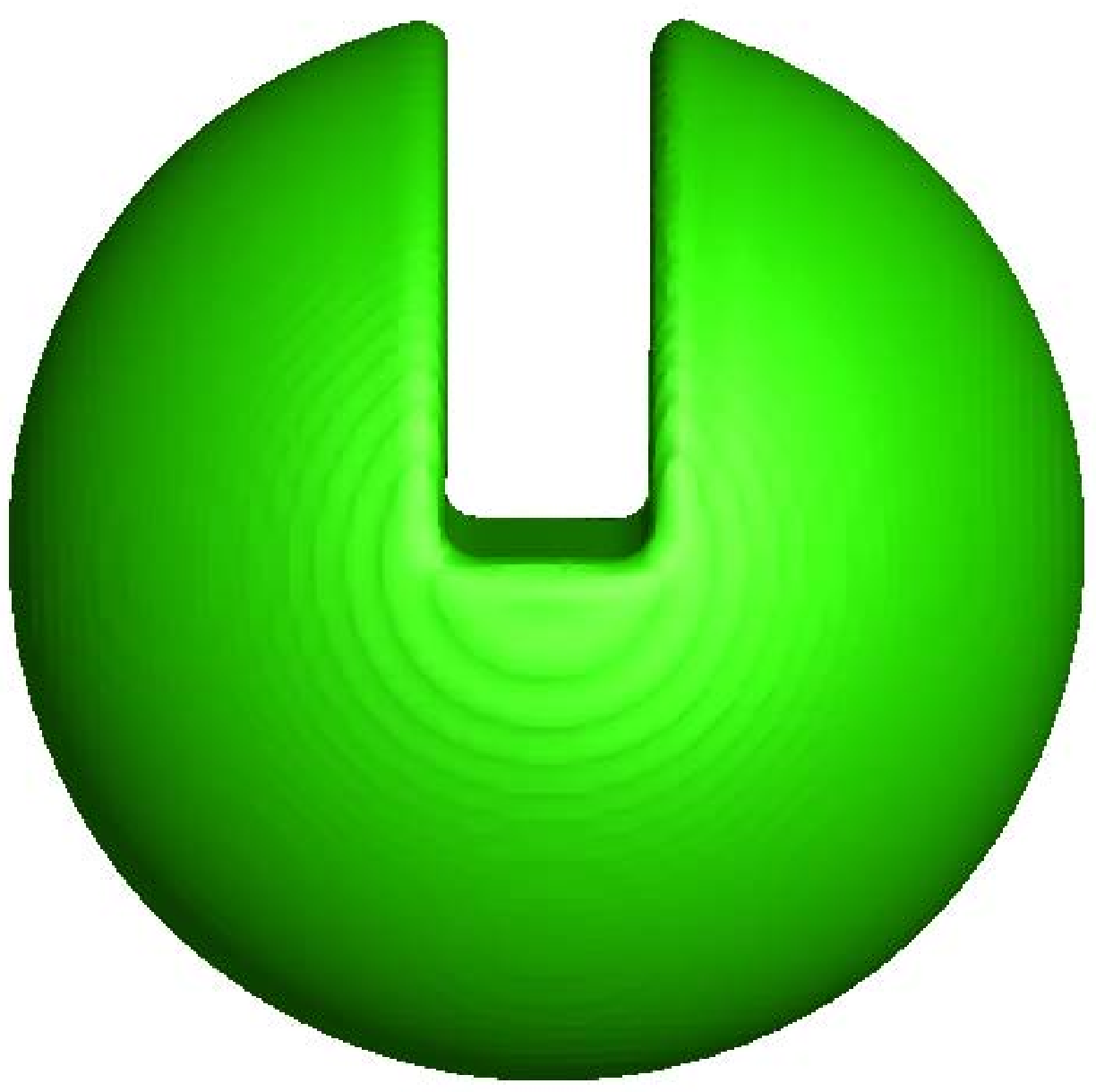}
\includegraphics[width=1.2in,height=1.212in]{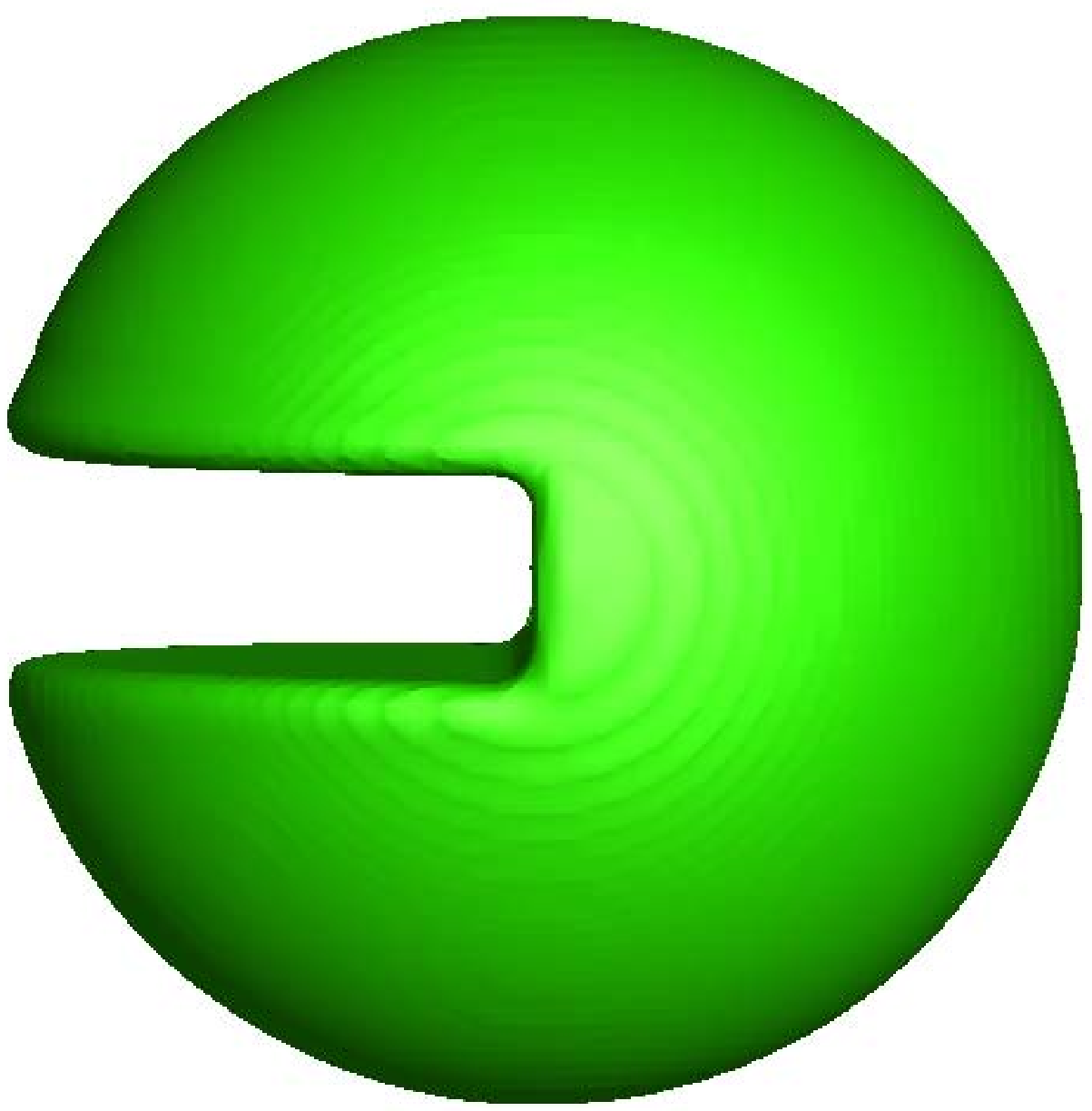}
\includegraphics[width=1.2in,height=1.212in]{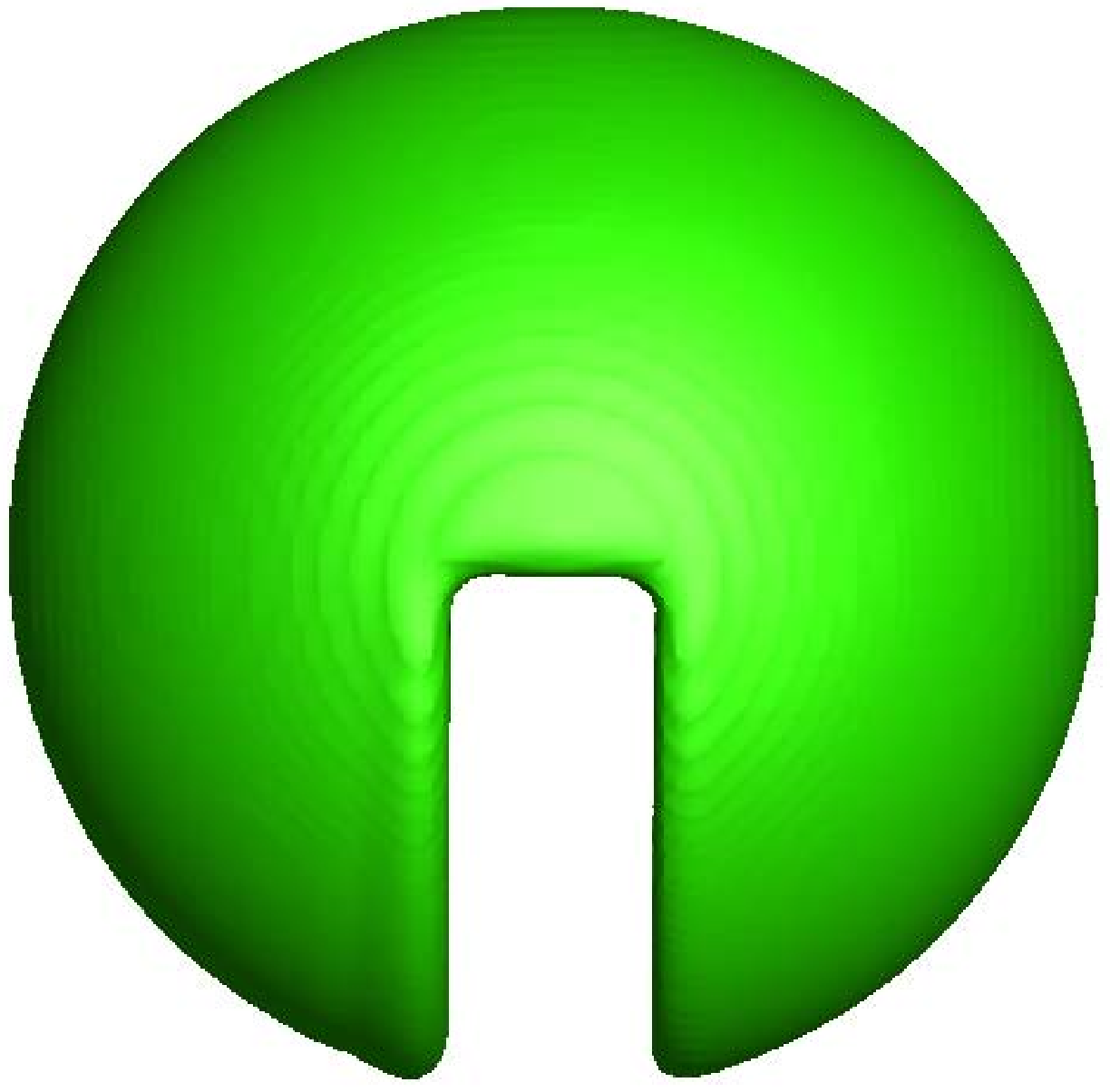}\\
\hspace{10pt}(a)\hspace{180pt}\\
\includegraphics[width=1.2in,height=1.212in]{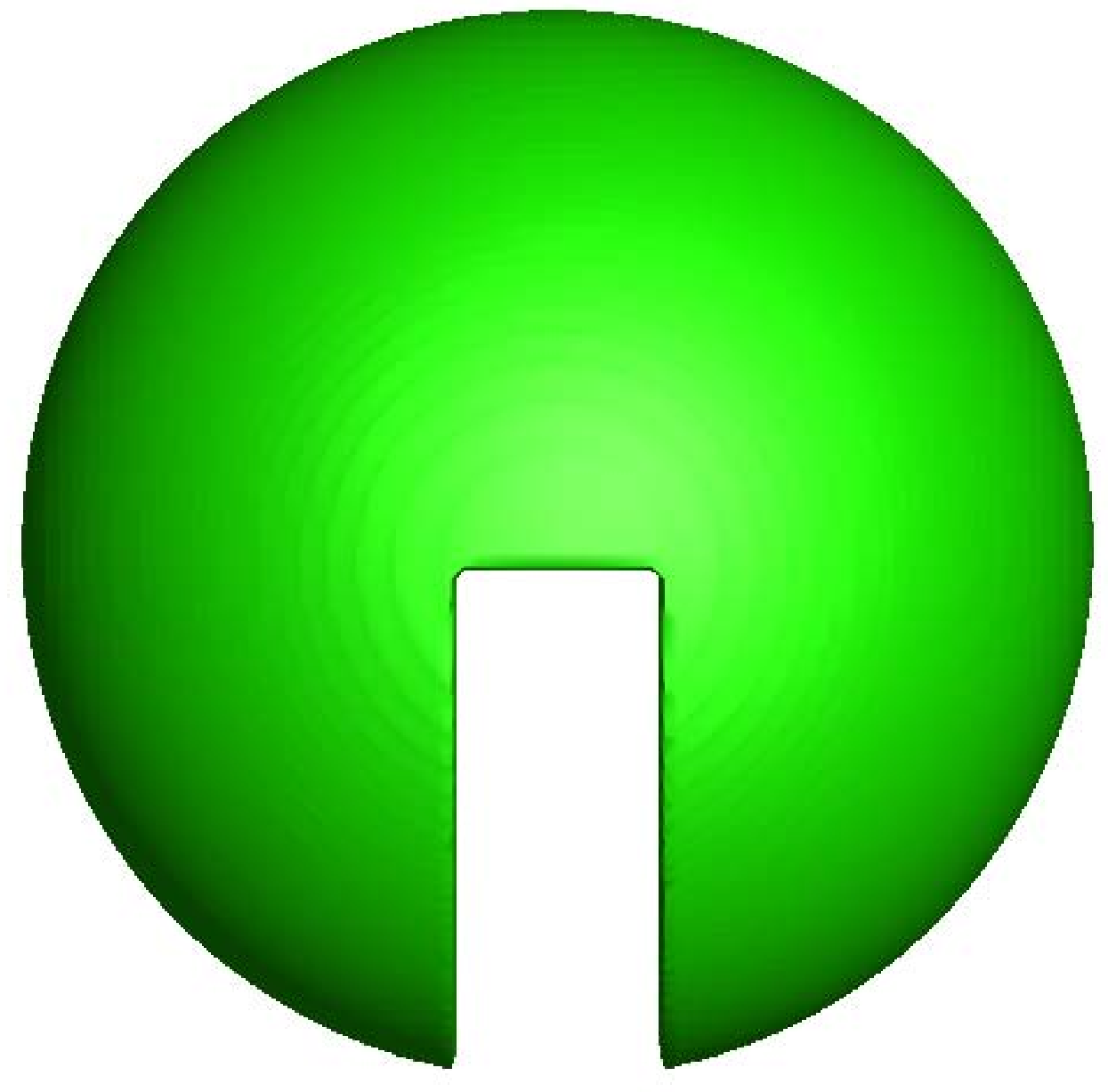}
\includegraphics[width=1.2in,height=1.212in]{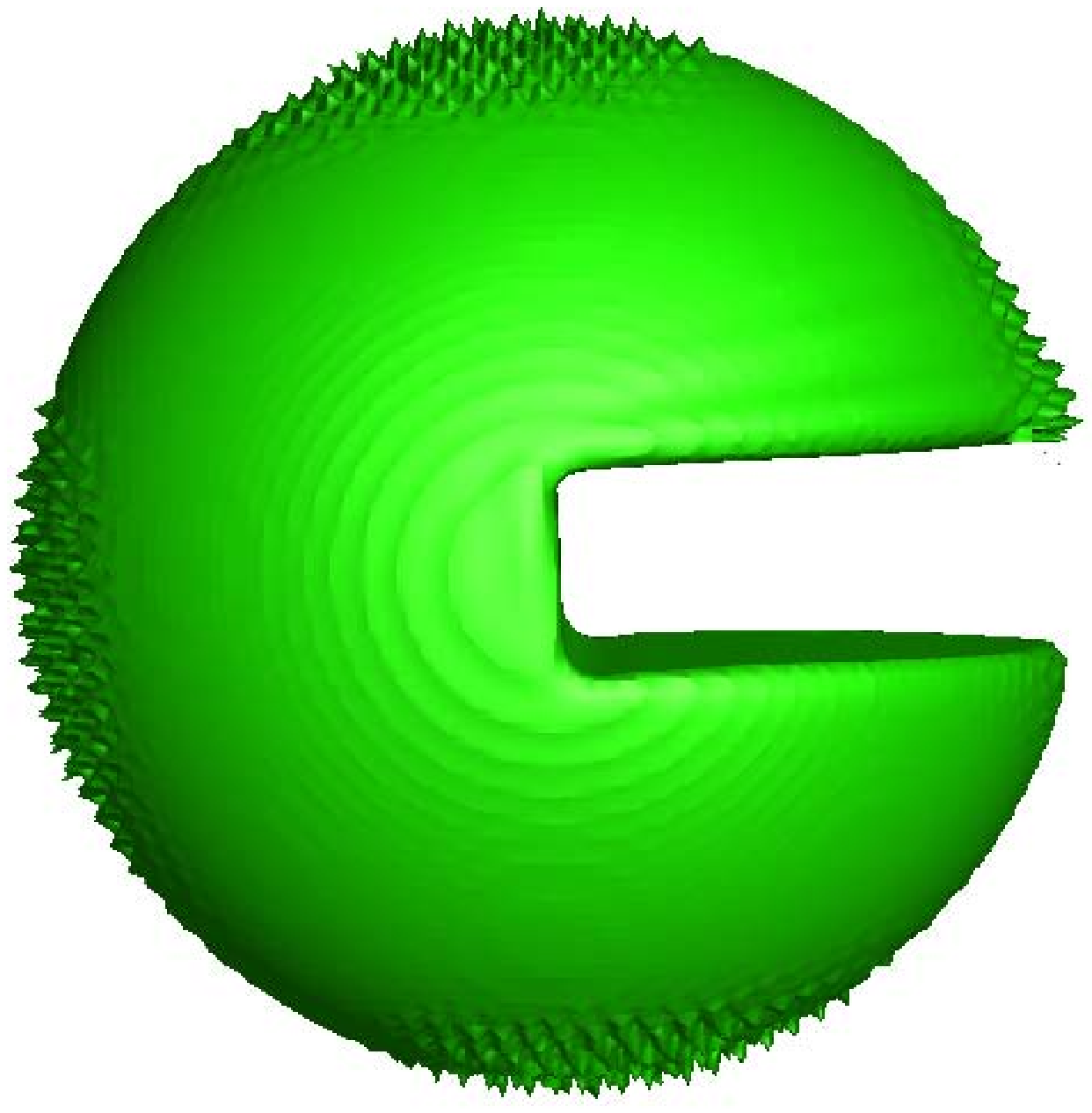}
\includegraphics[width=1.2in,height=1.212in]{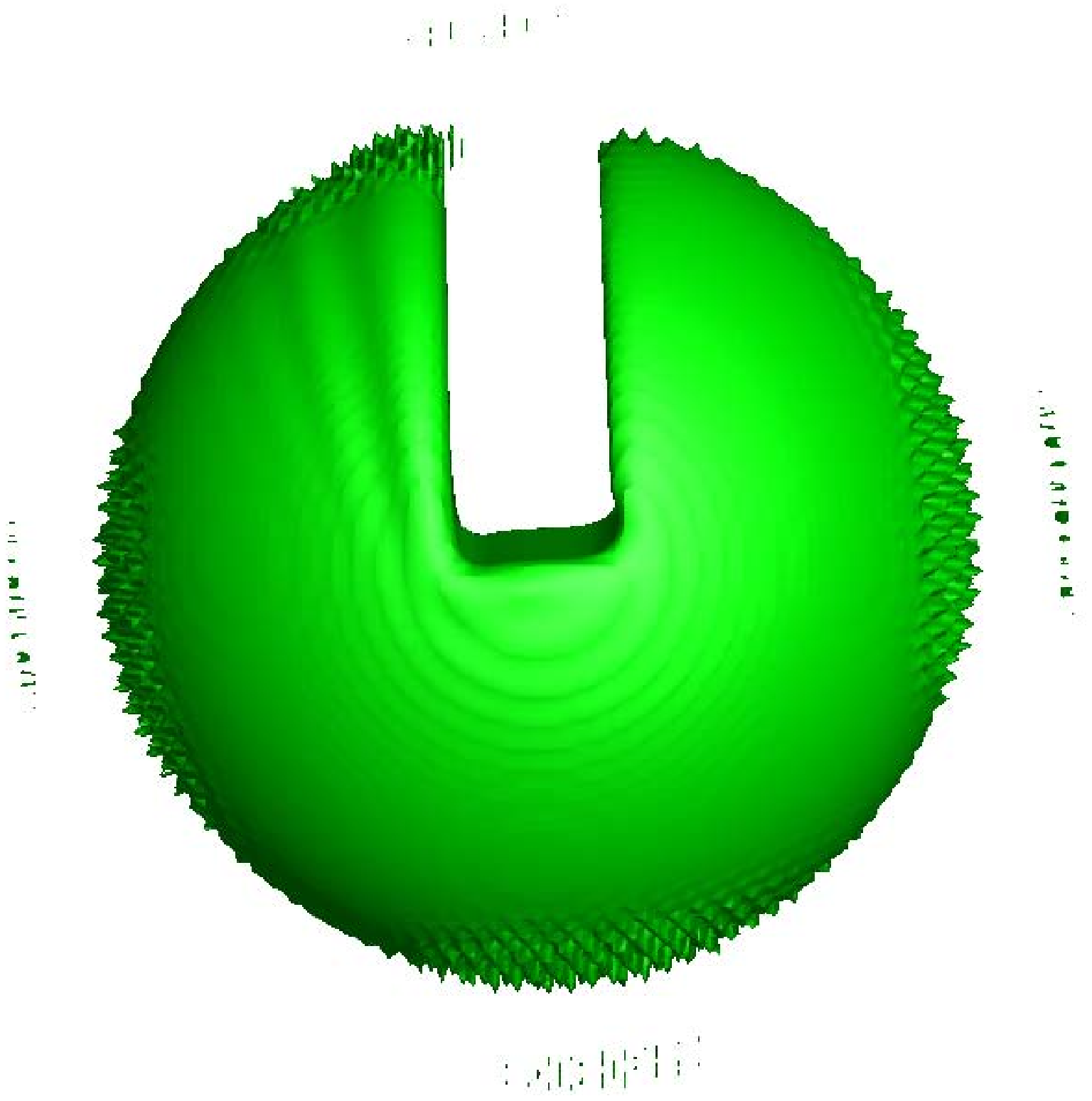}
\includegraphics[width=1.2in,height=1.212in]{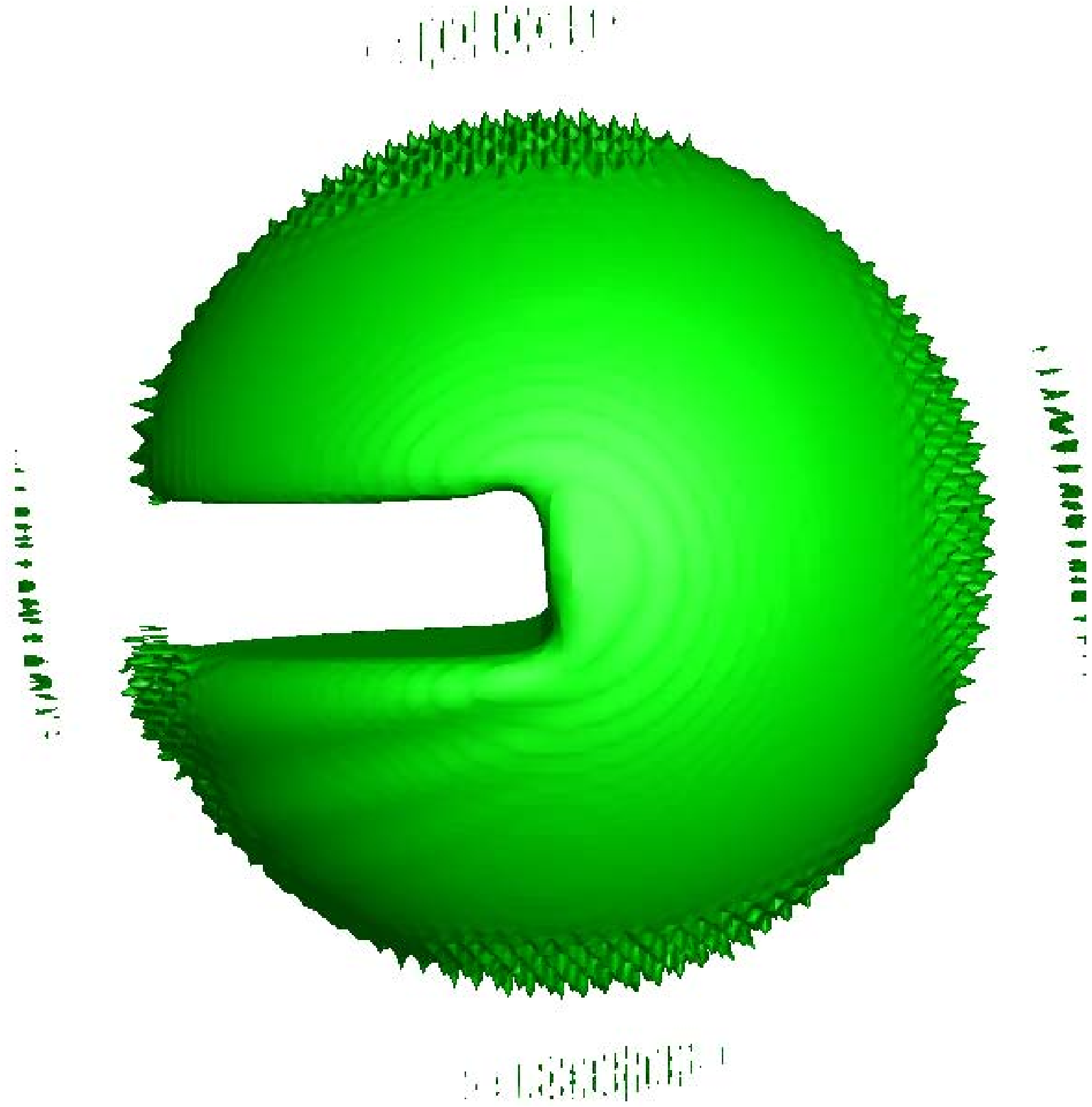}
\includegraphics[width=1.2in,height=1.212in]{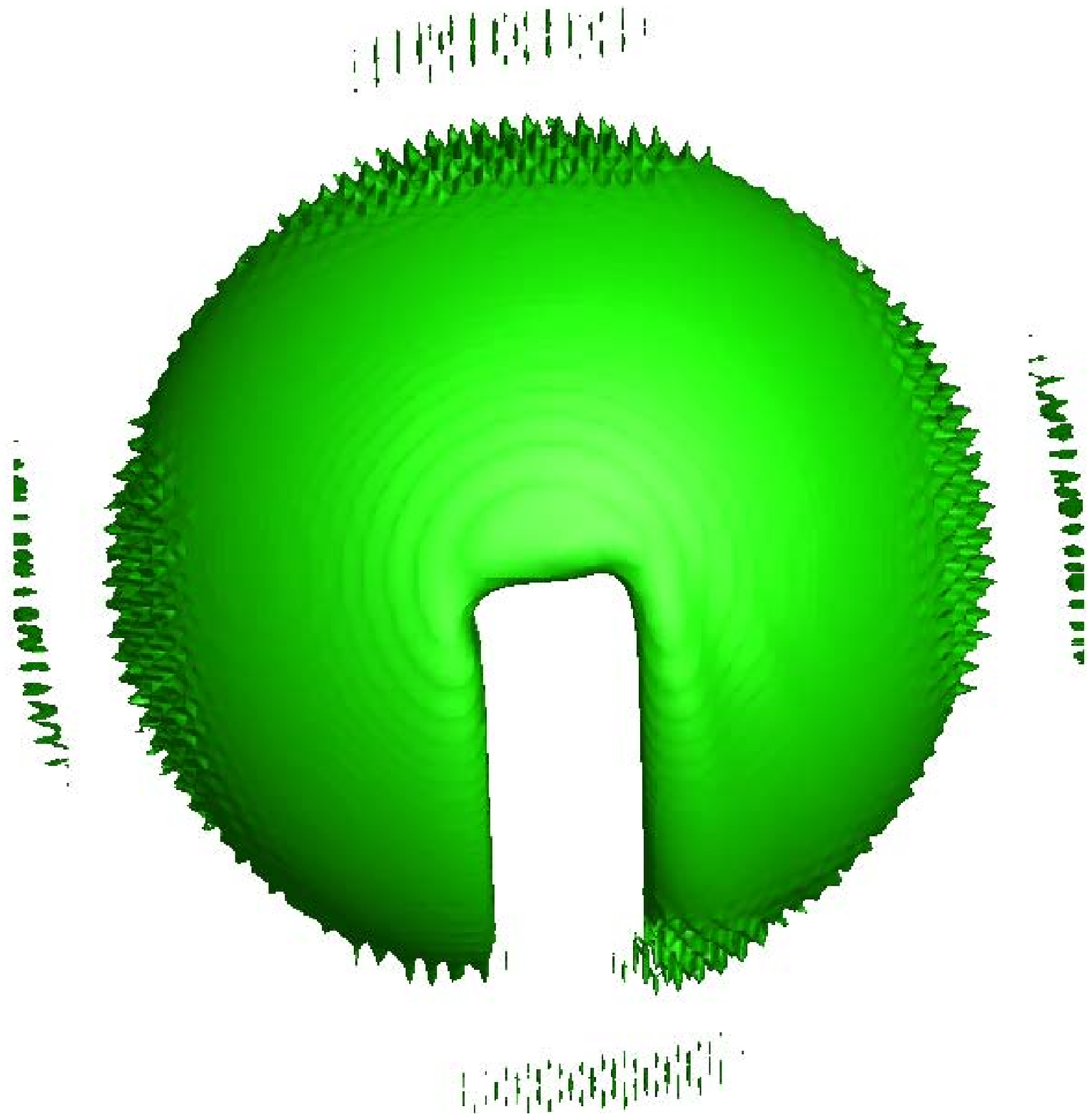}
\hspace{10pt}(b)\hspace{180pt}\\
 \tiny\caption{The rotation of Zalesak's sphere during one period at $Pe=200$: (a) the present MRT model; (b) the previous three-dimensional LB model~\cite{Zheng3}.}
\end{figure}

\begin{figure}
\centering
\includegraphics[width=1.2in,height=1.212in]{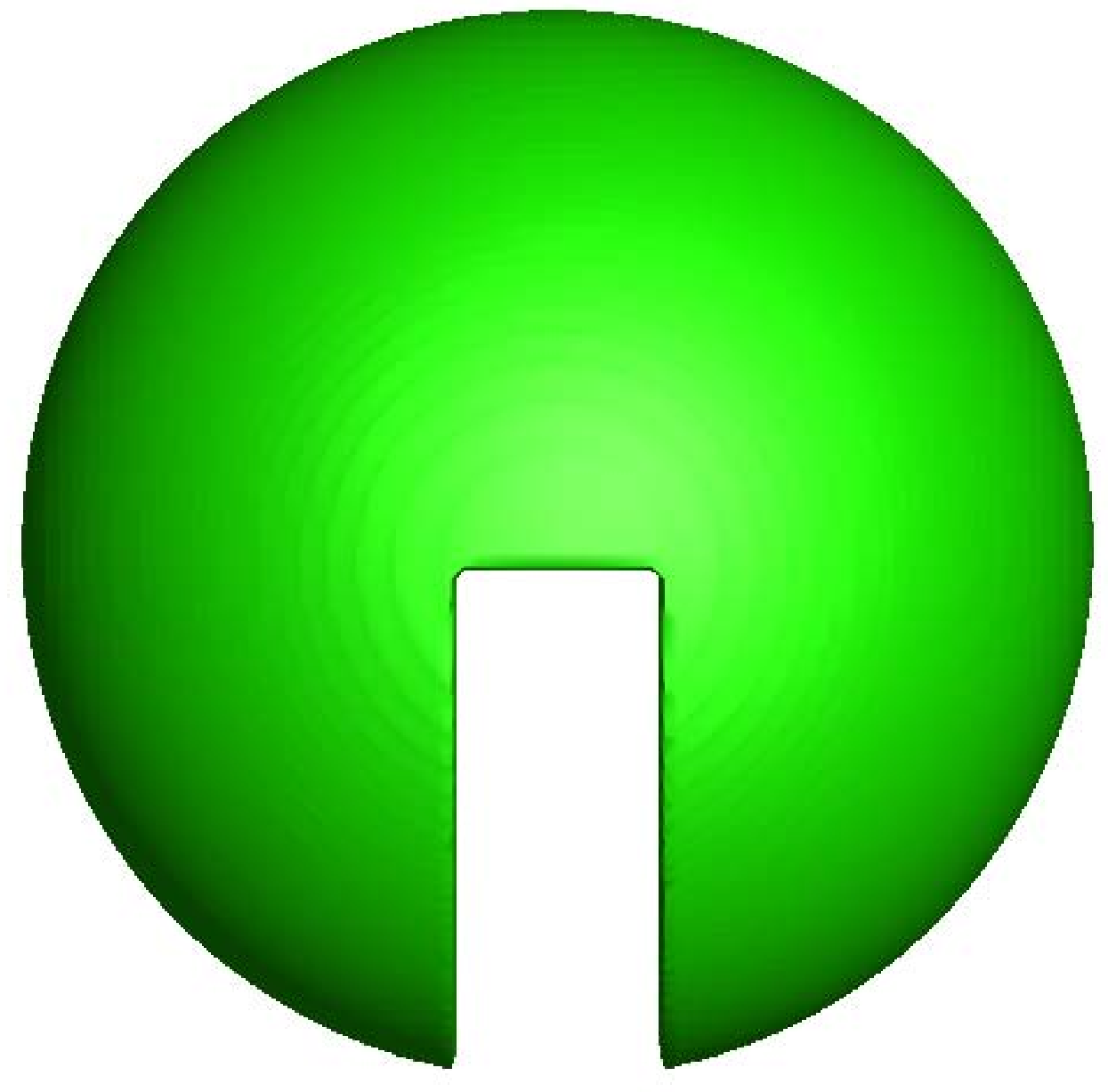}
\includegraphics[width=1.2in,height=1.212in]{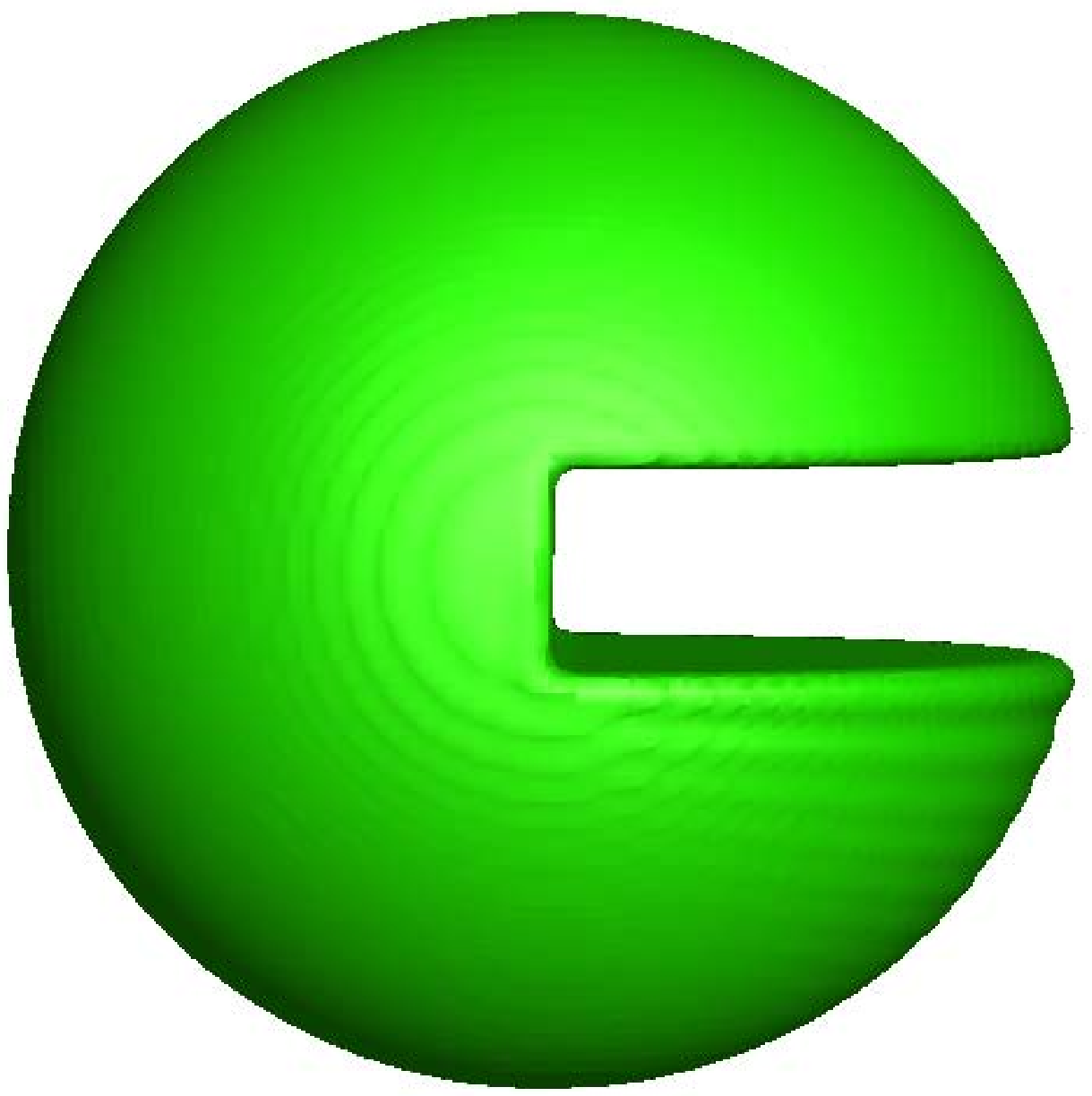}
\includegraphics[width=1.2in,height=1.212in]{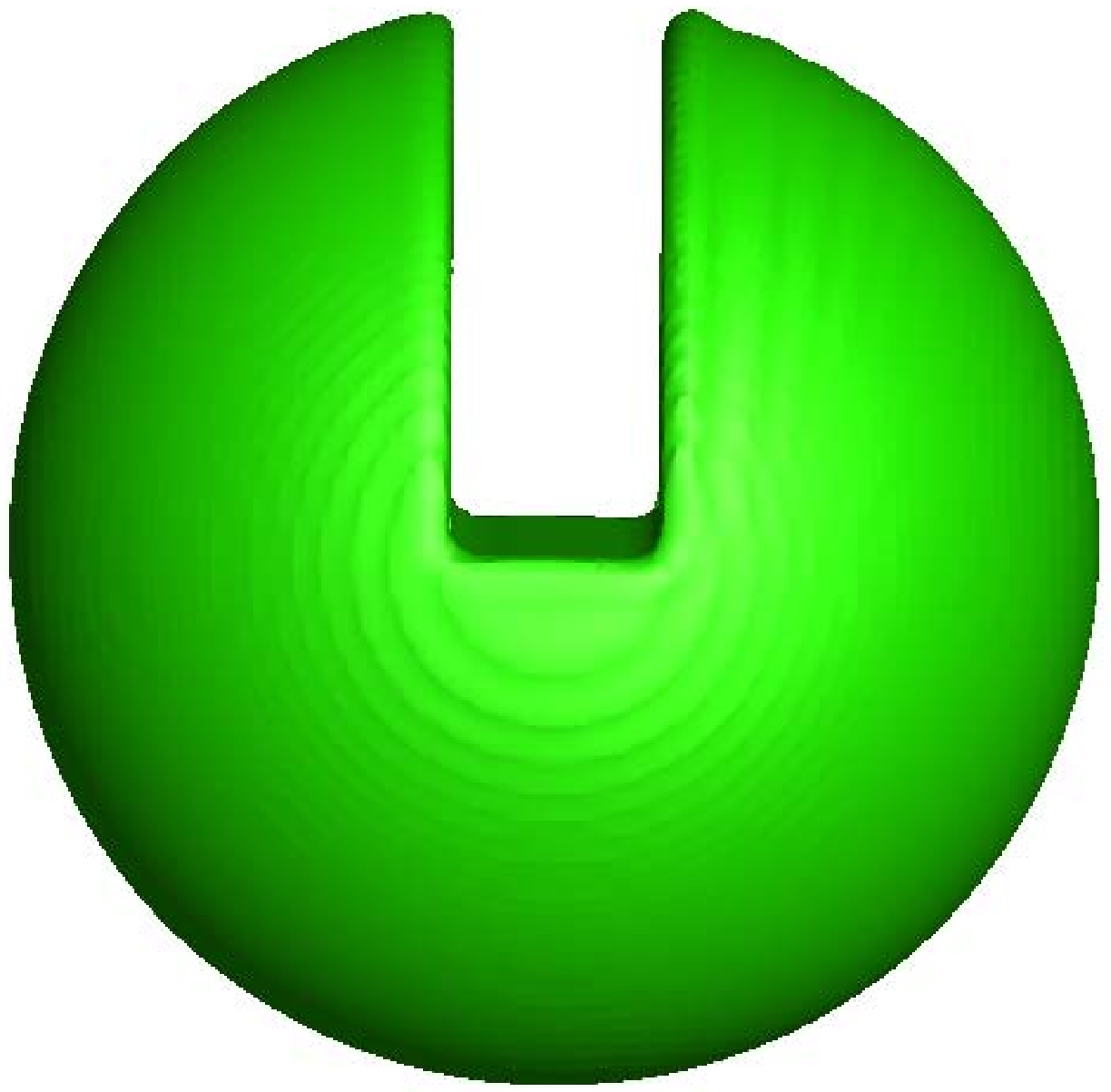}
\includegraphics[width=1.2in,height=1.212in]{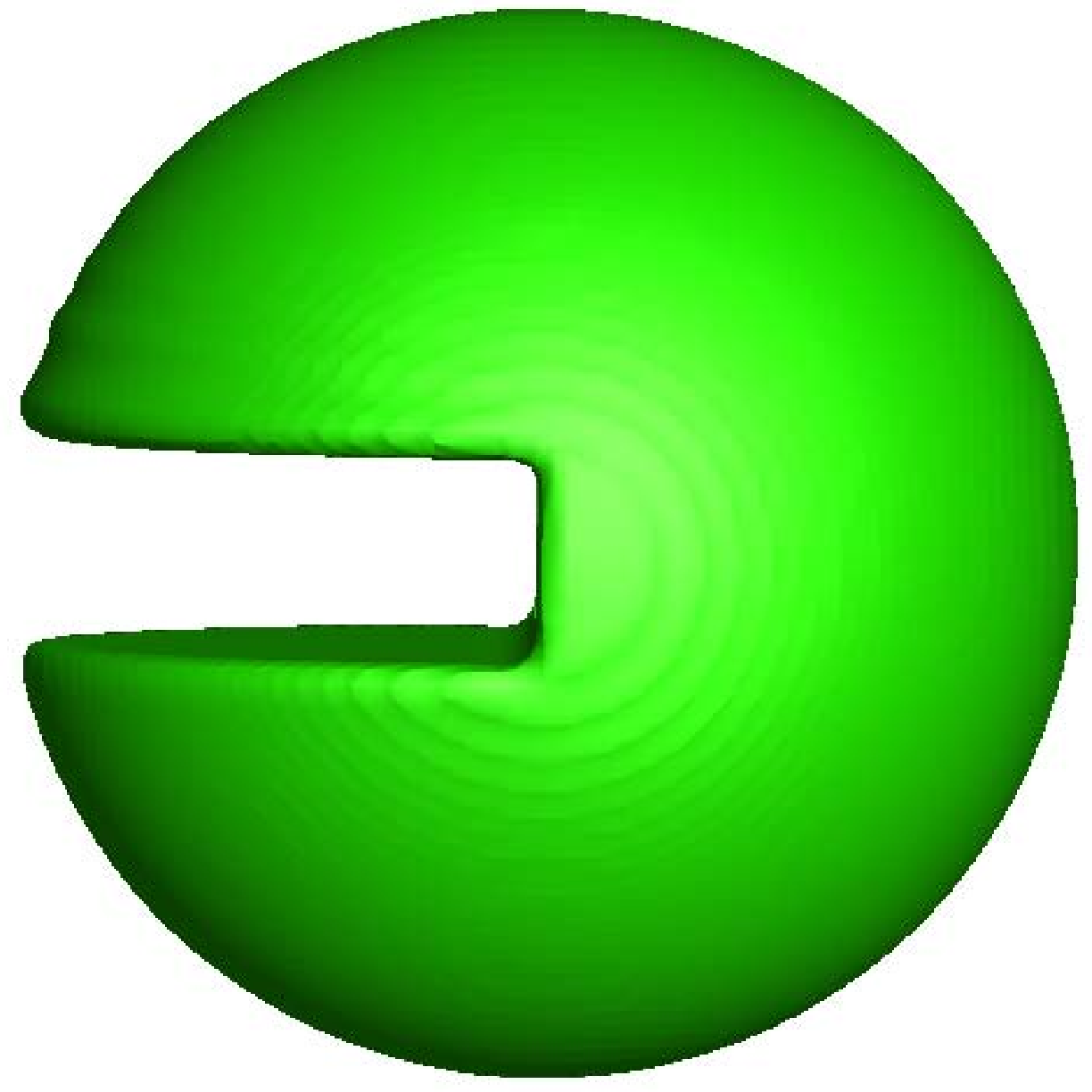}
\includegraphics[width=1.2in,height=1.212in]{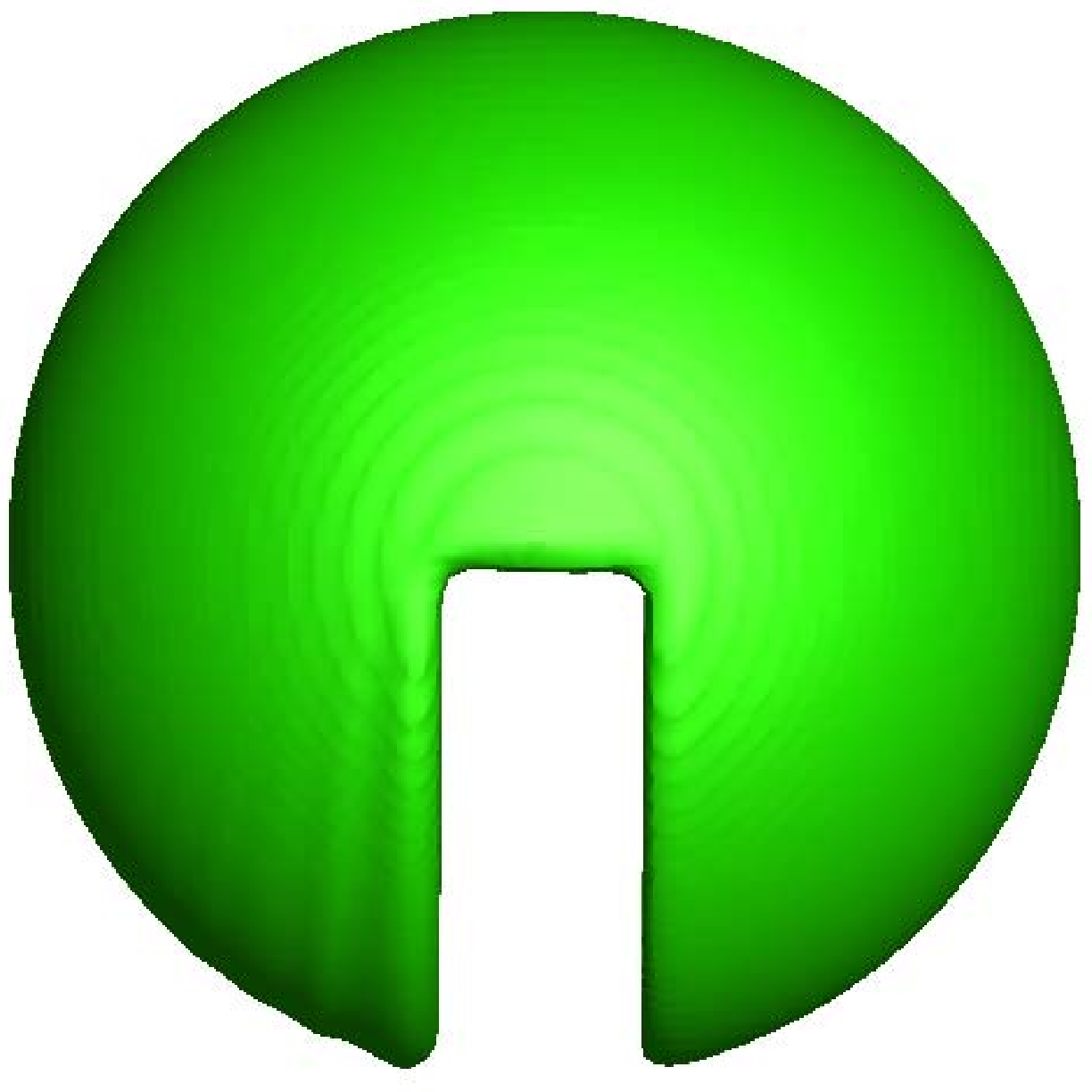}\\
\hspace{10pt}(a)\hspace{180pt}\\
\includegraphics[width=1.2in,height=1.212in]{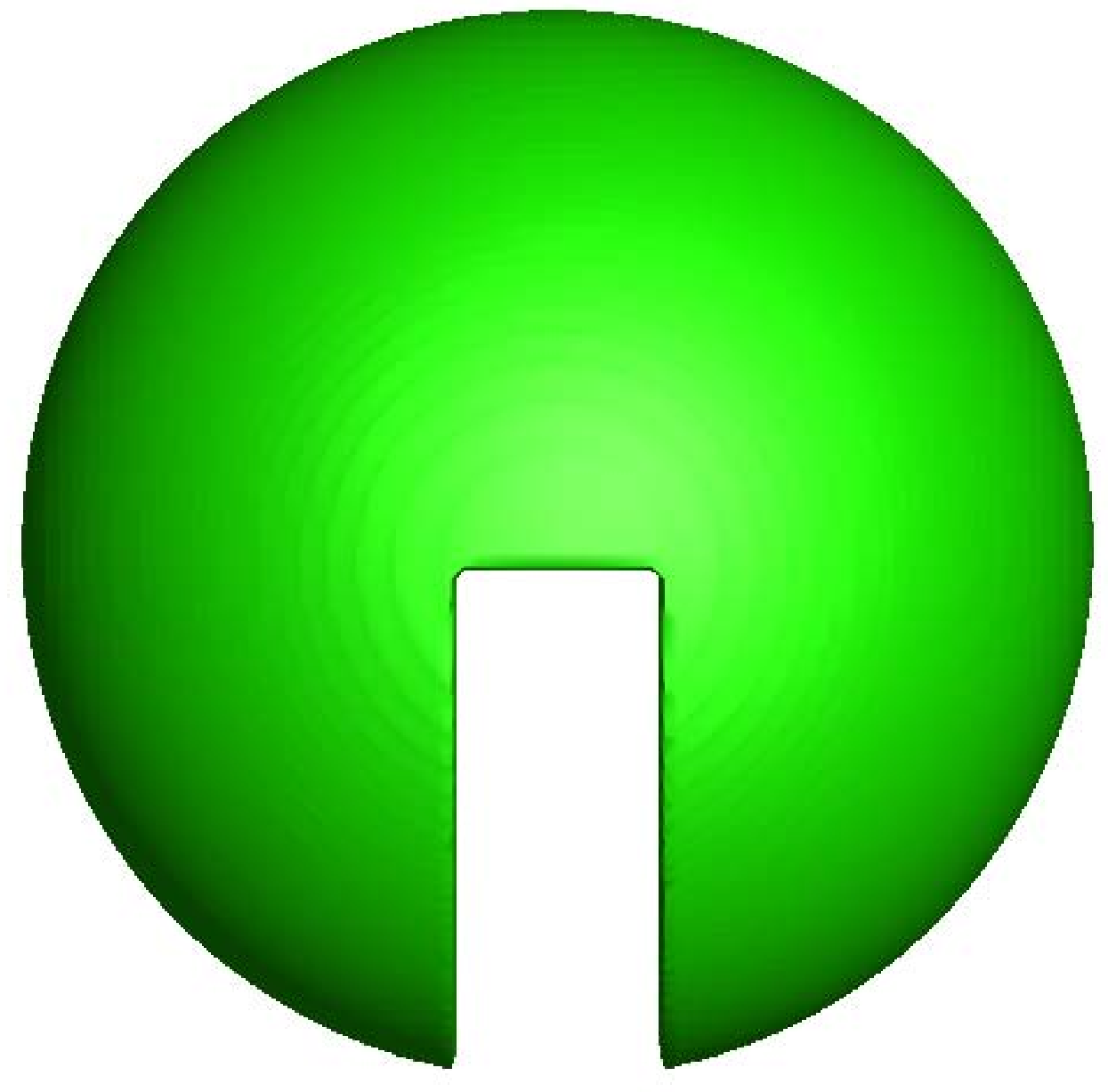}
\includegraphics[width=1.2in,height=1.212in]{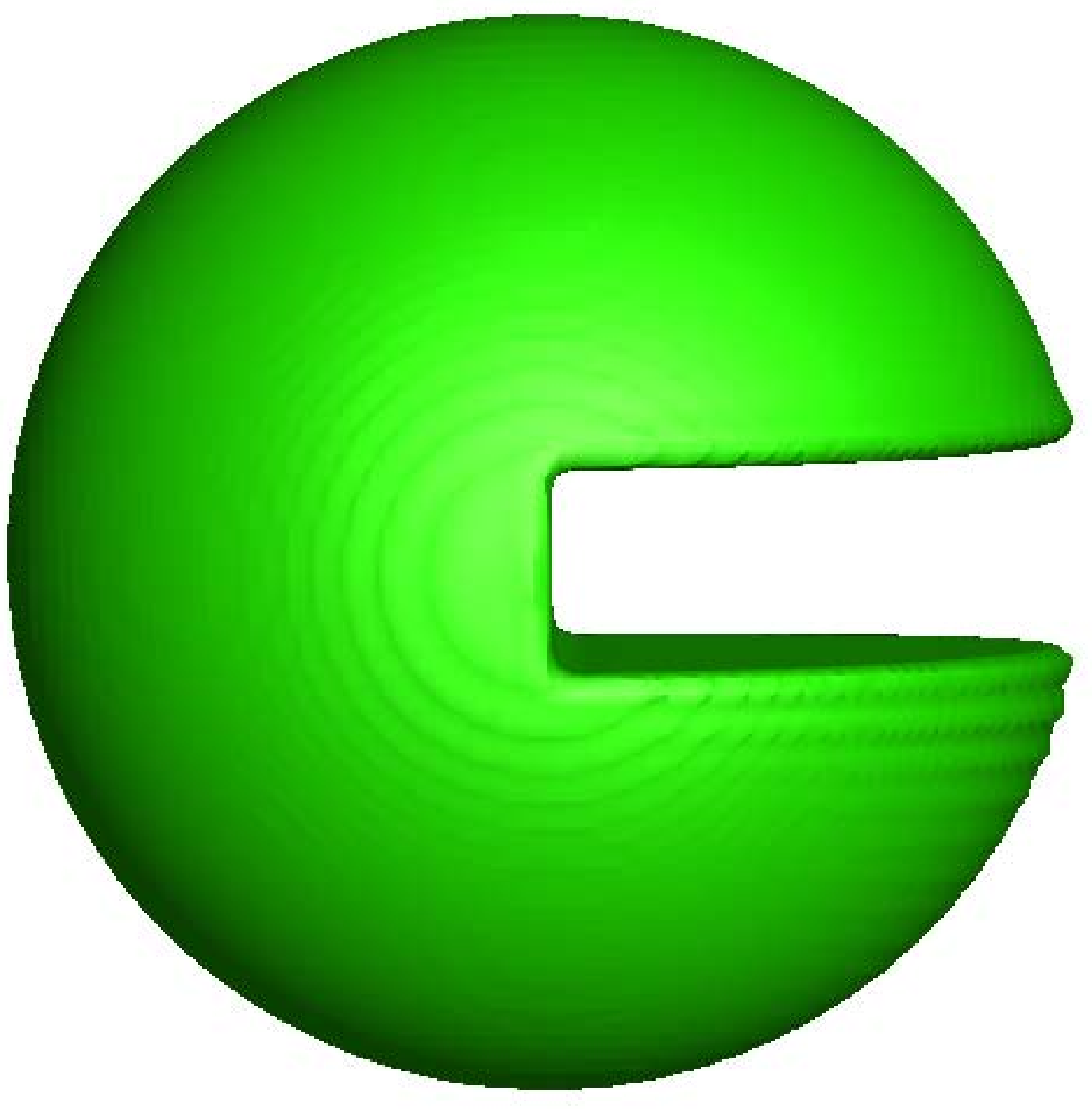}
\includegraphics[width=1.2in,height=1.212in]{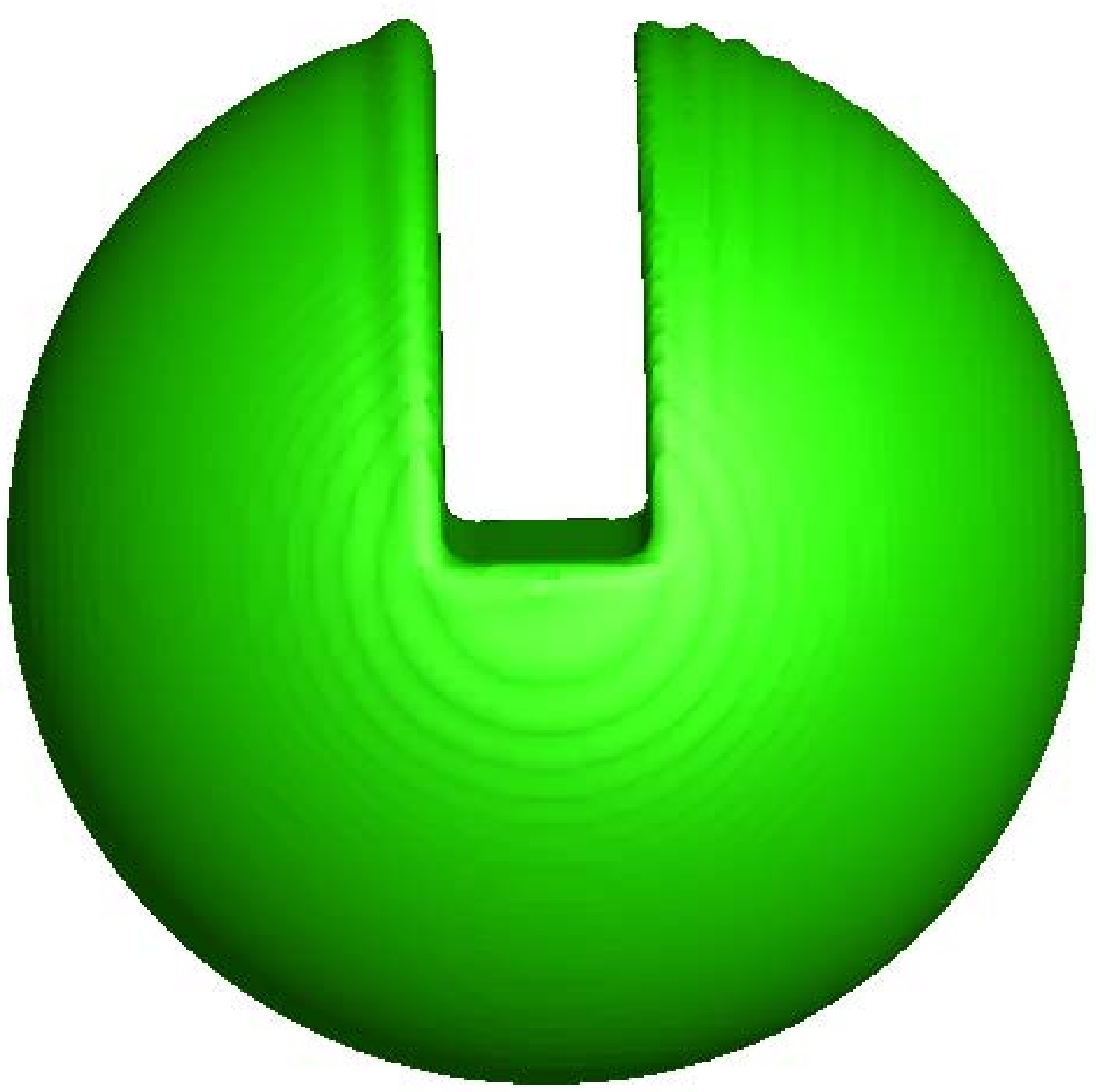}
\includegraphics[width=1.2in,height=1.212in]{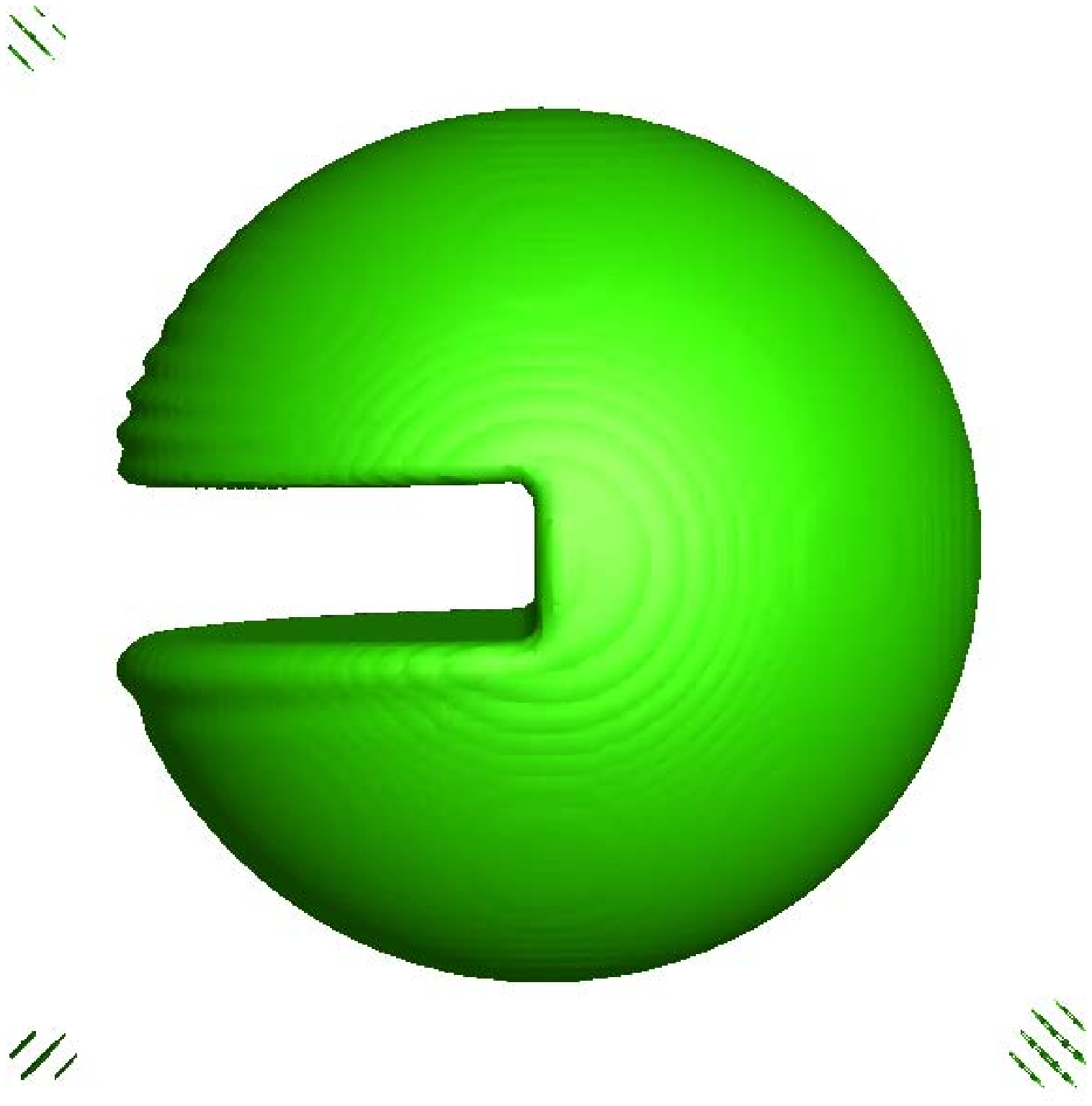}
\includegraphics[width=1.2in,height=1.212in]{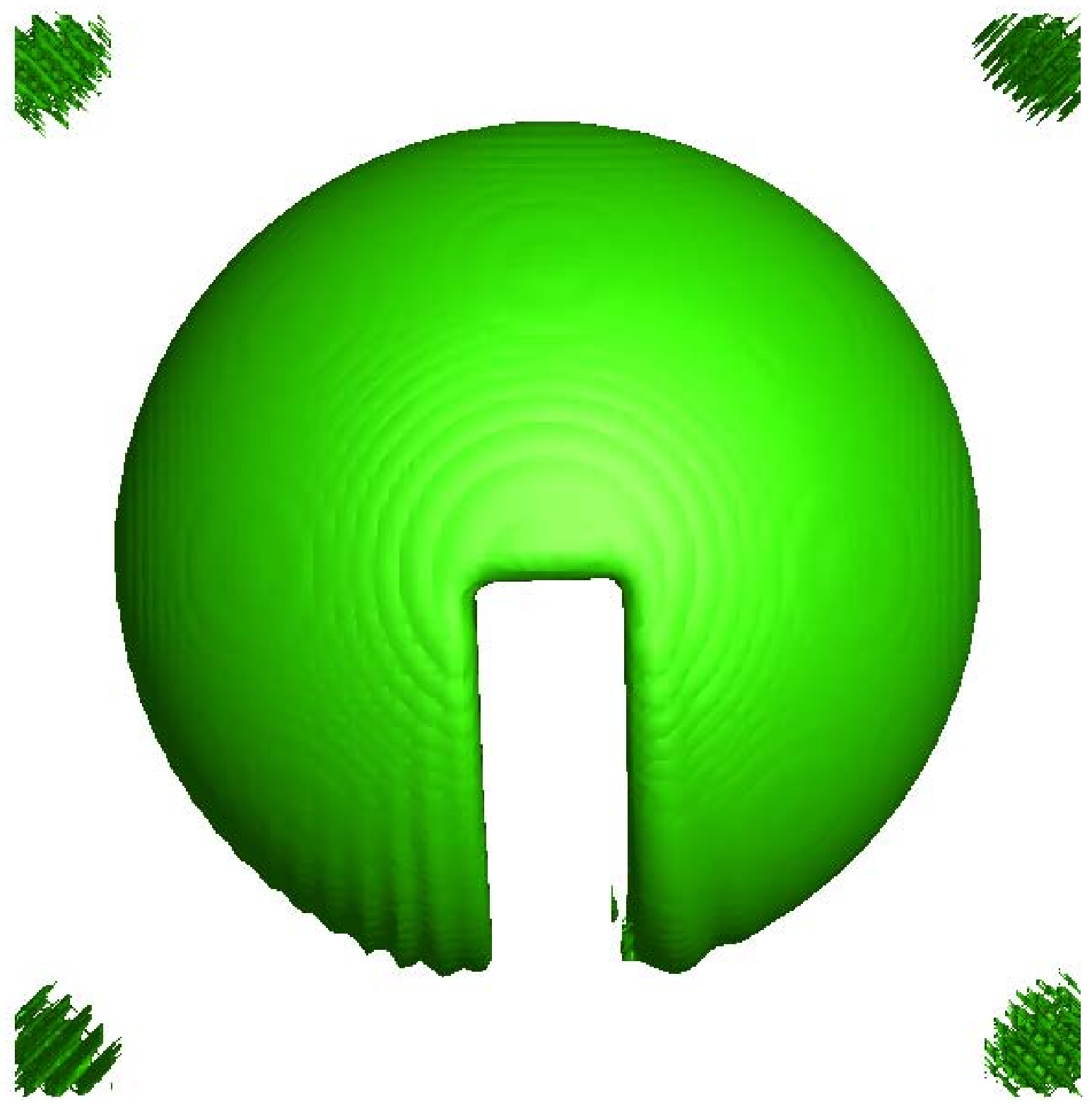}
\hspace{10pt}(b)\hspace{180pt}\\
 \tiny\caption{The rotation of Zalesak's sphere during one period at $Pe=600$: (a) the present MRT model; (b) the present SRT model.}
\end{figure}
\subsection{Rotation of the Zalesak's sphere}
Rotation has been widely used in the literatures to test interface tracking methods~\cite{Rudman,Enright,Liang1}. Here we consider the rotation of the Zalesak's sphere which has a
slot of 16 lattice units width in a $100\times100\times100$ domain. This sphere is initially centered at $(50,~50,~50)$ and the radius $R$ occupies 40 lattice units.
The revolution of the sphere is driven by a constant vorticity velocity field,
\begin{equation}
 \begin{split}
 u(x,y,z)&=0, \\
 v(x,y,z)&= (U\pi/100)(50-z),\\
 w(x,y,z)&= (U\pi/100)(y-50),
\end{split}
\end{equation}
where $U$ takes a value of $0.02$ so that the sphere complete one cycle every $10^4$ time steps. In our simulations,
the interface thickness $D$ and surface tension $\sigma$ are fixed as 2.0 and 0.04, respectively; the relaxation matrix is
set as
\begin{equation}
\mathbf{S}^f=diag(1.0, \frac{4}{3}, \frac{4}{3}, \frac{4}{3}, 1.2, 1.2, 1.2).
\end{equation}
The periodic boundary condition is applied at all boundaries. To examine the mobility effect, we introduce the dimensionless Peclet number,
which is defined as,
\begin{equation}
Pe=\frac{LU}{M},
\end{equation}
where $L$ is the characteristic length, and set as the value of $R$. Different mobilities can be obtained by changing the value of $Pe$.
Figure 1 shows the revolution processes of the sphere during one period at $Pe=200$. It is seen from Fig. 1 that the present model can precisely capture the interface shape in one period, and the sphere returns to its initial configuration at time $T$, which is accordance with the expected results. For a comparison,
the results obtained by the previous three-dimensional LB model~\cite{Zheng3} are also presented in Fig. 1. It is observed that the previous model
not only produces some obvious sawteeth on the sphere surface, but also induces some unphysical disturbances around the computational domain.
This implies that the present model can obtain a more accurate and stable interface. It has been mentioned above that the
SRT model can be deemed as a special case of the MRT model when all the relaxation parameters equal to each other.
Due to the more adjustable relaxation factors, the MRT model should show more potential to achieve a better numerical stability
against the SRT model. To illustrate this point, we present a comparison between them using this case. Figure 2 depicts
snapshots of rotating sphere during one period at $Pe=600$ by the MRT and SRT models. It is clearly seen that the results of the
SRT model are unstable. The slot of the sphere is slightly distorted and there produces some extra jetsam at the corners of the
computational domain. In contrast, the present MRT model can capture the moving interface of the sphere correctly.
\begin{figure}
\centering
\includegraphics[width=4.0in,height=3.0in]{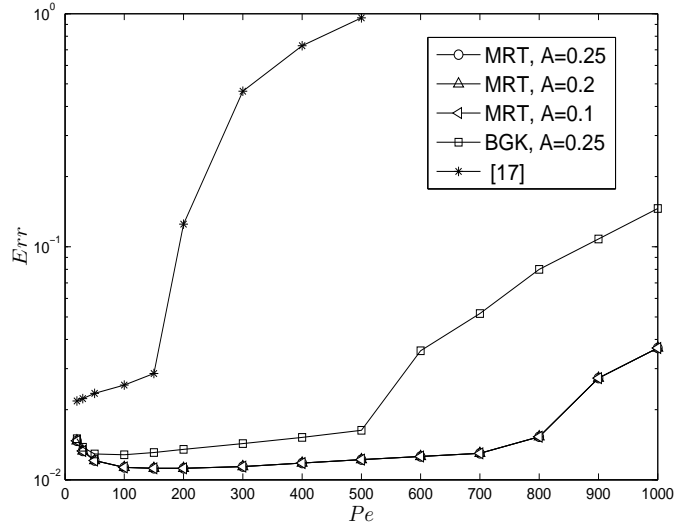}
 \tiny\caption{Comparison of the relative errors produced by different LB models in Zalesak's sphere rotation tests with various Peclet numbers.}
\end{figure}
To further quantitatively describe the accuracy of the interface capturing, the global relative error on the order parameter is
introduced as
\begin{equation}
{Err } = \frac{{\sum {_{\bf{x}}} \left| {\phi ({\bf{x}},T) -
{\phi }({\bf{x}},0)} \right|}}{{\sum {_{\bf{x}}\left| {{\phi
}({\bf{x}},{\rm{0}})} \right|} }},
\end{equation}
where $\phi({\bf{x}},T)$ is the value of the order parameter at position $\mathbf{x}$ and time $T$, and ${{\phi}({\bf{x}},{\rm{0}})}$
is the exact solution at initial time. We computed the relative errors generated by the LB previous model~\cite{Zheng3} and the present MRT and BGK models
at various Peclet numbers, and show the results in Fig. 3. As seen from this figure, one can find that the MRT model is more accurate than
the SRT model, which is further more accurate than the previous model, especially at large Pelcet number. The effect of the model
parameter $A$ is also examined, and it is found from Fig. 3 that the parameter $A$ has almost no effect on the accuracy of the
MRT model. Without loss of generality, in the following simulations we fix the parameter $A$ at $0.25$.

\begin{figure}
\centering
\includegraphics[width=1.2in,height=1.24in]{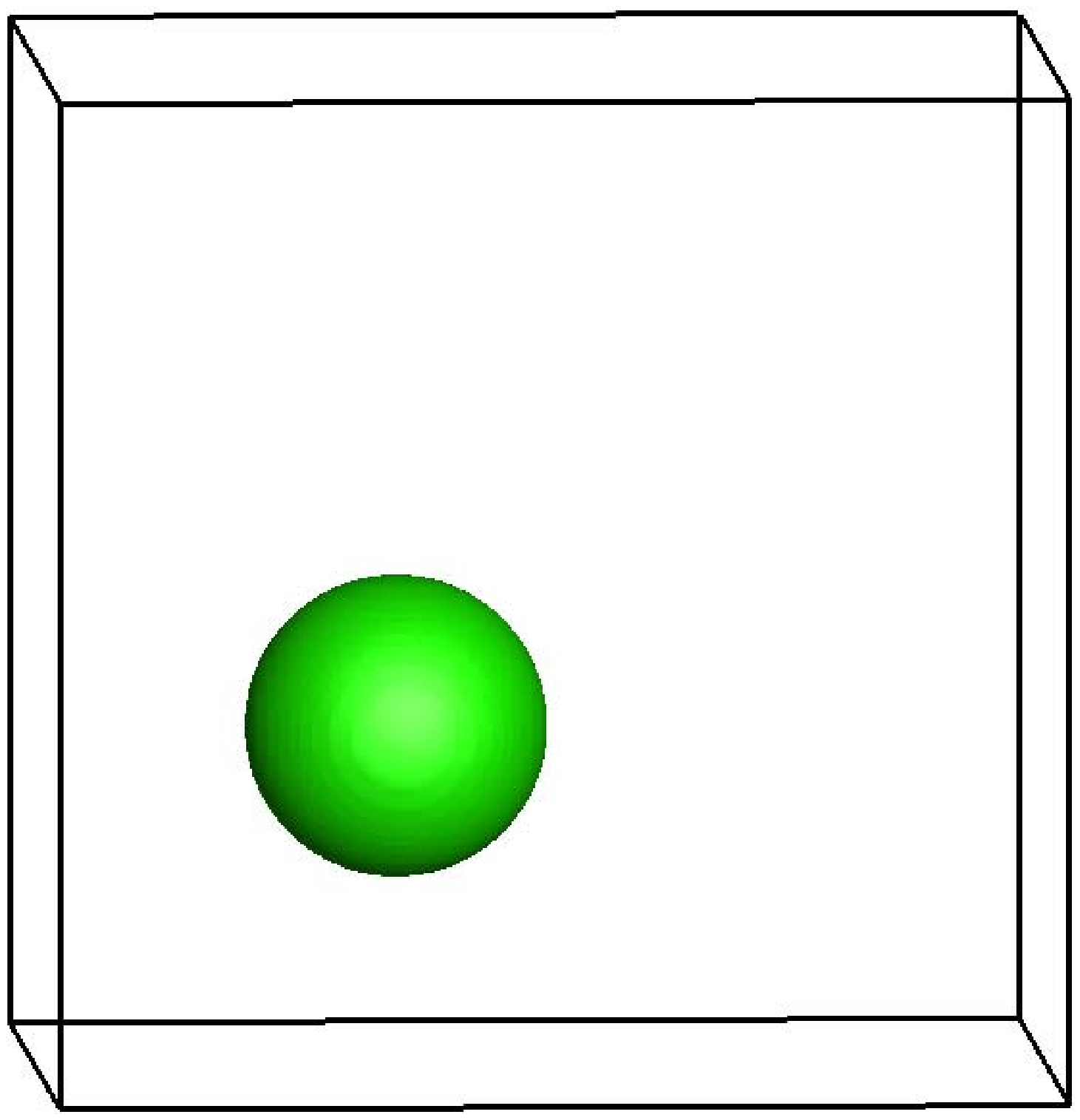}
\includegraphics[width=1.2in,height=1.24in]{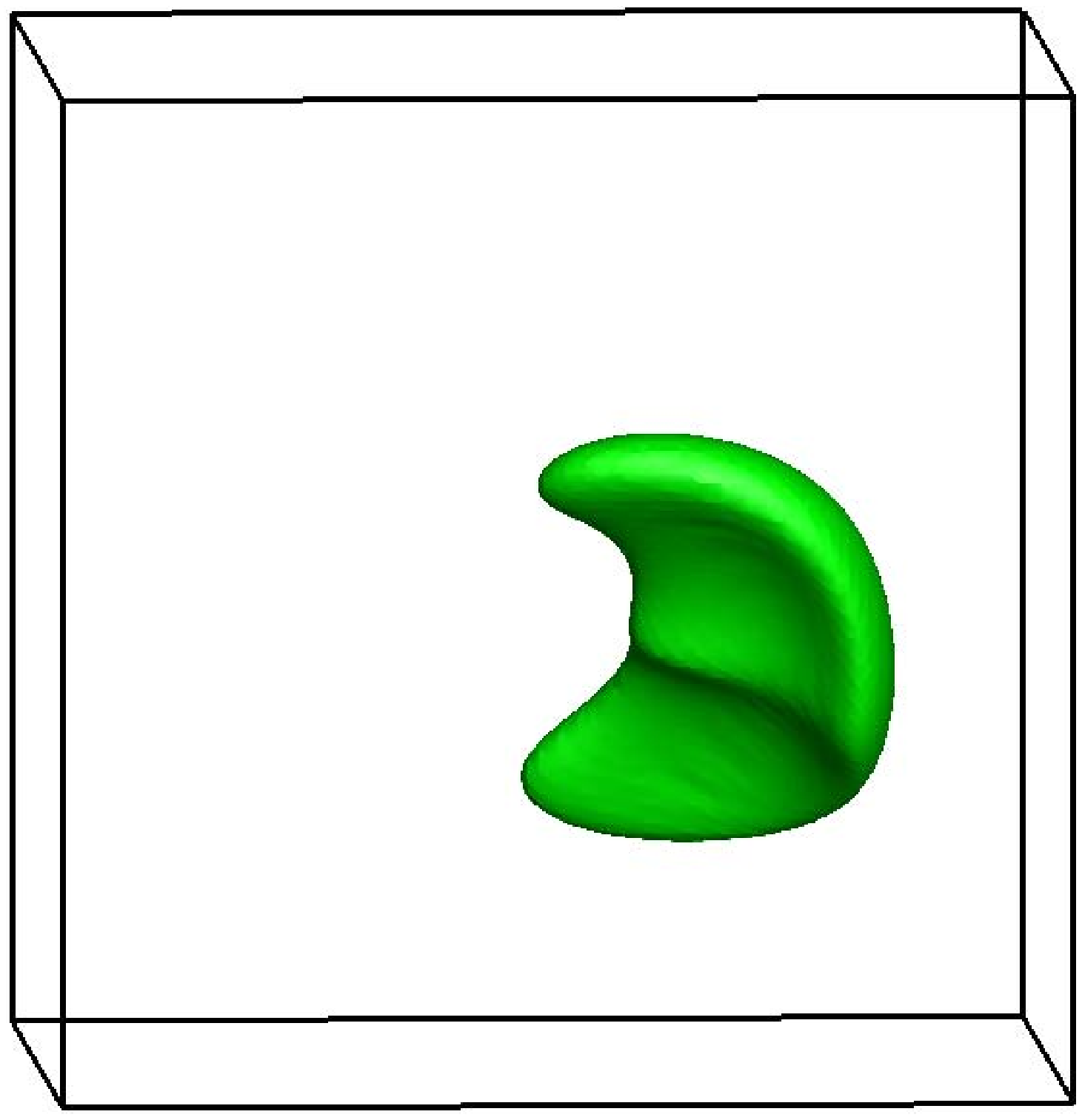}
\includegraphics[width=1.2in,height=1.24in]{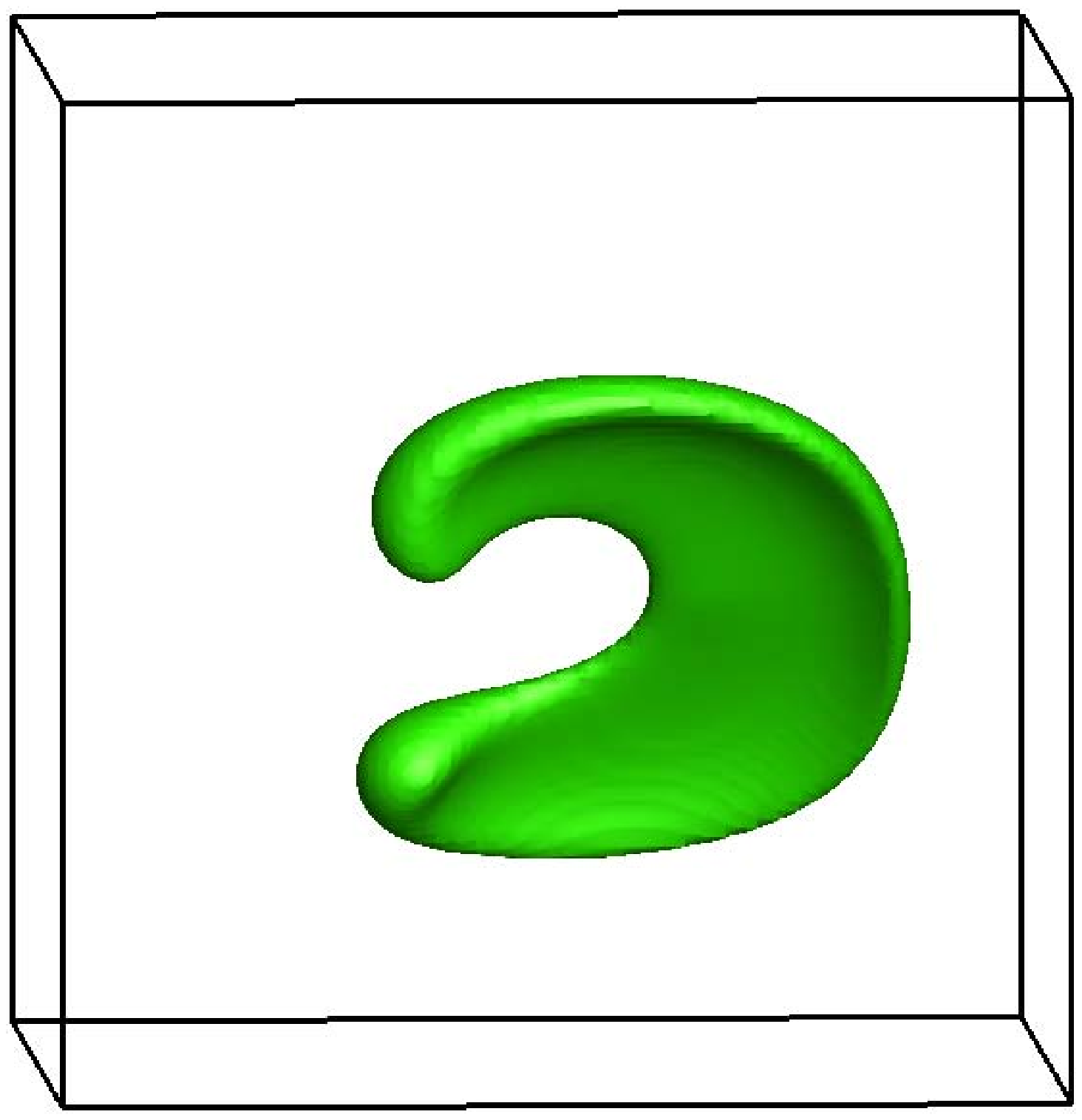}
\includegraphics[width=1.2in,height=1.24in]{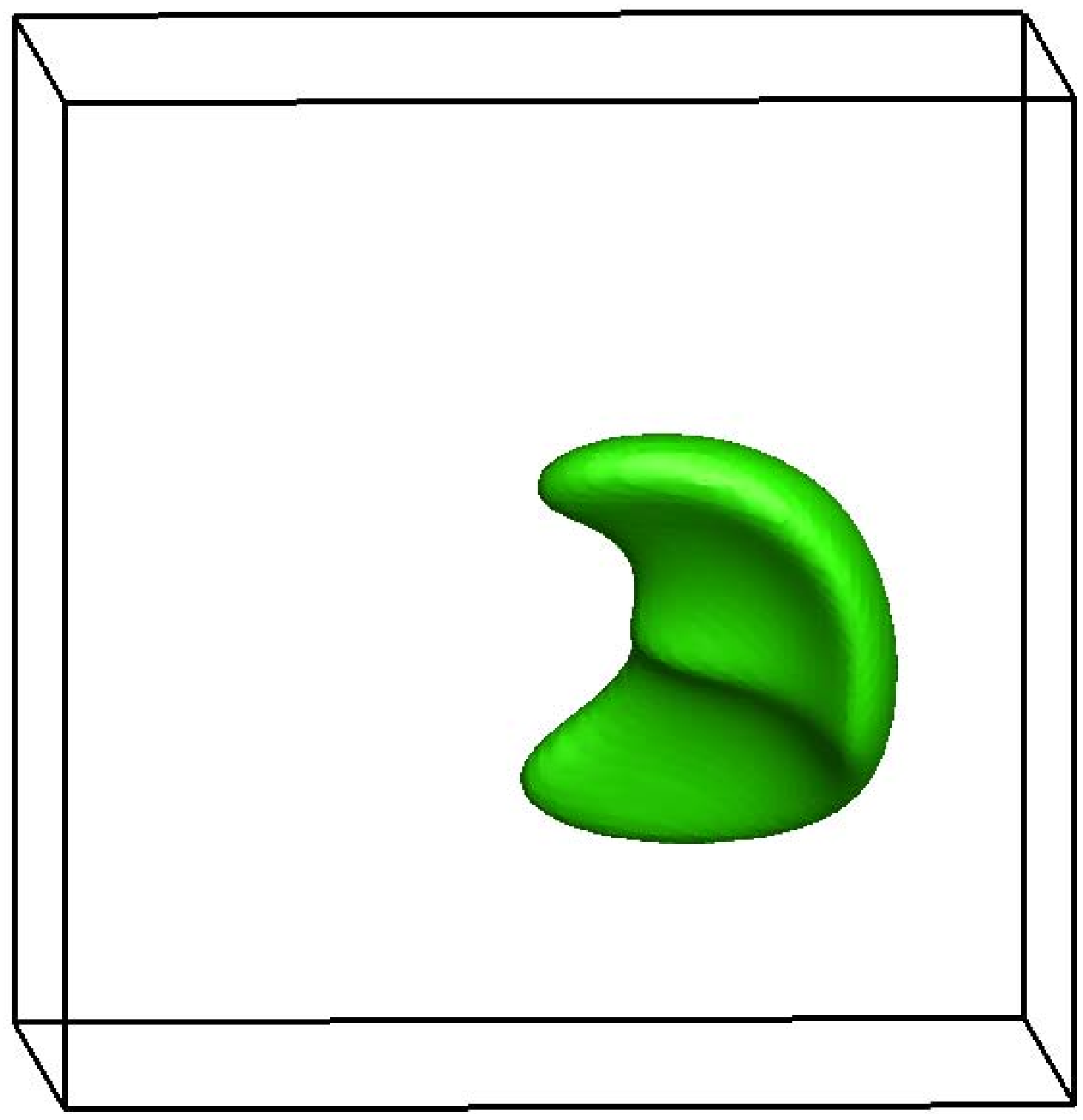}
\includegraphics[width=1.2in,height=1.24in]{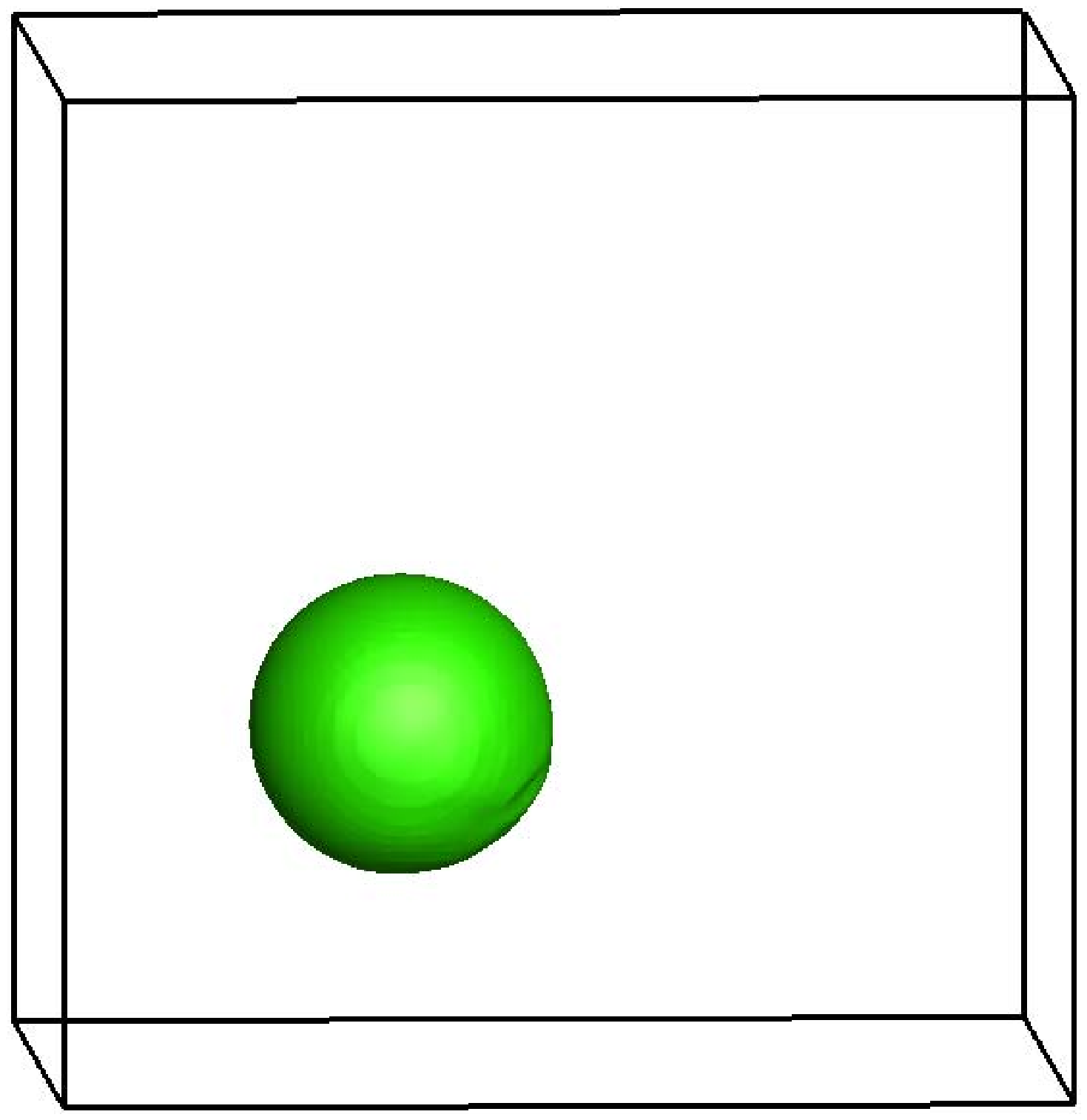}\\
\hspace{10pt}(a)\hspace{180pt}\\
\includegraphics[width=1.2in,height=1.24in]{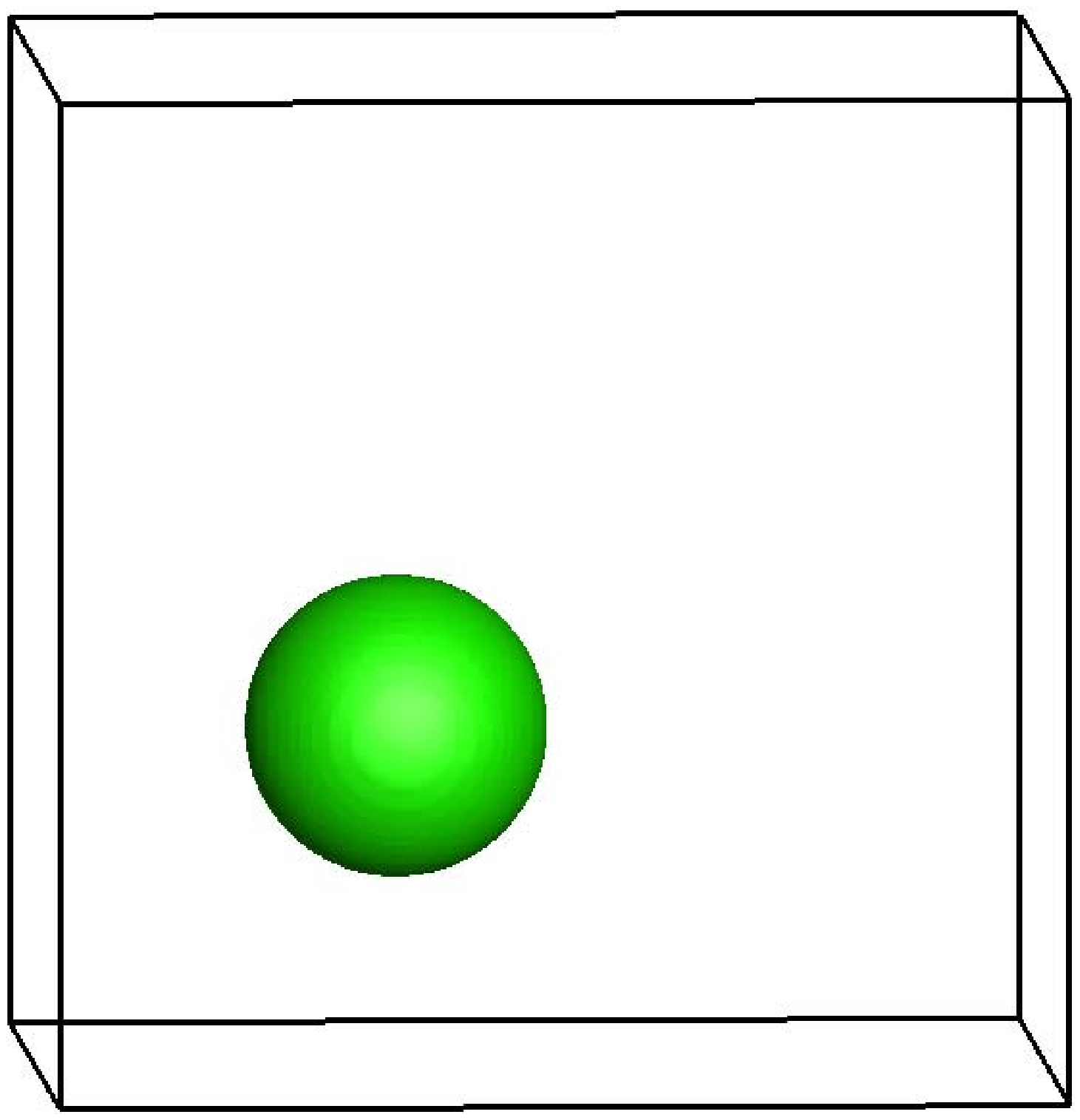}
\includegraphics[width=1.2in,height=1.24in]{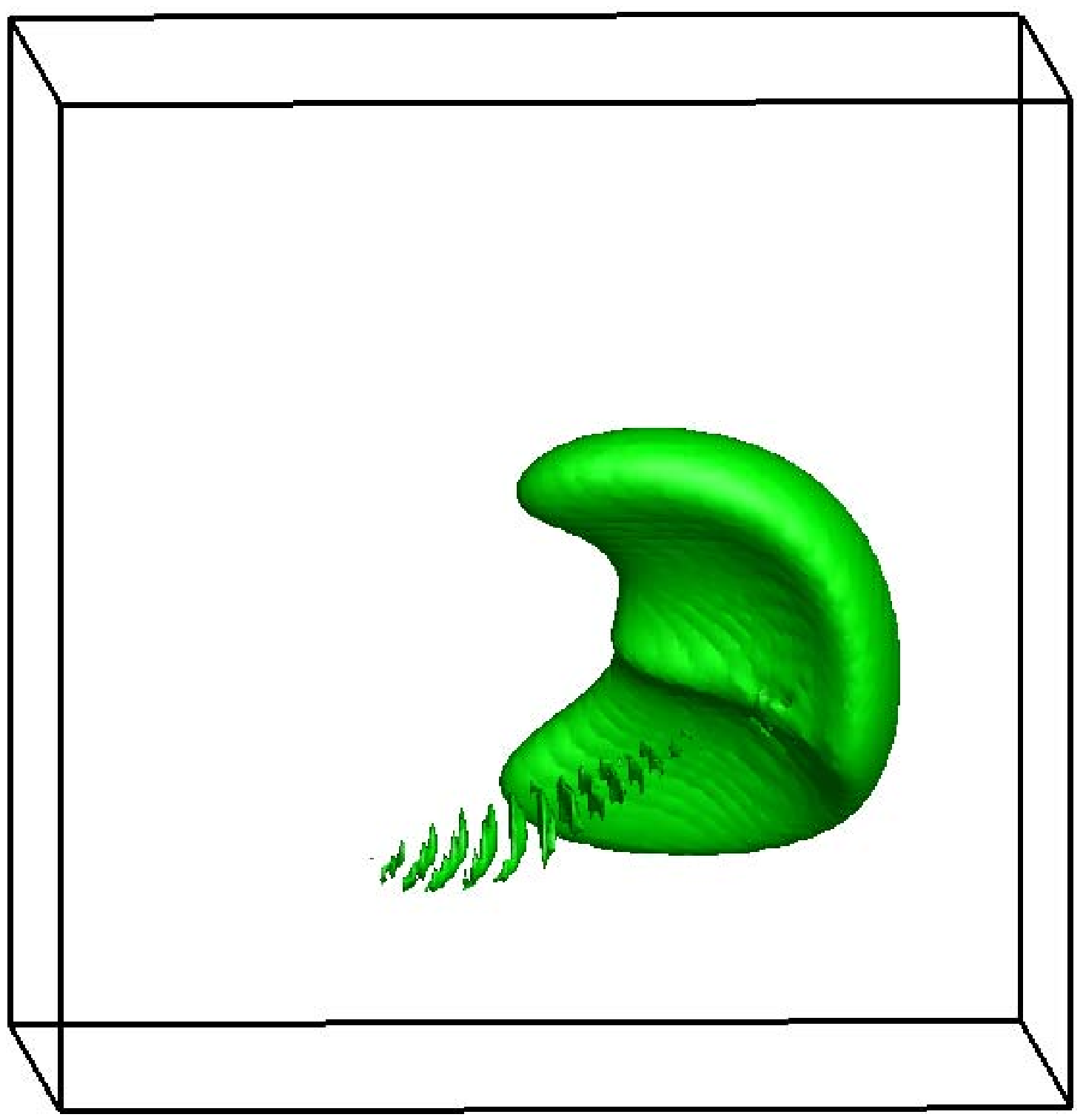}
\includegraphics[width=1.2in,height=1.24in]{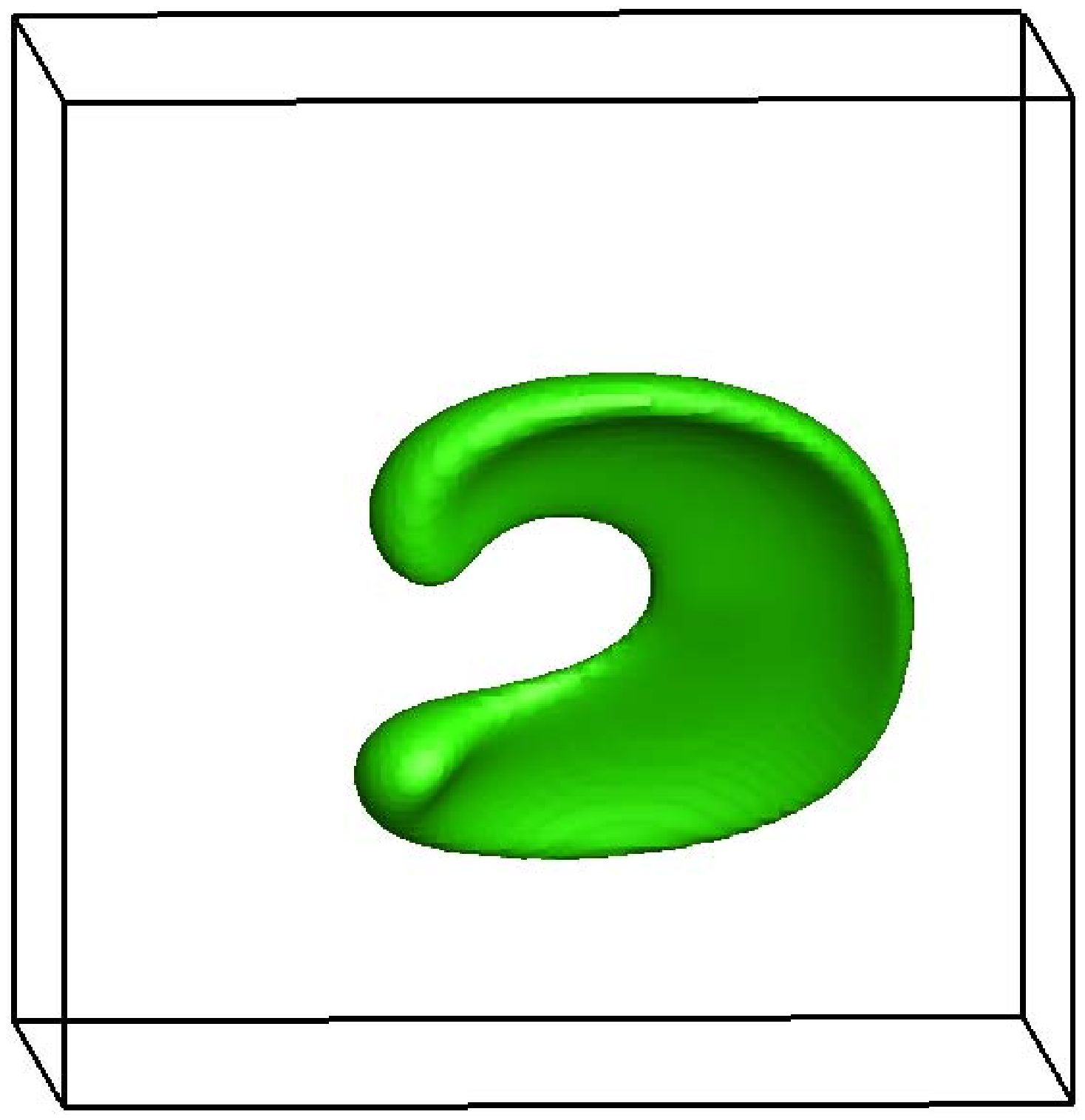}
\includegraphics[width=1.2in,height=1.24in]{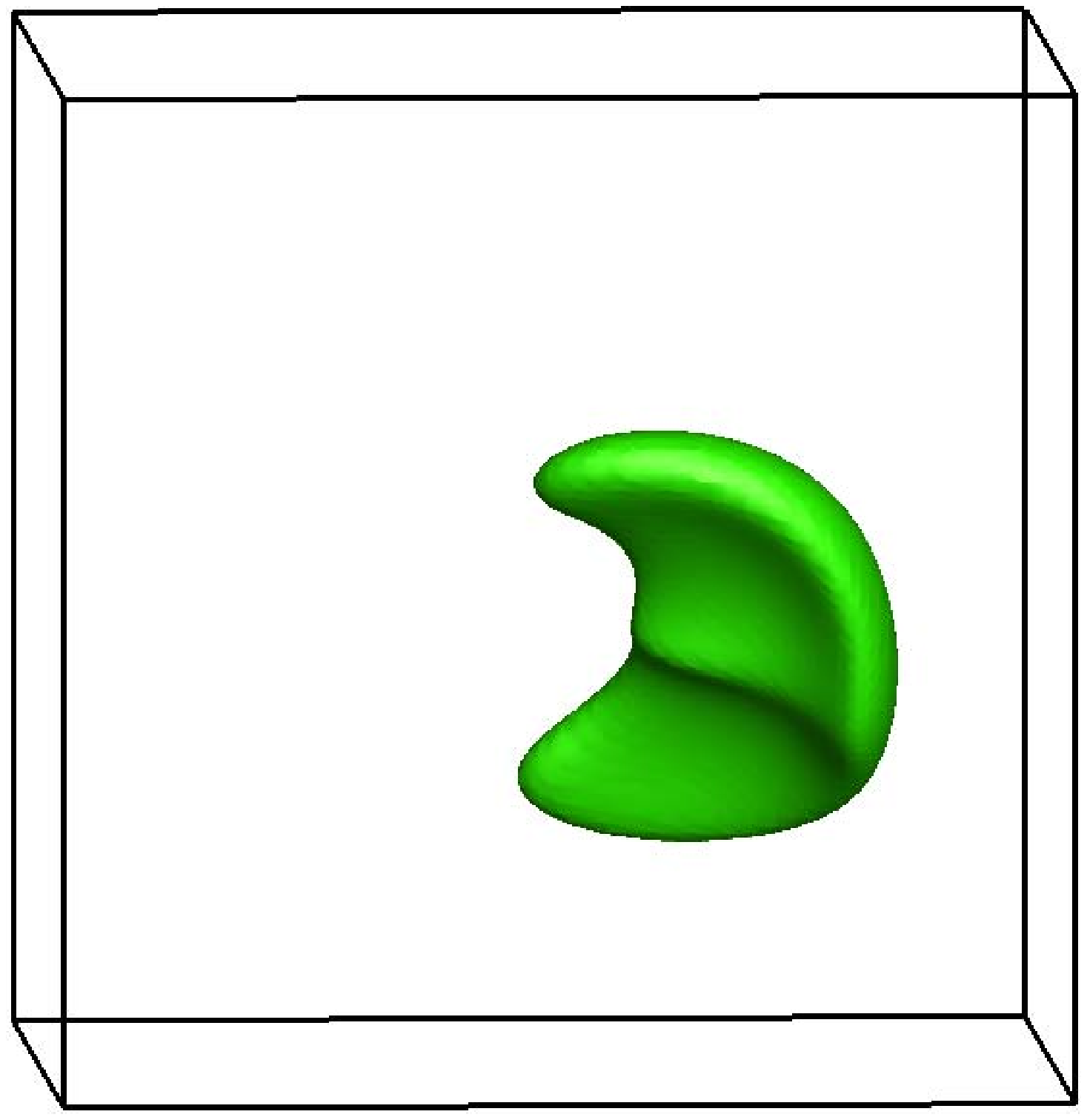}
\includegraphics[width=1.2in,height=1.24in]{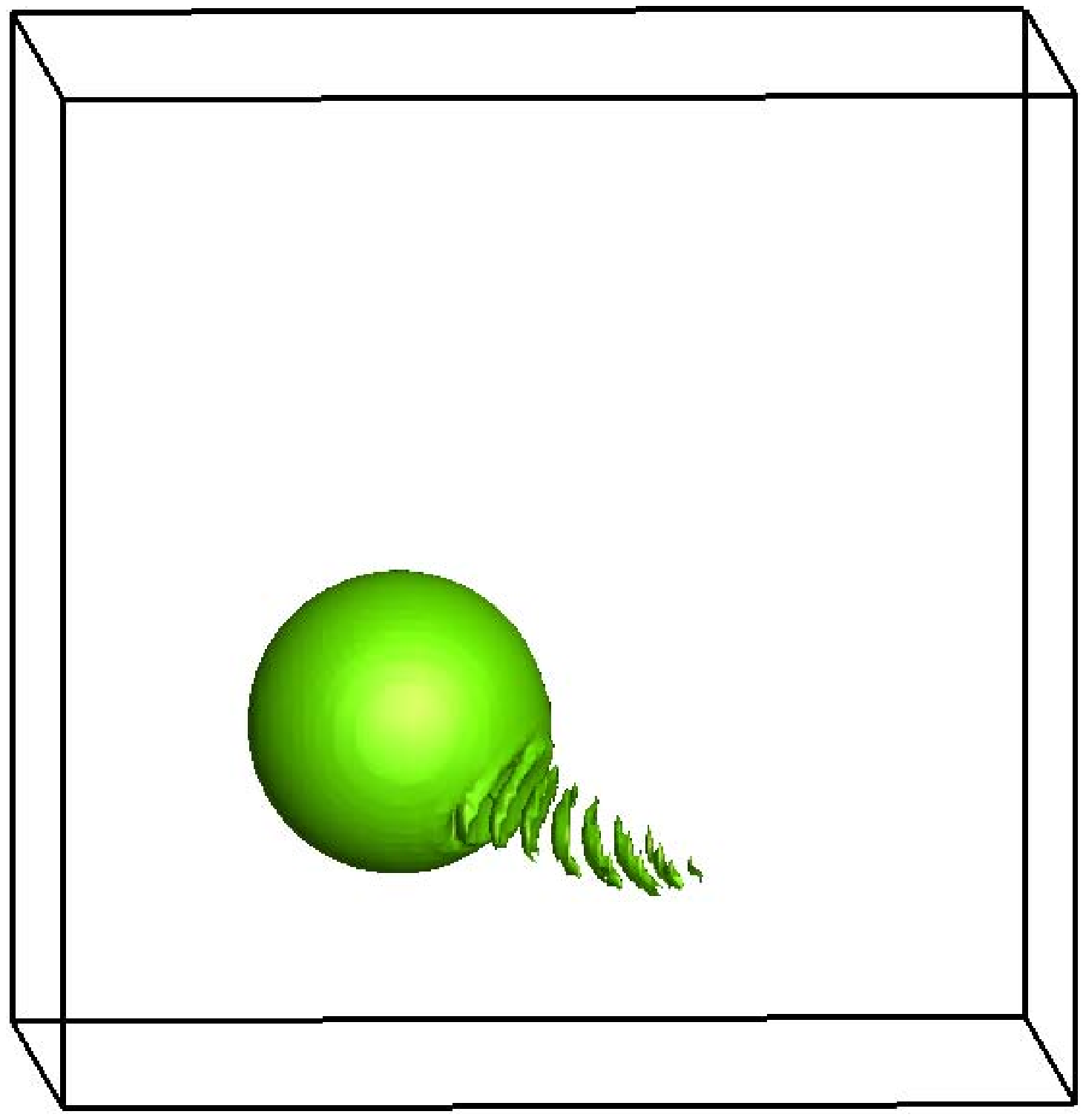}
\hspace{10pt}(b)\hspace{180pt}\\
\includegraphics[width=1.2in,height=1.24in]{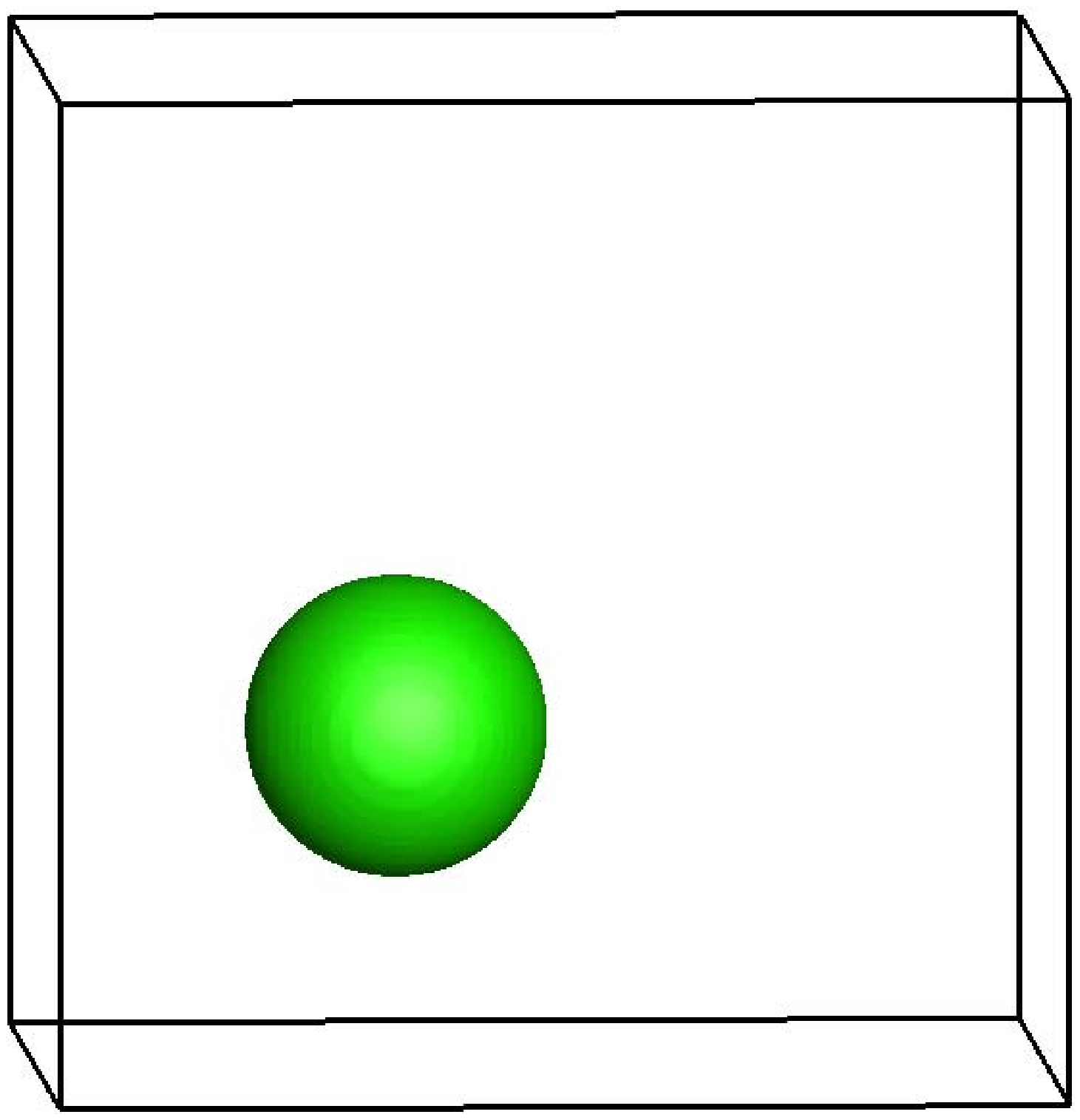}
\includegraphics[width=1.2in,height=1.24in]{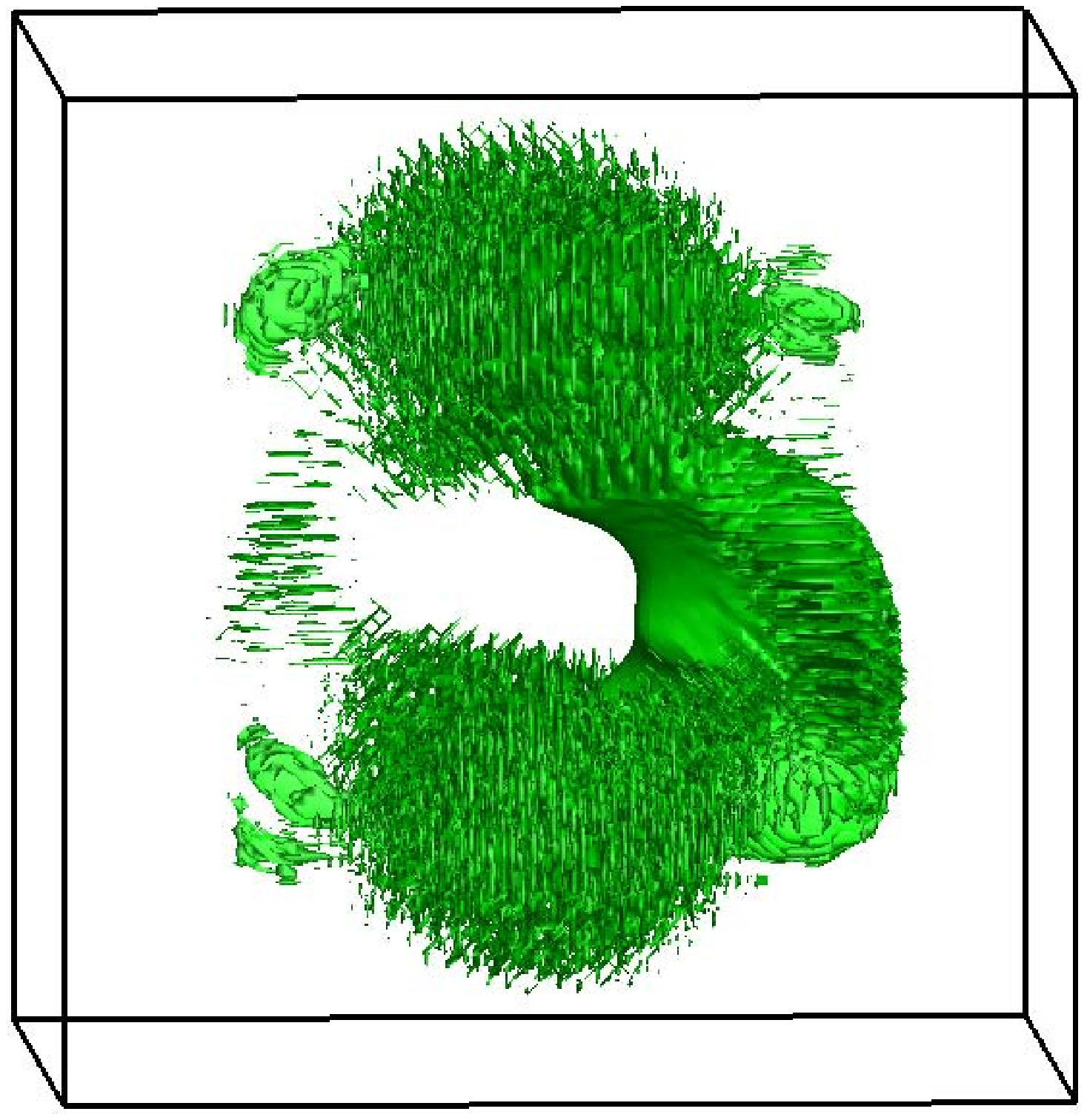}
\includegraphics[width=1.2in,height=1.24in]{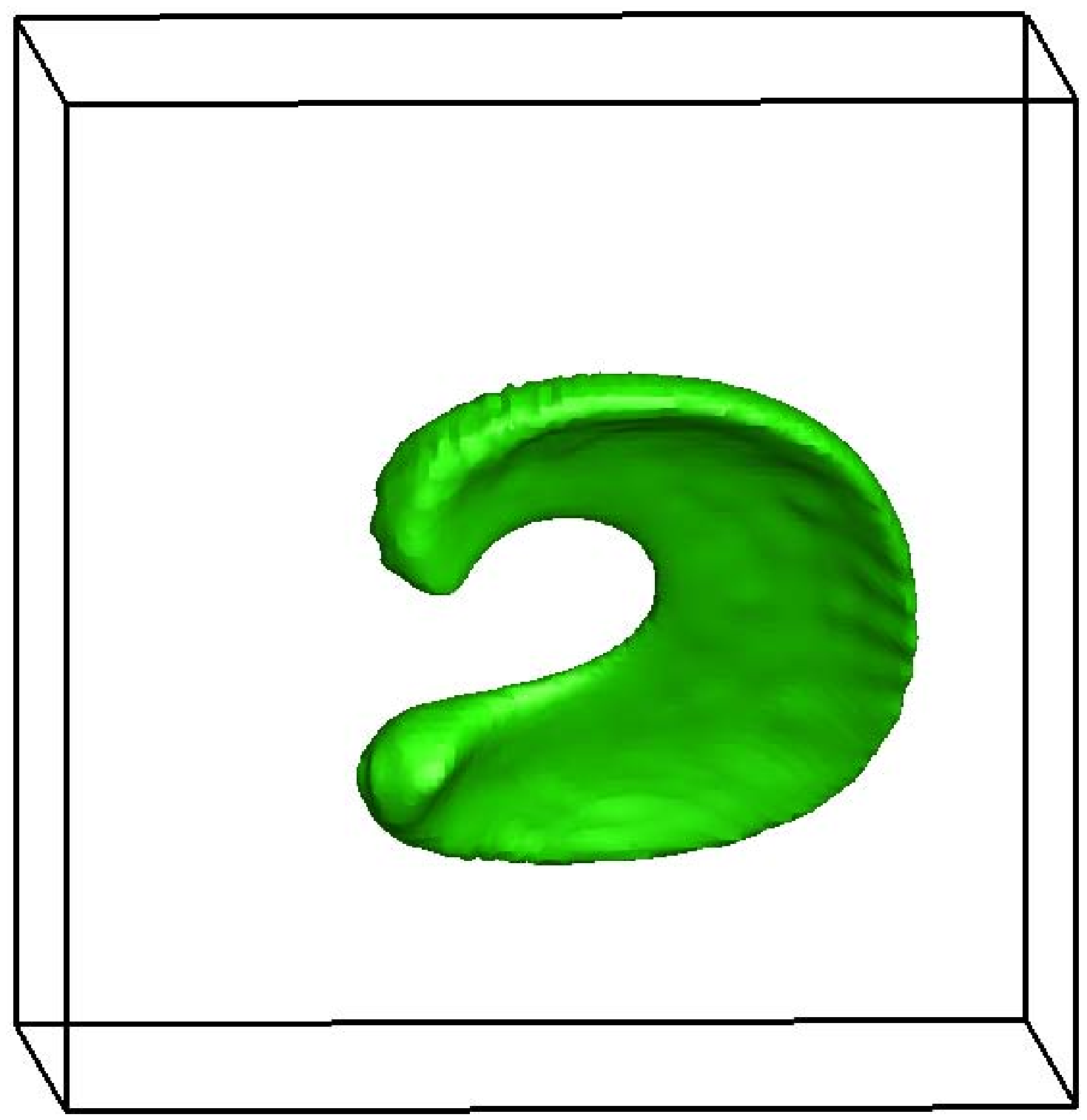}
\includegraphics[width=1.2in,height=1.24in]{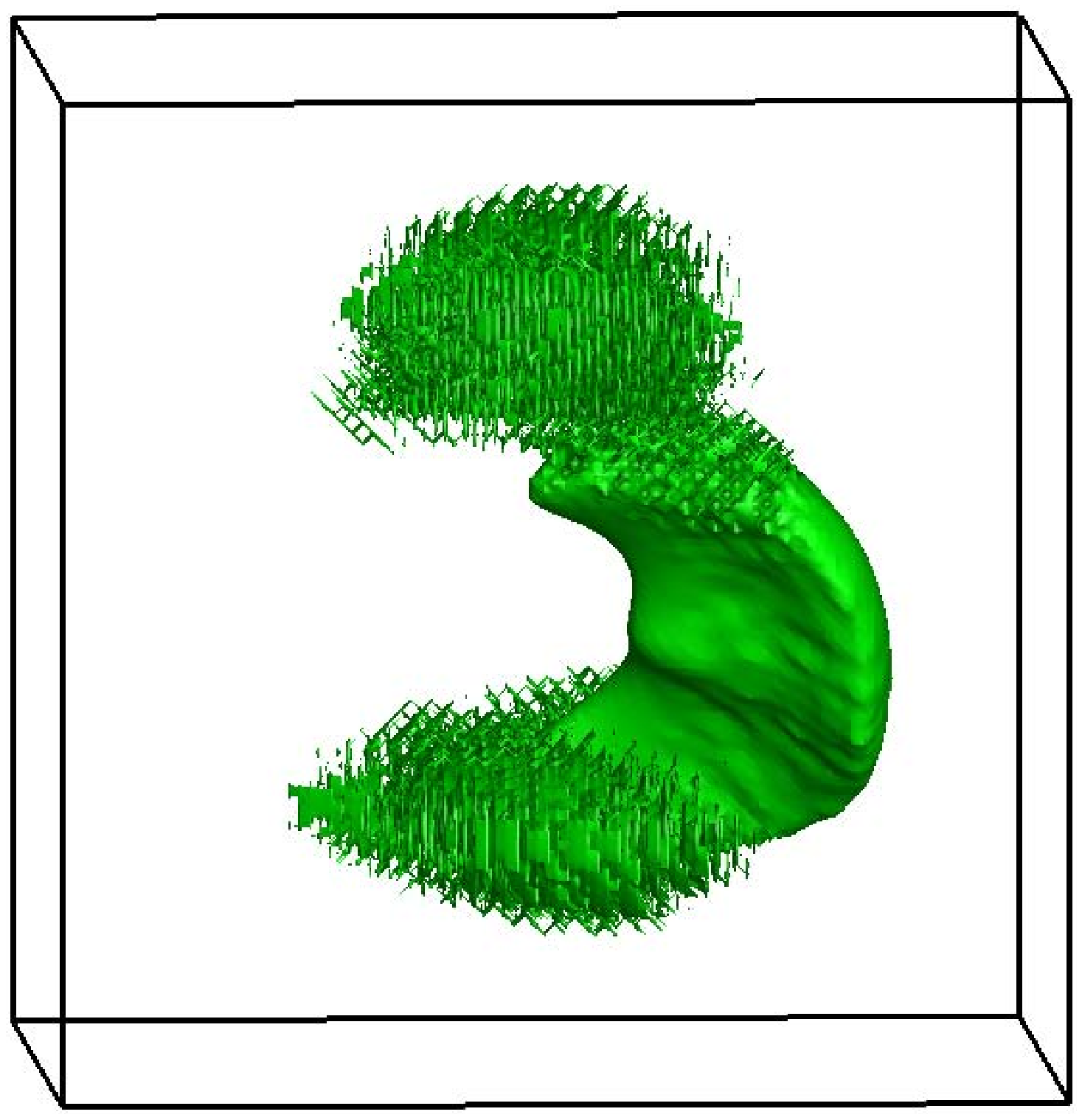}
\includegraphics[width=1.2in,height=1.24in]{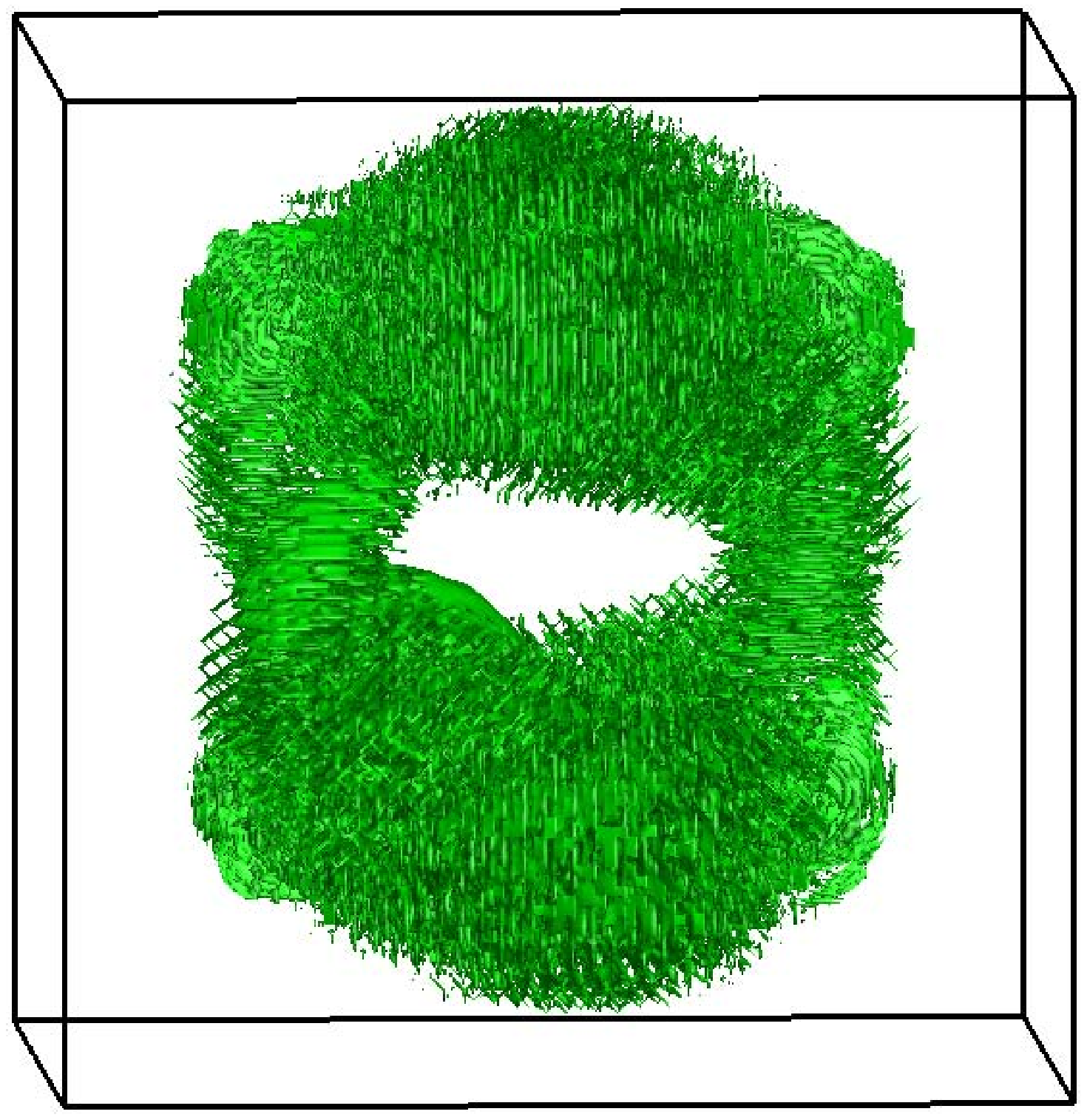}
\hspace{10pt}(c)\hspace{180pt}\\
 \tiny\caption{Snapshots of deformation field test at $Pe=400$: (a) the present MRT model; (b) the present SRT model; (c) the previous LB model~\cite{Zheng3}.
 The times from the left pattern to the right are $0$, $T/4$, $T/2$, $3T/4$ and $T$.}
\end{figure}

\begin{figure}
\centering
\includegraphics[width=4.0in,height=3.0in]{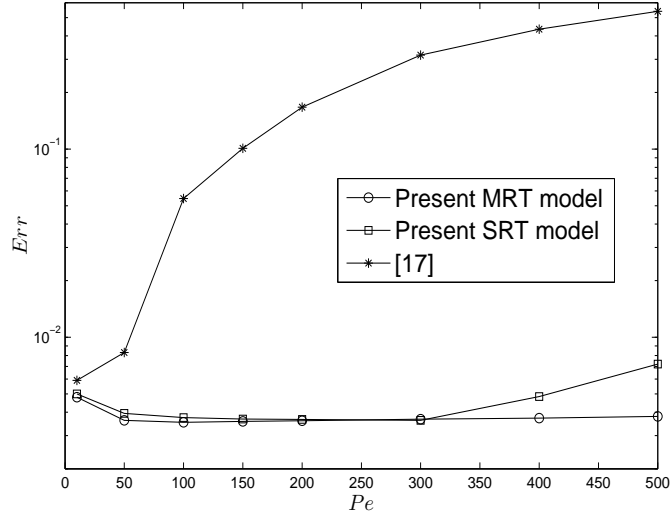}
 \tiny\caption{Comparison of the relative errors produced by different LBMs in deformation field tests with various Peclet numbers.}
\end{figure}

\subsection{Deformation field flow}
The above test does not induce large changes of the interface. To show the capacity of the present MRT model, in this subsection
we will consider a rather challenging problem of deformation field flow, in which the interface could undergo a large deformation
\cite{Enright,Leveque}. The initial setup of this problem is described as follows. A sphere with a radius of $15$ is placed in
a $100\times100\times100$ computational domain centered at $(35,~35,~35)$. The velocity field of this flow is strongly nonlinear and is given by
\begin{equation}
 \begin{split}
 u(x,y,z)&=2U{\sin }^2({\pi}x/100){\sin}(2{\pi}y/100){\sin}(2{\pi}z/100),\\
 v(x,y,z)&=-U{\sin}(2{\pi}x/100){\sin}^2({\pi}y/100){\sin}(2{\pi}z/100),\\
 w(x,y,z)&=-U{\sin}(2{\pi}x/100){\sin}(2{\pi}y/100){\sin}^2({\pi}z/100),
\end{split}
\end{equation}
and we postmultiply a time-dependent function $\cos({\pi}t/T)$ to make the flow periodic,
where $t$ is the iteration step divided by $100/U$. In our simulations, $U$ and $T$ are given as $0.02$ and $2$;
the other physical parameters and boundary condition are set as those in the last test. In theory, the sphere will undergo
the deformation continuously until time $T/2$, and the velocity field then is revised in time, the sphere goes back and returns
to its original position. Figure 4 shows the evolution of the interface pattern at $Pe=400$ obtained by the use of
the MRT model, the SRT model and the previous LB model~\cite{Zheng3}. It can be clearly observed that the present MRT model can give an
accurate prediction in the evolution of the interface: it has a largest deformation at time $T/2$ and moves back to
the initial configuration at time $T$. The behaviors of the interface conform to the theoretical results. In contrast, the
results of the SRT model and the previous three-dimensional LB model are unstable. At time $T/4$ or $T$, some jagged shapes
are produced by the SRT model in the vicinity of the interface. At the same time, the previous LB model is worst in tracking the interface.
A massive amount of unphysical disturbances can be clearly observed in the system. We also perform simulations with different Peclet numbers,
and compare the relative errors generated by these LB models. The results are presented in Fig. 5. From this figure, one can find that
the present MRT model is more accurate than the SRT model and the previous LB model.

\subsection{Rayleigh-Taylor instability}
At last, we simulated a benchmark problem of the Rayleigh-Taylor instability (RTI). RTI is a classical and common
instability phenomenon, which occurs at perturbed interface between two different fluids, where the gravity force is applied.
RTI studies are useful since it has particular relevance and importance in fields including inertial confinement fusion~\cite{Lindl} and astrophysics~\cite{Remington}.
For this reason, RTI has become a subject of intensive researches in the past 60 years using theoretical analysis~\cite{Chandrasekhar},
experimental methods~\cite{Waddell, Wilkinson} and also numerical approaches~\cite{He1, He2, Ding, Liang1, Tryggvason, Li, Celani, Wei}. However, to our best knowledge, most of previous numerical studies are limited to the two-dimensional case~\cite{He1, Ding, Liang1, Celani, Wei}, and relatively few attention has been paid on the three-dimensional RTI~\cite{He2, Tryggvason, Li}. On the other hand, the simulated Reynolds numbers considered in previous studies are small in general. In this subsection, we applied the present MRT model to study the three-dimensional Rayleigh-Taylor instability, and the effect of Reynolds number on the
evolution of the interface was examined.
\begin{figure}
\includegraphics[width=0.7in,height=2.4in]{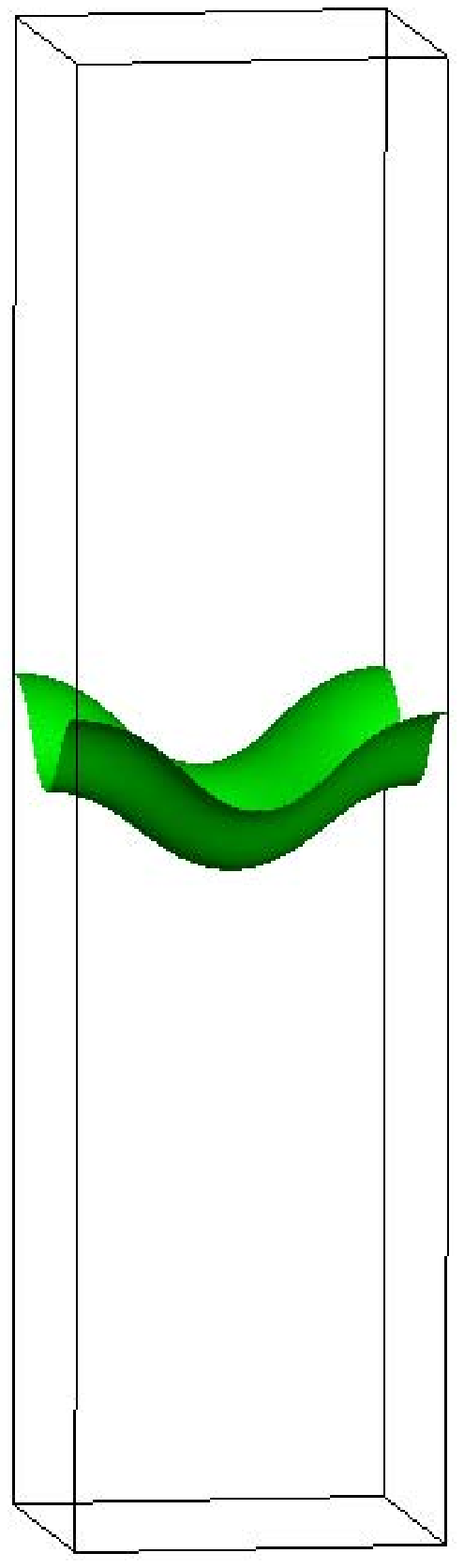}
\includegraphics[width=0.7in,height=2.4in]{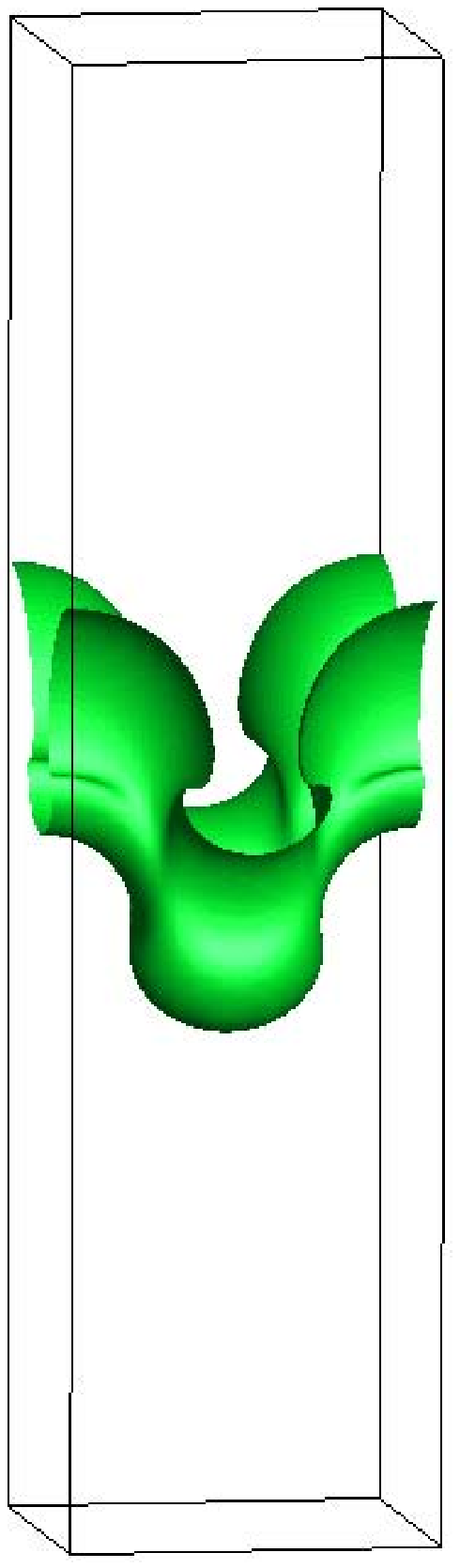}
\includegraphics[width=0.7in,height=2.4in]{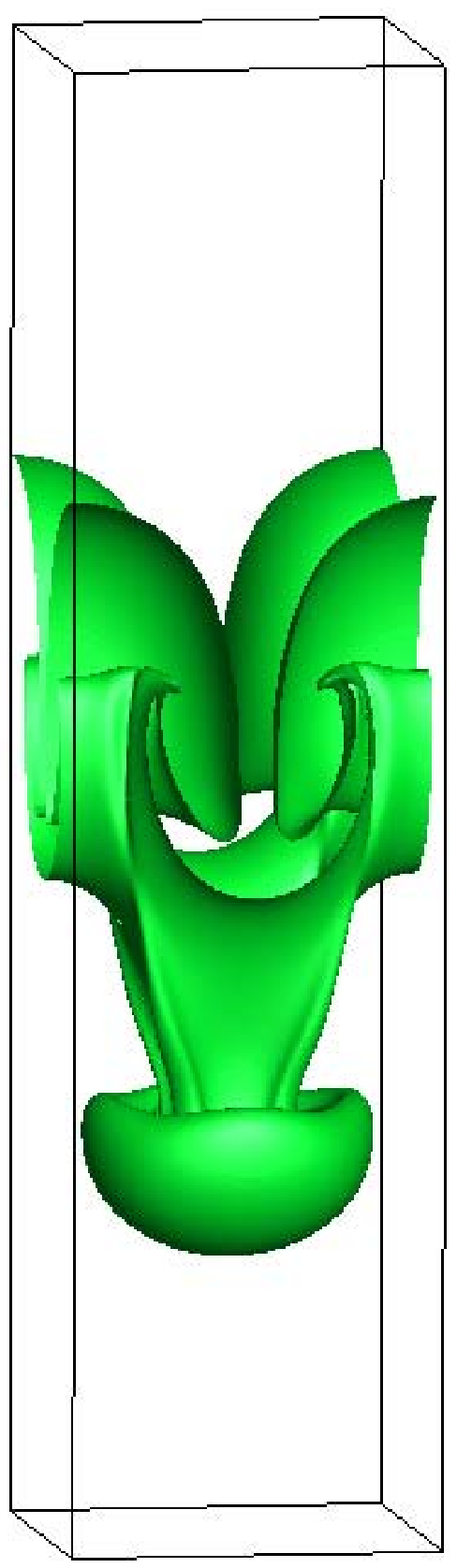}
\includegraphics[width=0.7in,height=2.4in]{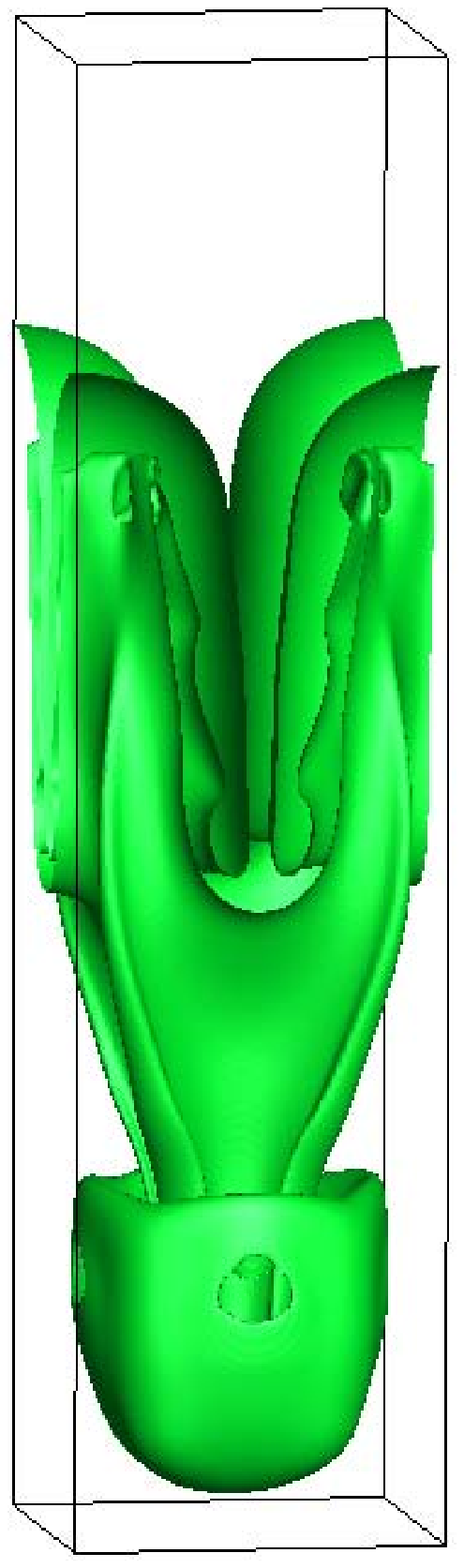}~~~~~~
\includegraphics[width=0.7in,height=2.4in]{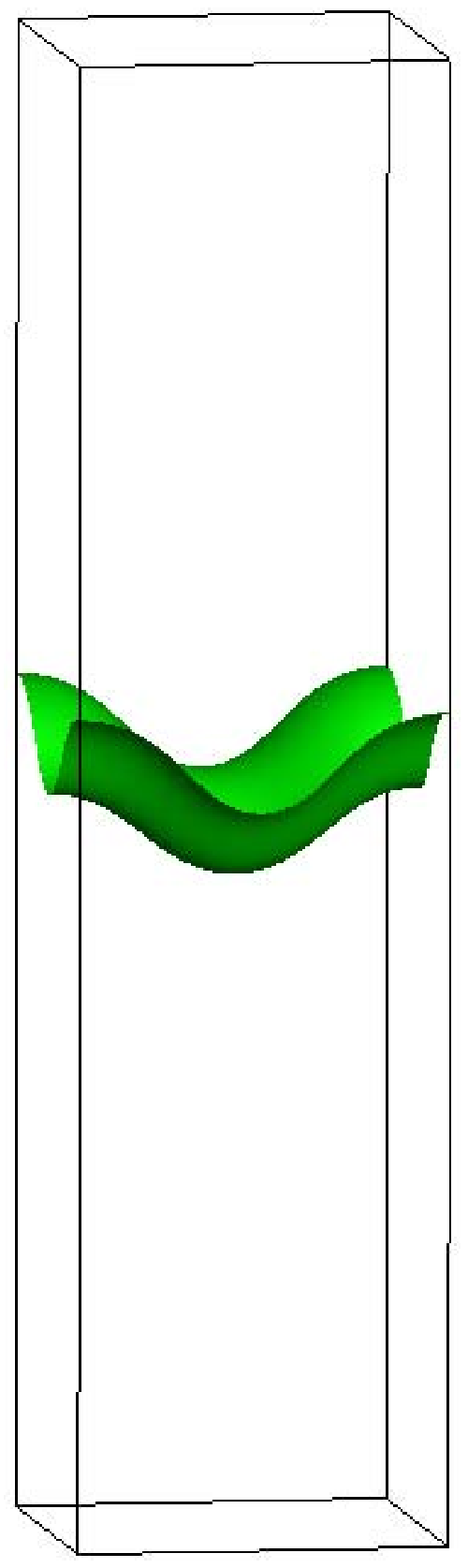}
\includegraphics[width=0.7in,height=2.4in]{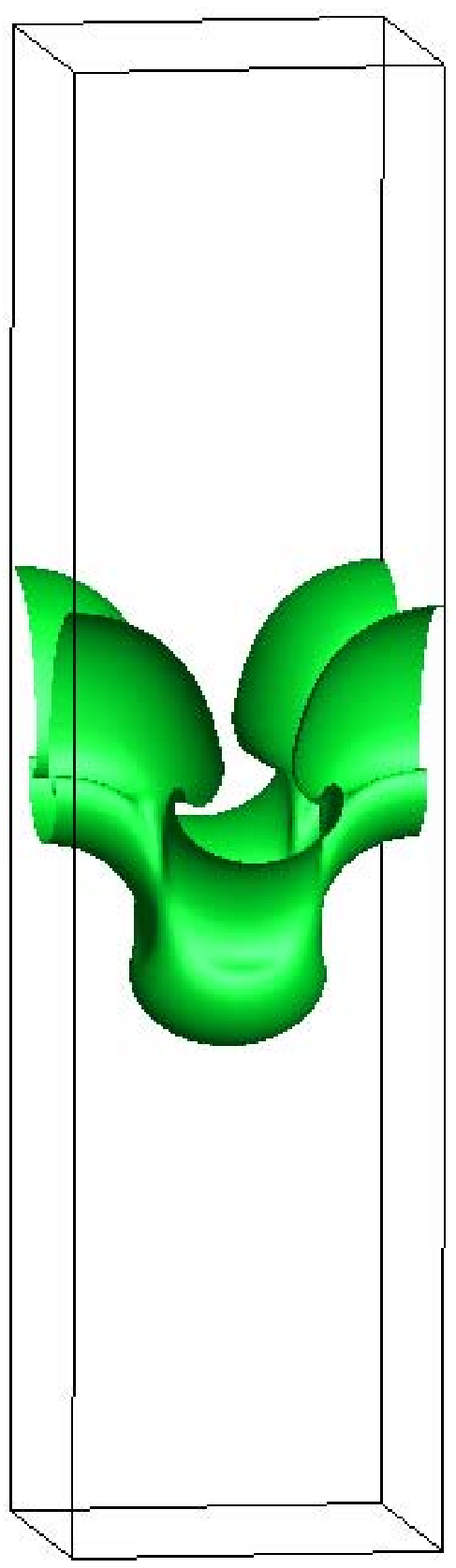}
\includegraphics[width=0.7in,height=2.4in]{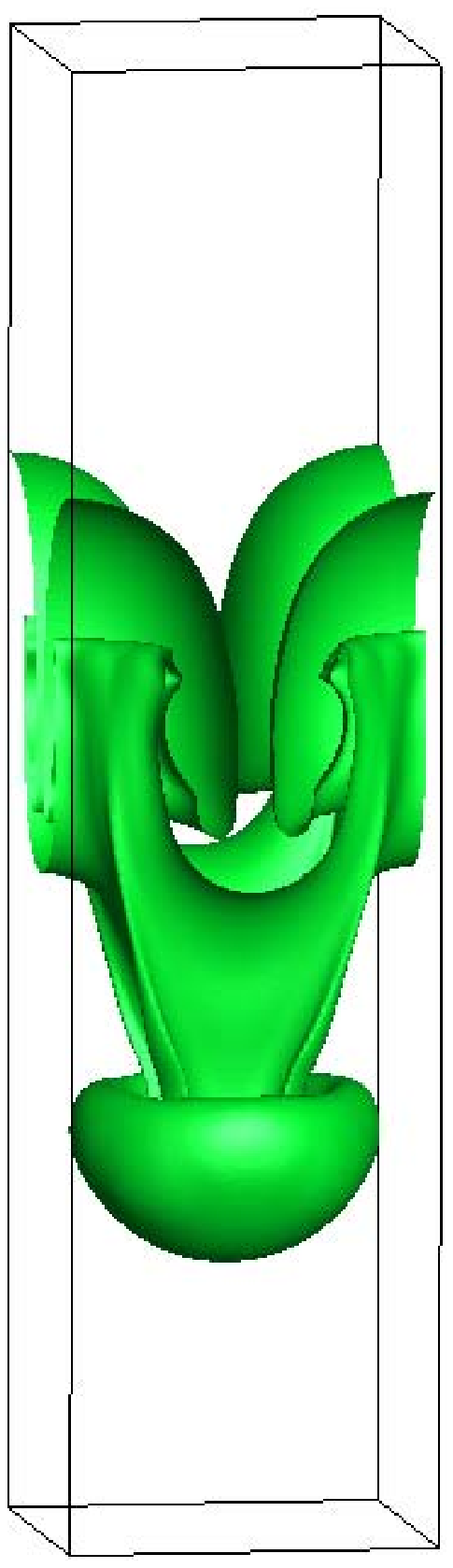}
\includegraphics[width=0.7in,height=2.4in]{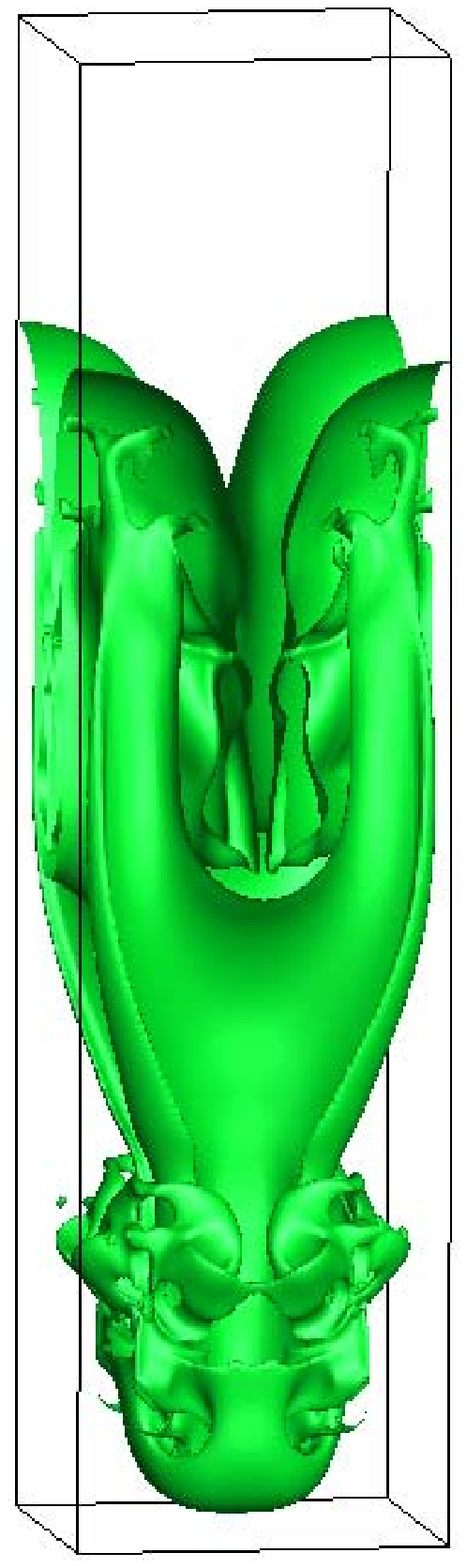}\\
\hspace{10pt}(a)\hspace{180pt}\hspace{10pt}(b)\hspace{180pt}\\
\tiny\caption{Evolution of the fluid interface in immiscible RTI at different Reynolds numbers: (a) Re=1024; (b) Re=4000. The corresponding times from the left pattern to the right are t=1.0, 2.0, 3.0, 4.0, where t is normalized by the characteristic time $\lambda/\sqrt{g\lambda}$.}
\end{figure}

The physical problem we considered here is a rectangular box with an aspect ratio of $4\lambda\times\lambda\times\lambda$, where $\lambda$ is
the box width. The initial interface is located at the midplane ($z=2\lambda$), with an imposed square-mode perturbation
\begin{equation}
h(x,y)=0.05\lambda[\cos(\frac{2\pi x}{\lambda})+\cos(\frac{2\pi y}{\lambda})],
\end{equation}
and the initial order distribution then can be given by
\begin{equation}
\phi(x,y,z)=\tanh \frac{2(z-h-2\lambda)}{D}.
\end{equation}
According to \cite{He2}, the Reynolds numbers (Re) characterizing RTI can be defined as
\begin{equation}
Re=\frac{\sqrt{g\lambda}\lambda}{\nu},
\end{equation}
where $g$ is a gravitational acceleration, and $\nu$ is the kinematic viscosity. In our simulations,
we take the densities of liquid and gas phases as 3.0 and 1.0, corresponding to an Atwood number of 0.5;
some physical parameters are given as: $\lambda=128$, $\sqrt{g\lambda}=0.04$, $\lambda=128$ $D=4.0$, $\sigma=0.0001$; the
Peclet number defined in [10,11] is fixed at 50.0; the relaxation matrix is set to be
\begin{equation}
\mathbf{S}^f=diag(1.0, 1.25, 1.25, 1.25, 1.2, 1.0, 1.0),
\end{equation}
and the relaxation factors in $\mathbf{S}^g$  are chosen to be unity expect for the ones related to the kinematic viscosity.
The periodic boundary conditions in the lateral directions and no-slip boundary condition in the vertical direction are applied in our studies.
Figure 6(a) shows the evolution of the density contours in immiscible RTI at a Reynolds number of 1024. It can be seen that, due to the gravity effect,
a heavy fluid and a light one penetrate into each other at early time, and then forms the spike and bubble, respectively.
After that, the heavy fluid rolls up along the flank of the spike, and a mushroom-like structure appears (t=3.0), which can be attributed
to Kelvin-Helmholtz instabilities providing the rolling motion of the interface. Finally, the mushroom develops further and becomes much bigger.
The patterns of the liquid interface obtained by the present model compare well with previous results~\cite{Tryggvason, He2, Zu}.
We also simulated the three-dimensional RTI at a high Reynolds number of 4000, which has not been considered in Ref. [15],
and presented the results in Fig. 6(b). It can be observed that the interface takes on almost similar behaviors before time 3.0.
However, it subsequently presents a distinct manner. Due to larger shear interaction between different layers, the interface
becomes more complex, which eventually undergoes a breakup inducing some tiny dissociative drops in the system. To observe the evolution of
the interface more clearly, we plotted in Fig. 7 the interface patterns at the diagonal vertical plane ($x=y$) with above two Reynolds numbers.
It is found that for both $Re=1024$ and $Re=4000$, the development of the initial mode follows the pattern similar from two-dimensional simulations~\cite{Liang1}.
In the following, two pairs of counter-rotating vortices are formed at the spike tip and the saddle point (see time 3.0), which is significantly
different from the two-dimensional results~\cite{Liang1}. The vortices grow with time for $Re=1024$, while they become unstable for $Re=4000$, resulting in
the mixing of two different fluids at the vicinity of the spike tip. We also gave a quantitative study on the Reynolds number effect. Figure 8 depicts the
evolution of the positions of the bubble front and spike tip obtained by the present MRT model and the previous numerical results in Ref. [15]. It is shown
that the results of $Re=1024$ obtained by our model agree well with those in Ref. [15], which verifies the numerical accuracy of the present MRT model in dealing
with complex interfacial flows. From Fig. 8, one can also find that there is no evident difference in trajectory of bubble front at two different
Reynolds numbers, while the spike tip moves slightly faster at a larger Reynolds number.

\begin{figure}
\includegraphics[width=0.7in,height=1.95in]{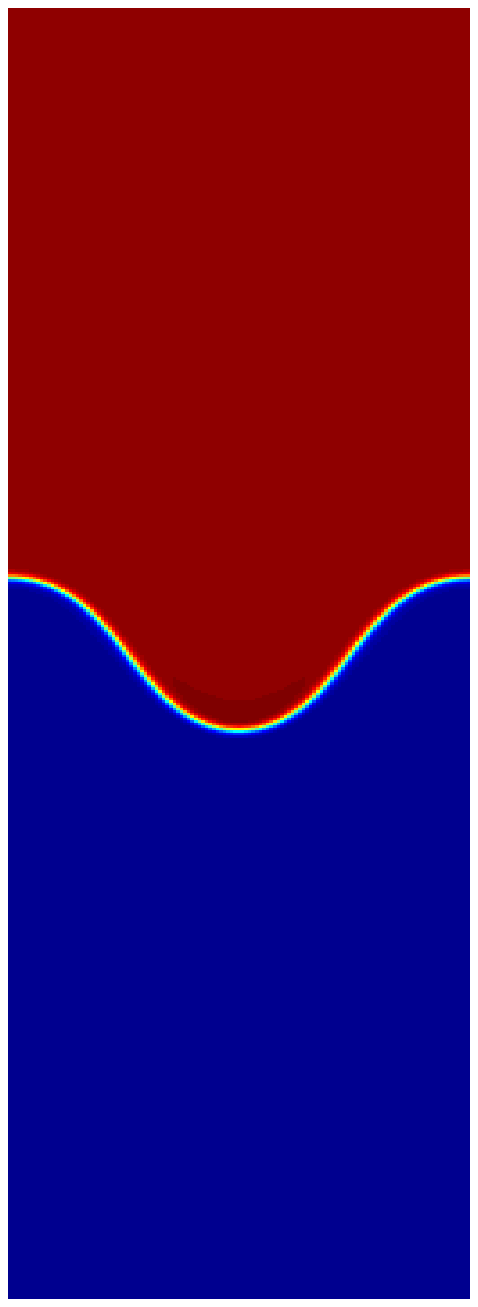}
\includegraphics[width=0.7in,height=1.95in]{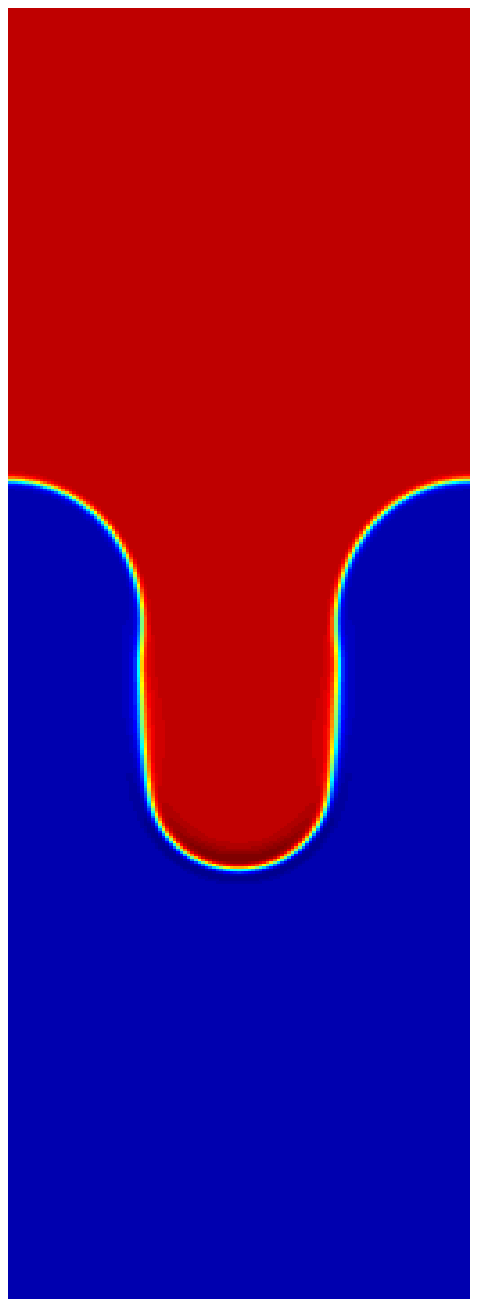}
\includegraphics[width=0.7in,height=1.95in]{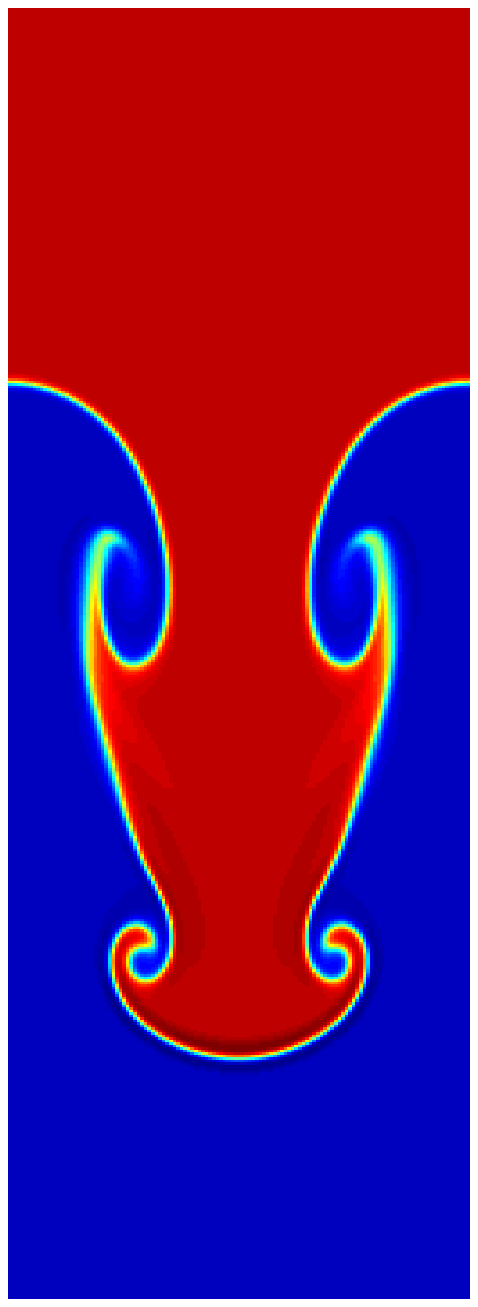}
\includegraphics[width=0.7in,height=1.95in]{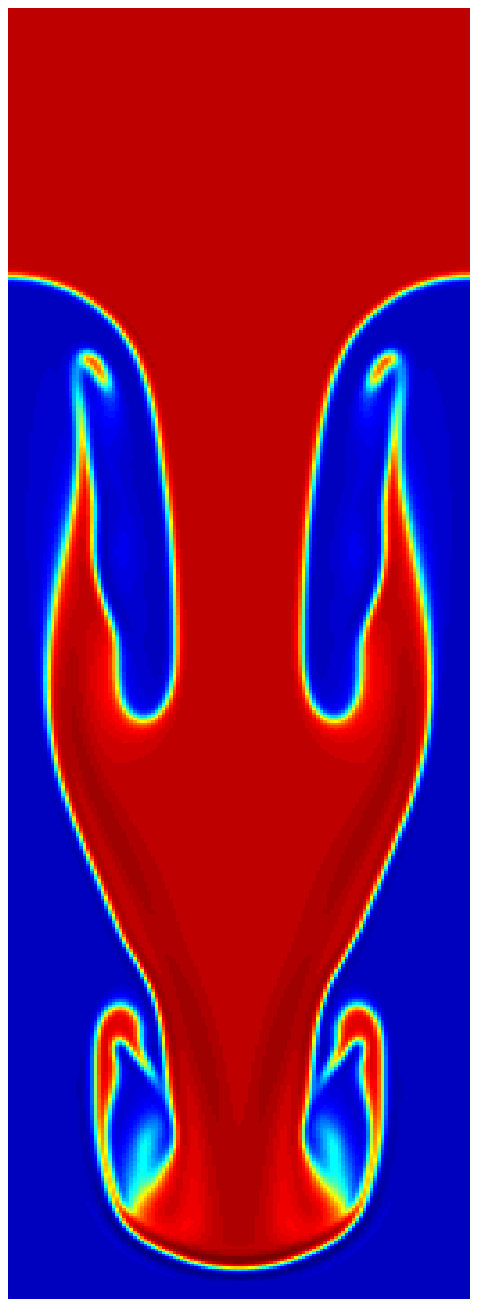}~~~~~~
\includegraphics[width=0.7in,height=1.95in]{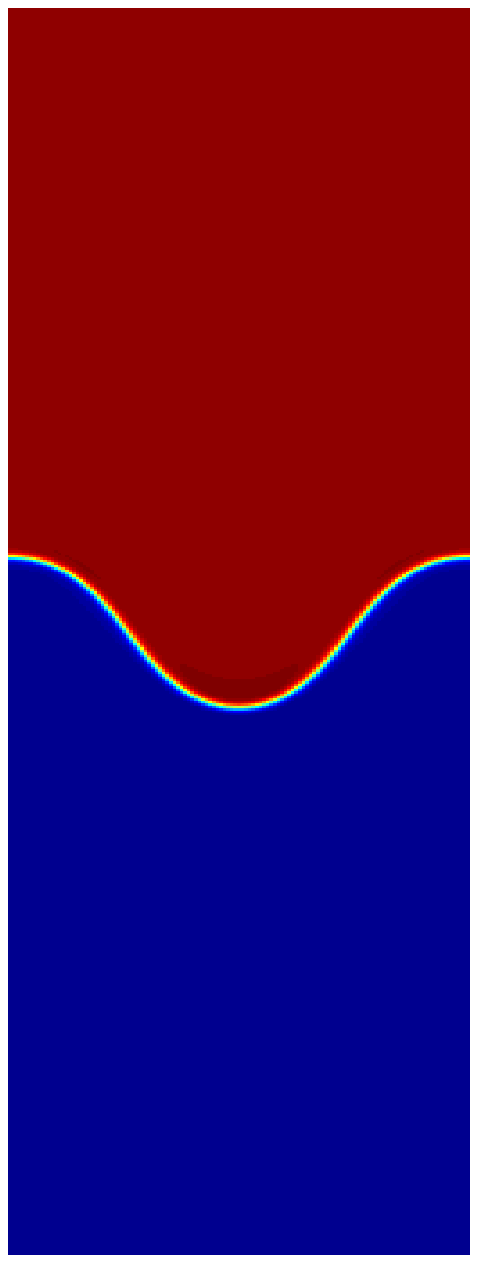}
\includegraphics[width=0.7in,height=1.95in]{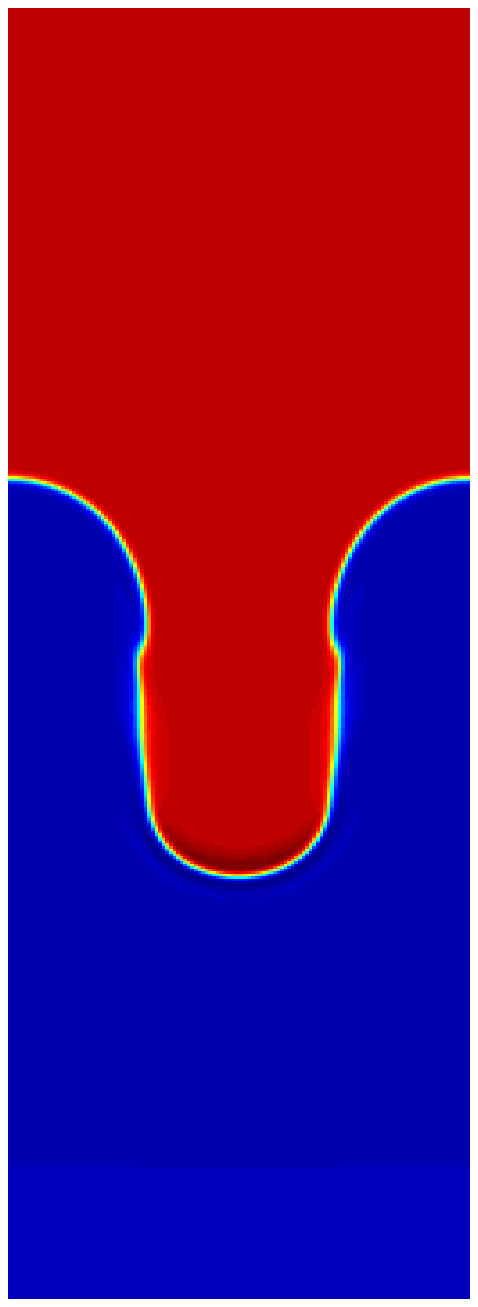}
\includegraphics[width=0.7in,height=1.95in]{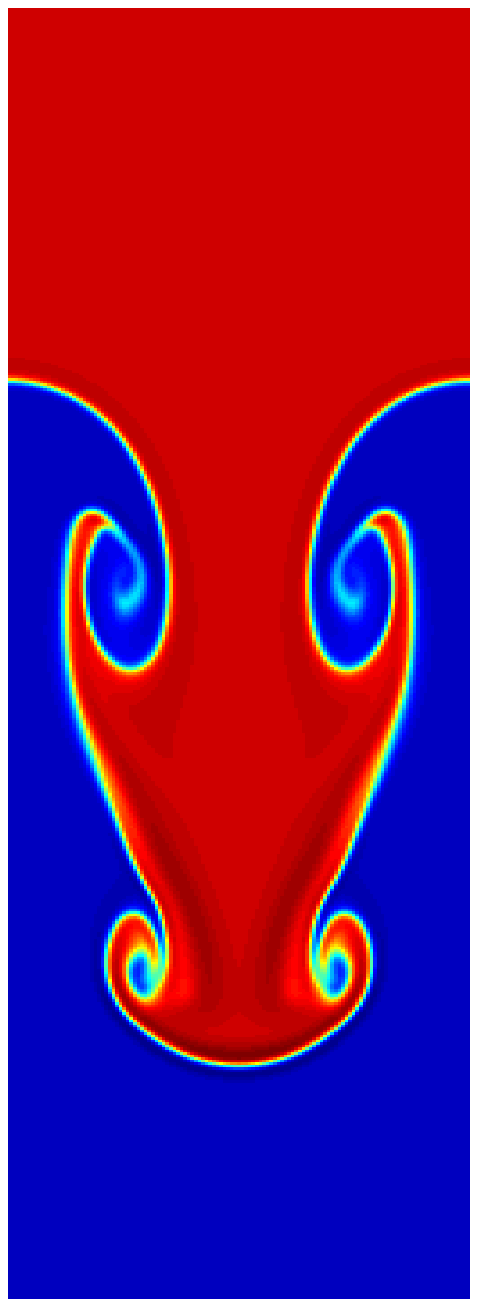}
\includegraphics[width=0.7in,height=1.95in]{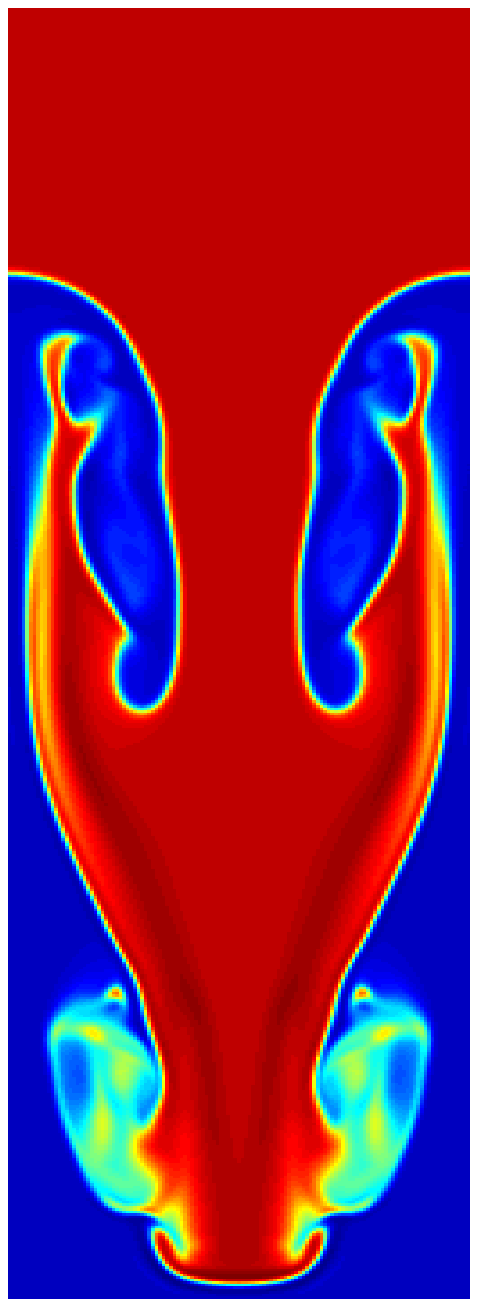}\\
\hspace{10pt}(a)\hspace{180pt}\hspace{10pt}(b)\hspace{180pt}\\
\tiny\caption{Evolution of the fluid interface at the diagonal vertical plane ($x=y$) with different Reynolds numbers: (a) Re=1024; (b) Re=4000. The corresponding times from the left pattern to the right are t=1.0, 2.0, 3.0, 4.0, where t is normalized by the characteristic time $\lambda/\sqrt{g\lambda}$.  }
\end{figure}

\begin{figure}
\centering
\includegraphics[width=4.0in,height=3.0in]{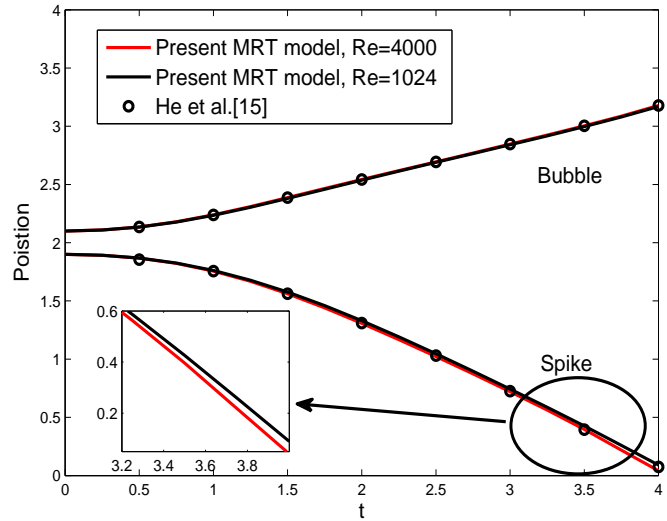}
 \tiny\caption{The time evolution of the positions of the bubble front, spike tip. The length and time are normalized by $\lambda$ and $\lambda/\sqrt{g\lambda}$, respectively.}
\end{figure}

%%%%%%%%%%%%%%%%%%%%%%%%%%%%%%%%%%%%%%%%%%%%%%%%%%%%%%%%%%%%%%%%%%%%%%%%%%%%%%%%%%%%%%%%
%%
%% Section 5 Summary
%%
%%%%%%%%%%%%%%%%%%%%%%%%%%%%%%%%%%%%%%%%%%%%%%%%%%%%%%%%%%%%%%%%%%%%%%%%%%%%%%%%%%%%%%%%
\section{Summary}\label{sec: sum}
In this work, an efficient three-dimensional lattice Boltzmann model based on the MRT collision method is proposed for multiphase flow systems. This model
is a straightforward extension of our previous two-dimensional model~\cite{Liang1} to the three-dimensions. The present model for the CHE only utilizes
seven discrete velocities in three dimensions, while most of previous LB models need at least fifteen discrete velocities. As a result, the computational
efficiency of the present model can be greatly improved in simulating three-dimensional multiphase flows. In addition, the advanced MRT collision model
is adopted in LB equations for both the CHE and the NSEs, which has a better stability than the SRT model. Two classical interface-capturing problems
including rotation of the Zalesak's sphere and deformation field flow were conducted to test the model, and the results show that the present MRT model
is more stable and accurate than the SRT model and the previous LB model in tracking the interface. Finally, the present MRT model is applied
to study the three-dimensional Rayleigh-Taylor instability at various Reynolds numbers. It is found that the numerical results at low Reynolds numbers
agree well with previous data, while the instability at a high Reynolds numbers induces a more complex structure of the interface.

%%%%%%%%%%%%%%%%%%%%%%%%%%%%%%%%%%%%%%%%%%%%%%%%%%%%%%%%%%%%%%%%%%%%%%%%%%%%%%%%%%%%%%%%
%%
%% Section 5 Summary
%%
%%%%%%%%%%%%%%%%%%%%%%%%%%%%%%%%%%%%%%%%%%%%%%%%%%%%%%%%%%%%%%%%%%%%%%%%%%%%%%%%%%%%%%%%
\section*{Acknowledgments}
This work is financially supported by the National
Natural Science Foundation of China(Grant Nos. 11272132), and the Fundamental Research Funds for the Central
Universities (Grant No. 2014TS065).

%%%%%%%%%%%%%%%%%%%%%%%%%%%%%%%%%%%%%%%%%%%%%%%%%%%%%%%%%%%%%%%%%%%%%%%%%%%%%%%%%%%%%%%%
%%
%%   Bibliography
%%
%%%%%%%%%%%%%%%%%%%%%%%%%%%%%%%%%%%%%%%%%%%%%%%%%%%%%%%%%%%%%%%%%%%%%%%%%%%%%%%%%%%%%%%%
%%%% Bibliography  %%%%%%%%%%


\begin{thebibliography}{99}

\bibitem{Guo1}
 Z. L. Guo, C. Shu, {\it Lattice Boltzmann method and its applications in engineering} (World Scientific Singapore, 2013).

\bibitem{Gunstensen}
 A.~K. Gunstensen, D.~H. Rothman, S. Zaleski, and G. Zanetti, Phys. Rev. A, 43, 4320 (1991).

\bibitem{Shan1}
 X. Shan and H. Chen,  Phys. Rev. E 47, 1815 (1993).

\bibitem{Shan2}
 X. Shan and H. Chen,  Phys. Rev. E 49, 2941 (1994).

\bibitem{Swift1}
 M. Swift, W. Osborn, and J. Yeomans, Phys. Rev. Lett. 75, 830 (1995).

\bibitem{Swift2}
 M. Swift, S. Orlandini, W. Osborn, and J. Yeomans, Phys. Rev. E 54, 5041 (1996).

\bibitem{He1}
 X. He, S. Chen, and R. Zhang, J. Comput. Phys. 152, 642 (1999).

\bibitem{Lee1}
 T. Lee and L. Liu, J. Comput. Phys. 229, 8045 (2010).

 \bibitem{Zhengl}
 L. Zheng, S. Zheng, Q. Zhai, Phys. Rev. E 91, 013309 (2015).

 \bibitem{Zu}
 Y. Q. Zu and S. He, Phys. Rev. E 87, 043301 (2013).

 \bibitem{Liang1}
 H. Liang, B. C. Shi, Z. L. Guo, Z. H. Chai, Phys. Rev. E 89, 053320 (2014).

 \bibitem{Jacqmin}
 D. Jacqmin, J. Comput. Phys. 155, 96 (1999).

 \bibitem{Ding}
 H. Ding, P. D. M. Spelt, and C. Shu, J. Comput. Phys. 226, 2078 (2007).

 \bibitem{Zheng1}
 H. W. Zheng, C. Shu, and Y. T. Chew, J. Comput. Phys. 218, 353 (2006).

\bibitem{He2}
 X. He, R. Zhang, S. Chen, and G. D. Doolen, Phys. Fluids 11, 1143 (1999).

\bibitem{Zheng2}
 H. W. Zheng, C. Shu, and Y. T. Chew, Phys. Rev. E 72, 056705 (2005).

\bibitem{Zheng3}
H. W. Zheng, C. Shu, Y. T. Chew, and J. H. Sun, Int. J. Numer. Methods Fluids 56, 1653 (2008).

 \bibitem{Liang2}
 H. Liang, Z.H. Chai, B.C. Shi, Z.L. Guo, and T. Zhang, Phys. Rev. E 90, 063311 (2014).

\bibitem{Shi1}
 B. C. Shi and Z. L. Guo, Phys. Rev. E 79, 016701 (2009).

\bibitem{Yoshida}
 H. Yoshida and M. Nagaoka, J. Comput. Phys. 229, 7774 (2010).

 \bibitem{Guo2}
 Z. L. Guo and C. G. Zheng, Int. J. Comput. Fluid Dyna. 22, 465 (2008).

 \bibitem{Qian1}
 Y. Qian, D. $\mathrm{d^{'}Humieres}$, P. Lallemand, Europhys. Lett. 17, 479 (1992).

 \bibitem{Humieres}
D. $\mathrm{d^{'}Humieres}$, I. Ginzburg, M. Krafczyk, P. Lallemand, L.S. Luo, Proc. Roy. Soc. Lond. A 360, 367 (2002).

\bibitem{Shi2}
B. C. Shi, B. Deng, R. Du, and X. W. Chen, Comput. Math. Appl. 55, 1568 (2008 ).

\bibitem{Lou}
 Q. Lou, Z. L. Guo, and B. C. Shi, Europhys. Lett. 99, 64005 (2012).

\bibitem{Enright}
D. Enright, R. Fedkiw, J. Ferziger, and I. Mitchell, J. Comput. Phys. 183, 83 (2002).

\bibitem{Leveque}
R. LeVeque, SIAM J. Numer. Anal. 33, 627 (1996).

\bibitem{Rudman}
 M. Rudman, Int. J. Numer. Methods Fluids 24, 671 (1997).

\bibitem{Lindl}
J. D. Lindl et al. , Phys. Plasmas 11, 339 (2004).

\bibitem{Remington}
 B. A. Remington, R. P. Drake, and D. D. Ryutov, Rev. Mod. Phys. 78, 755 (2006).

\bibitem{Chandrasekhar}
 S. Chandrasekhar, {\it Hydrodynamic and Hydromagnetic Stability} (Oxford University Press, Oxford, 1961).

\bibitem{Waddell}
J. T. Waddell, C. E. Niederhaus, and J. W. Jacobs, Phys. Fluids 13, 1263 (2001).

\bibitem{Wilkinson}
J. P. Wilkinson and J. W. Jacobs, Phys. Fluids 19, 124102 (2007).

\bibitem{Tryggvason}
G. Tryggvason and S. O. Unverdi, Phys. Fluids 2, 656 (1990).

\bibitem{Li}
X. L. Li, B. X. Jin, and J. Glimm, J. Comput. Phys. 126, 343 (1996).

\bibitem{Celani}
 A. Celani, A. Mazzino, P. Muratore-Ginanneschi, L. Vozella, J. Fluid Mech. 622, 115 (2009).

\bibitem{Wei}
T. Wei, and D. Livescu, Phys. Rev. E 86, 046405 (2012).

\end{thebibliography}
\end{document}